\documentclass[11pt,a4paper]{article}

\usepackage[utf8]{inputenc}
\usepackage[T1]{fontenc}
\usepackage[UKenglish]{babel}
\usepackage[autostyle=false,style=american]{csquotes}

\usepackage{amsmath,mathtools,amssymb,amsfonts,bbm,comment}
\usepackage[top=3.11cm,bottom=4.11cm,right=2.9cm,left=2.9cm]{geometry}
\usepackage{physics}
\usepackage{slashed}
\usepackage{tensor}
\usepackage{mathrsfs}
\usepackage{amsthm}
\usepackage{xcolor,graphicx}
\usepackage{bbold}
\usepackage{appendix}
\usepackage{simplewick}
\usepackage{textpos}
\usepackage{stackrel}
\numberwithin{equation}{section}
\usepackage[style=ieee, citestyle=numeric-comp, backend=biber]{biblatex}
\usepackage{orcidlink}

\usepackage{hyperref} 
\usepackage[capitalise]{cleveref}

\usepackage{tabularx}
\newcolumntype{C}{>{\centering\arraybackslash}X}
\newcolumntype{R}{>{\raggedleft\arraybackslash}X}
\newcolumntype{L}{>{\raggedright\arraybackslash}X}
\usepackage{multirow, bigdelim}

\allowdisplaybreaks

\newcommand{\Dintx}{\int d^Dx}
\newcommand{\projR}{\mathbb{P}_{\mathrm{R}}}
\newcommand{\projL}{\mathbb{P}_{\mathrm{L}}}

\newcommand{\myTrbig}[1]{\mathrm{Tr}\big(#1\big)}

\newcommand{\trYY}[2]{\mathrm{Tr}\big(Y_{#1}Y_{#2}\big)}
\newcommand{\trYbarYbar}[2]{\mathrm{Tr}\big(Y_{#1}^{\dagger}Y_{#2}^{\dagger}\big)}

\newcommand{\vgb}{\mathcal{V}}
\newcommand{\myGenLe}{T}
\newcommand{\myGenQu}{\mathcal{T}}
\newcommand{\myGenScl}{\Theta}
\newcommand{\myxi}{\xi_{F}}
\newcommand{\myNL}{N_{L}}

\usepackage{etoolbox}
\makeatletter
\newcommand{\@preprint}{}
\newcommand{\preprint}[1]{\gdef\@preprint{#1}}
\patchcmd{\@maketitle}
  {\vskip 2em}
  {\vskip -2em{\hfill\footnotesize\@preprint\par}\vskip 2em}
  {}{\PackageWarning{preprint}{Could not patch \string\@maketitle}}
\makeatother
\preprint{\texttt{DESY-26-148}}

\title{The Standard Model with Non-Anticommuting $\gamma_5$\\ in the BMHV Scheme}

\newcommand{\email}{}
\author{
Paul Kühler,$^a$\thanks{\email{paul.kuehler@tu-dresden.de}}\quad
Dominik  Stöckinger,$^a$\thanks{\email{dominik.stoeckinger@tu-dresden.de}}\quad
Matthias Weißwange\,\orcidlink{0000-0003-3852-4355}\,$^b$\thanks{\email{matthias.weisswange@desy.de}}\\[2em]
$^a$Institut für Kern- und Teilchenphysik, TU Dresden,\\ Zellescher Weg 19, 01069 Dresden, Germany\\[0.5em]
$^b$Deutsches Elektronen-Synchrotron DESY,\\ Notkestr.\ 85, 22607 Hamburg, Germany
}

\begin{document}
\thispagestyle{empty}

\maketitle

\setcounter{footnote}{0}
\vspace{2ex}
\begin{abstract}
The Breitenlohner-Maison/'t~Hooft-Veltman (BMHV) scheme  provides a treatment of $\gamma_5$ in dimensional regularisation (DReg) that is guaranteed to be self-consistent at all orders, but it breaks gauge and BRST invariance at the regularised level, which must be restored by suitable counterterms. 
We present the complete 1-loop renormalisation of the full Standard Model in the BMHV scheme in general $R_\xi$-gauge, including all singular and all finite symmetry-restoring counterterms. 
We further analyse the impact of evanescent gauge interactions, which allow the photon and gluon currents to be treated either 4- or $D$-dimensionally. 
We show that QCD always contributes to the symmetry breaking, so that the Standard Model must be treated as a whole, and we recommend a purely 4-dimensional treatment of electroweak gauge interactions. 
Our results provide a necessary step towards multi-loop electroweak precision calculations with a fully consistent treatment of $\gamma_5$.
\end{abstract}

\newpage
\setcounter{page}{2}

\tableofcontents\newpage


\section{Introduction}

Parity violation and chiral fermion--gauge boson interactions are fundamental observations of nature and defining properties of the Standard Model (SM) of particle physics. 
At the quantum level, chiral fermions give rise to rich and subtle effects, most prominently the chiral anomalies, see Refs.~\cite{Adler:1969gk,Bell:1969ts,Bardeen:1969md}, whose required cancellation imposes stringent constraints on the fermion content of any consistent chiral gauge theory.
In general, their quantum treatment is highly nontrivial.
In non-perturbative lattice formulations, the Nielsen--Ninomiya theorem~\cite{Nielsen:1980rz,Nielsen:1981xu} prevents a local, Hermitian, and doubler-free lattice regularisation of chiral fermions that preserves exact chiral symmetry.
In perturbation theory, dimensional regularisation (DReg) --- and any other regularisation scheme that features linearity, numerator-denominator consistency and shift invariance of loop momenta (see Ref.~\cite{Bruque:2018bmy}) --- is affected by the well-known $\gamma_5$-problem: as already noted in Ref.~\cite{tHooft:1972tcz}, there is no extension of $\gamma_5$ to $D$ dimensions that simultaneously preserves the cyclicity of $\gamma$-traces, the anticommutation with all $\gamma^\mu$-matrices, and the nonzero relation of traces involving $\gamma_5$ to the $\varepsilon$-tensor with correct 4-dimensional limit. 
Hence, at least one of these three properties must be abandoned in $D$ dimensions.

For reviews and further literature on the $\gamma_5$-problem and its consequences, we refer to Refs.~\cite{Jegerlehner:2000dz,Belusca-Maito:2023wah} and references therein. 
The problem is of direct phenomenological relevance, as it requires a careful application of DReg to the SM whenever electroweak interactions enter at the multi-loop level.
Example electroweak calculations with detailed discussions of the $\gamma_5$ treatment include the 2-loop electroweak corrections to 
the $W$-boson mass $M_W$~\cite{Freitas:2000gg,Freitas:2002ja,Awramik:2003ee}, 2-loop $Z$-pole observables such as the $Z b\bar{b}$ vertex and the $Z$ branching ratios~\cite{Awramik:2004ge,Awramik:2006uz,Awramik:2008gi,Freitas:2014hra}, and the electroweak 2-loop corrections to
the anomalous magnetic moment of the muon~\cite{Czarnecki:2002nt,Heinemeyer:2004yq}.
The issue also arises in the recent study of electroweak 2-loop corrections to Drell-Yan and related processes in Ref.~\cite{Freitas:2025vax}, and in higher-order calculations such as the 3-loop analysis of $M_W$ in Ref.~\cite{Dubovyk:2026nhx} and 3-loop corrections to further electroweak precision observables discussed in Refs.~\cite{Chen:2020xot,Chen:2020xzx}.
At the 4-loop level, the treatment of $\gamma_5$ has been a central challenge in the determination of the SM $\beta$-functions, see Refs.~\cite{Chetyrkin:2012rz,Zoller:2015tha,Bednyakov:2015ooa,Poole:2019txl,Poole:2019kcm,Davies:2019onf,Davies:2021mnc,Herren:2021vdk,Herren:2021yur}, where schemes with an anticommuting $\gamma_5$ lead to ambiguities that have to be resolved by additional external arguments such as Weyl consistency conditions, see also Refs.~\cite{Chen:2023lus,Chen:2024zju}.
Further discussions on $\gamma_5$ in electroweak phenomenology at future colliders can be found in Ref.~\cite{Blondel:2018mad}.

The only scheme that has been proven to be mathematically consistent at all orders of perturbation theory is the Breitenlohner-Maison/'t~Hooft-Veltman (BMHV) scheme, see Refs.~\cite{tHooft:1972tcz,Breitenlohner:1977hr}.
In this scheme full anticommutativity of $\gamma_5$ with all $\gamma^\mu$-matrices is abandoned; instead $\gamma_5$ remains a manifestly 4-dimensional object and anticommutes only with the 4-dimensional part of the $\gamma^\mu$-matrices.
Its consistency comes at the cost of manifest gauge and BRST invariance at the regularised level, which are spuriously broken by the modified algebra.
The associated Slavnov-Taylor identity must be re-established through dedicated finite symmetry-restoring counterterms.

In the series of Refs.~\cite{Belusca-Maito:2020ala,Belusca-Maito:2021lnk,Stockinger:2023ndm,Ebert:2024xpy,Kuhler:2025znv,vonManteuffel:2025swv}, a no-compromise approach to the $\gamma_5$-problem has been pursued, and the required counterterms have been determined in a range of chiral gauge theories for the BMHV scheme (for a compact overview see Ref.~\cite{Belusca-Maito:2022usw} and for detailed reviews see Refs.~\cite{Belusca-Maito:2023wah,Weisswange:2025tab}). The approach uses the standard $R_\xi$ gauge fixing and employs the regularised quantum action principle of DReg~\cite{Breitenlohner:1977hr} as a tool, and the counterterms cancel UV divergences and restore the Slavnov-Taylor identity.
In these references the BMHV scheme has been applied in generic non-Abelian chiral gauge theories at the 1-loop level~\cite{Belusca-Maito:2020ala} and at the 2-loop level~\cite{Kuhler:2025znv}. 
In Abelian chiral gauge theories, it has been examined at the 2-loop~\cite{Belusca-Maito:2021lnk}, the 3-loop~\cite{Stockinger:2023ndm} and the 4-loop level in Ref.~\cite{vonManteuffel:2025swv}, where the latter constitutes the highest-order application of the scheme to date. 
Conceptual aspects of the $D$-dimensional implementation of the BMHV scheme, in particular the treatment of fermions and evanescent gauge interactions, were analysed in Ref.~\cite{Ebert:2024xpy}. 
Further conceptual issues regarding renormalisation-group $\beta$ functions were addressed in Refs.~\cite{Belusca-Maito:2020ala,Belusca-Maito:2022wem}. For earlier references applying the same basic methodology we refer to Refs.~\cite{Martin:1999cc,Sanchez-Ruiz:2002pcf}.

In a series of parallel recent works, the BMHV scheme was also applied to chiral gauge theories quantised in background field gauge~\cite{Cornella:2022hkc,OlgosoRuiz:2024dzq,Fuentes-Martin:2025meq}, and 1-loop counterterms restoring the background field Ward identity were derived.
Additionally, the Abelian Higgs-Kibble model has been investigated at the 1-loop level in Refs.~\cite{Sanchez-Ruiz:2002pcf,Quadri:2026zwv}.

In recent years, the BMHV scheme has also increasingly been applied to effective field theories (EFTs), where chiral structures are ubiquitous, see Refs.~\cite{Naterop:2023dek,Naterop:2024cfx,Naterop:2025lzc,Naterop:2025cwg,Heinrich:2024ufp,Heinrich:2024rtg,DiNoi:2023ygk,DiNoi:2025uan,DiNoi:2025uhu,DiNoi:2025arz,Fuentes-Martin:2025meq,Solera:2026xnz,Bisht:2026qwe}, where Ref.~\cite{Solera:2026xnz} evaluates symmetry-restoring counterterms applying the same approach as Refs.~\cite{Belusca-Maito:2020ala,Belusca-Maito:2021lnk,Stockinger:2023ndm,Ebert:2024xpy,Kuhler:2025znv,vonManteuffel:2025swv}.
For the SM, the results of Refs.~\cite{OlgosoRuiz:2024dzq,Fuentes-Martin:2025meq} already provide symmetry-restoring finite counterterms valid for the chosen formulation in background field gauge and sufficient for physical 1-loop amplitudes. 

In the present work, we present the complete BMHV renormalisation of the full SM at the 1-loop level, based on the methodology of Refs.~\cite{Belusca-Maito:2020ala,Belusca-Maito:2021lnk,Stockinger:2023ndm,Ebert:2024xpy,Kuhler:2025znv,vonManteuffel:2025swv}.\footnote{%
A part of the results was already derived and presented in Refs.~\cite{Weisswange:2025tab,Weisswange:2026rya}.} This includes finite counterterms for physical amplitudes but also divergent and finite counterterms for off-shell Green functions that will be necessary for the subrenormalisation in multi-loop SM computations in the BMHV scheme.  
We work in general $R_\xi$-gauge, expressing results in terms of the gauge parameter $\xi_F=1-\xi$ (Feynman gauge corresponds to $\xi_F=0$), and parametrise DReg with $D=4-2\epsilon$.
The complete set of singular and finite symmetry-restoring counterterms is determined from off-shell Green functions.
The finite counterterms are obtained from special Green functions with a single insertion of the $\Delta$-operator, which, by the regularised quantum action principle, encodes the regularisation-induced symmetry breaking.

A central aspect where our analysis goes beyond the mentioned references is the inclusion of non-vanishing evanescent gauge interactions, which allow the vector-like photon and gluon currents to be treated either purely 4-dimensionally or fully $D$-dimensionally. 
We analyse their impact on the regularisation-induced symmetry breaking and on the counterterm structure in detail and derive a recommendation for their choice. Furthermore, left- and right-handed quark and lepton fields are combined to usual Dirac spinors as it is customary in phenomenological applications~\cite{Denner:2019vbn}, despite the expected additional complexity compared to the 1-loop discussions of e.g.\ Refs.~\cite{Belusca-Maito:2020ala,Fuentes-Martin:2025meq}.

To keep the calculation manageable, we formulate the SM in terms of gauge-field twelve\-plets $\vgb_\mu^A$ and generalised generators. 
This compact notation reduces both the number of Green functions with external adjoint indices to be computed and the number of terms generated within each Green function.
In addition, we provide a sector-wise representation of the counterterms in terms of the gauge fields $B_\mu$, $W^A_\mu$ and $G^A_\mu$ of each individual gauge group. 
It makes the structure of the gauge sectors transparent and allows for a more intuitive discussion of the results sector by sector, in particular of the dependence on the evanescent gauge interactions.

This paper is organised as follows:
In Sec.~\ref{Sec:SM-Definition-and-Setup}, we define the SM in $D$ dimensions and introduce the twelveplet formulation.
In Sec.~\ref{Sec:SymmetryBreaking_in_SM}, we discuss the possible choices for the evanescent fermion--gauge interactions and analyse the resulting tree-level symmetry breaking induced by the BMHV scheme.
Sec.~\ref{Sec:Procedure} describes the renormalisation procedure, including our method for symmetry restoration, together with its implementation and a discussion of cross-checks.
The complete 1-loop counterterm action is presented in compact form in Sec.~\ref{Sec:SM_1-Loop_Renormalisation}, and in a decomposed, sector-wise representation in Sec.~\ref{Sec:SM_1-Loop_Renormalisation_Expanded}, which also contains a detailed analysis of the impact of the evanescent gauge interactions, including a dedicated discussion of the QCD sector.
Finally, Sec.~\ref{Sec:Conclusion} contains our conclusions.


\section{Standard Model Lagrangian in the BMHV Scheme}\label{Sec:SM-Definition-and-Setup}

The Standard Model (SM) is a chiral gauge theory based on the gauge group $SU(3)_c\times SU(2)_L \times U(1)_Y $. Its complete tree-level Lagrangian can be decomposed as 
\begin{align}\label{Eq:SM-Lagrangian-Complete}
    \mathcal{L} = \mathcal{L}_\mathrm{fermion} + \mathcal{L}_\mathrm{gauge} + \mathcal{L}_\mathrm{Higgs} + \mathcal{L}_\mathrm{Yukawa} + \mathcal{L}_\mathrm{ghost+fix} + \mathcal{L}_\mathrm{ext}.
\end{align}
The following subsections will introduce each sector in detail. We work in the BMHV scheme of dimensional regularisation. For the full definition of the scheme we refer to the literature, e.g.\ the review in Ref.~\cite{Belusca-Maito:2023wah}. 
Most importantly, in this scheme any formally $D=4-2\epsilon$-dimensional quantity $a^\mu$ can be decomposed into its purely $4$-dimensional part $\bar{a}^\mu$ and the remaining $(-2\epsilon)$-dimensional, or evanescent part $\hat{a}^\mu$ as $a^\mu=\bar{a}^\mu+\hat{a}^\mu$. 
The $\gamma_5$ matrix anticommutes with purely $4$-dimensional $\overline{\gamma}^\mu$ matrices but commutes with the evanescent $\widehat{\gamma}^\mu$ matrices. 
For the SM, important relations involving the chiral projectors $\mathbb{P}_{L,R}=\frac12\left(1\mp\gamma_5\right)$ are
\begin{align}\label{Eq:PLPRrules}
    \mathbb{P}_{L}\overline{\gamma}^\mu & =
    \overline{\gamma}^\mu \mathbb{P}_{R},
    &
    \mathbb{P}_{L}\widehat{\gamma}^\mu & =
    \widehat{\gamma}^\mu \mathbb{P}_{L}
\end{align}
and similar for $(L\leftrightarrow R)$.
In the present section we introduce our notation and set up the SM Lagrangian in the BMHV scheme. 
We note that there is some freedom in the choice of the evanescent part of the Lagrangian (see also Ref.~\cite{Ebert:2024xpy}). 
We will leave some details unspecified here but discuss possible choices and their implications later in Sec.~\ref{Sec:SymmetryBreaking_in_SM}.

\subsection{Fermionic Sector}\label{Sec:LFermion}

The most critical part of the SM Lagrangian for the BMHV scheme is the fermionic sector, i.e.\ the kinetic and gauge interaction terms of quarks and leptons.
The general $D$-dimensional Ansatz for this sector may be written as
\begin{align}\label{Eq:L_fermion-in-the-SM}
\mathcal{L}_\mathrm{fermion} = 
    {\overline{l}}{}^a_I i \slashed{\partial} l_I^a - \myGenLe_{{\alpha}\beta,ab}^A {\overline{l}}{}^a_I \,\mathbb{P}_{\overline{\alpha}} {\slashed{\vgb}}{}^A \mathbb{P}_{\beta} \, l_I^b
    + {\overline{q}}{}^{i,a}_I i \slashed{\partial} q_I^{i,a} - \myGenQu_{{\alpha}\beta,ab,ij}^A {\overline{q}}{}^{i,a}_I \,\mathbb{P}_{\overline{\alpha}} {\slashed{\vgb}}{}^A \mathbb{P}_{\beta} \, q_I^{j,b}.
\end{align}

Here, the appearing fermion fields are the lepton and quark fields
\begin{align}\label{Eq:SM-Fermion-Doublets}
    l_I^{a} =
    {\begin{pmatrix}
        \nu_I\\
        e_I
    \end{pmatrix}}^a ,
    \qquad \quad
    q_I^{i,a} =
    {\begin{pmatrix}
        u^i_I\\
        d^i_I
    \end{pmatrix}}^a ,
\end{align}
where we denote generation indices by $I,J,K,\ldots\in\{1,2,3\equiv N_{f}\}$, fundamental $SU(3)_c$ colour indices by $i,j,k,\ldots\in\{1,2,3\}$ with $N_c=3$, and doublet indices of $SU(2)_L$ by $a,b,c,\ldots\in\{1,2\}$, setting $N_L=2$. 
The indices $\alpha,\beta$ denote projections onto left- or right-handed fermions, $\alpha,\beta\in\{L,R\}$ and we use the abbreviation $\overline{\alpha}=R$ if $\alpha=L$ and vice versa.
The generalised generators $\myGenLe^A$ and $\myGenQu^A$ are discussed below in detail.

The vector fields appearing in the Lagrangian are the twelve gauge fields of the SM that we collectively denote as 
$\vgb_\mu^A$ with adjoint colour indices $A,B,C,\ldots\in\{1,\ldots,12\}$. We choose the indices such that $A=1$ corresponds to $U(1)_Y$ and its gauge field $B_\mu$, $A\in\{2,3,4\}$ to $SU(2)_L$ and its gauge fields $W_\mu^A$, and $A\in\{5,\ldots,12\}$ to $SU(3)_c$ and its eight gluons $G_\mu^A$, i.e.
\begin{align}
    \left(\vgb_\mu^A\right) &= \left(B_\mu,W_\mu^1,\ldots,W_\mu^3,G_\mu^1,\ldots,G_\mu^{8}\right).
\end{align}

In the following we discuss details of the fermionic Lagrangian and its 4-dimensional and evanescent parts. For general investigations of different treatments of evanescent terms we refer to Ref.~\cite{Ebert:2024xpy}, see also Refs.~\cite{Belusca-Maito:2020ala,Belusca-Maito:2023wah,OlgosoRuiz:2024dzq,Weisswange:2025tab}.

\paragraph{Discussion of fermion kinetic terms:}

The fermion kinetic terms in Eq.~(\ref{Eq:L_fermion-in-the-SM}) are fully $D$-dimensional to guarantee $D$-dimensionally regularised propagators in Feynman diagrams. This means that each fermion field $f$ can be decomposed as $f=f_L+f_R=\mathbb{P}_{L}f+\mathbb{P}_{R}f$ and the kinetic terms have the structure
\begin{align}
    \overline{f}i \slashed{\partial} f &= \overline{f_L} i\overline{\slashed{\partial}} f_L + \overline{f_R} i\widehat{\slashed{\partial}} f_L + (L\leftrightarrow R).
\end{align}
Because of Eq.~(\ref{Eq:PLPRrules}), in the BMHV scheme the kinetic terms inevitably mix left-handed and right-handed spinors. This is at the heart of the breaking of gauge invariance in chiral gauge theories. Especially for the neutrinos this mixing also implies that the BMHV SM Lagrangian necessarily contains fictitious, sterile right-handed neutrino fields $\mathbb{P}_{R}\nu_I\equiv{\nu^\mathrm{st}_{R,I}}$. We will require that these fictitious right-handed neutrino fields do not appear anywhere in the Lagrangian except in the kinetic terms.

\paragraph{Discussion of fermion--gauge interaction terms:}

Suppressing indices, each fermion--gauge interaction current in Eq.~(\ref{Eq:L_fermion-in-the-SM}) has the structure
\begin{align}
    t_{\alpha\beta} \overline{f} \, \mathbb{P}_{\overline{\alpha}}\gamma^\mu \mathbb{P}_{\beta} \, f
    &=
    t_{LL}\overline{f_L} \, \overline{\gamma}^\mu \, f_L + t_{RL} \overline{f_R} \, \widehat{\gamma}^\mu \, f_L + (L\leftrightarrow R)    .
\end{align}
Hence the prefactors $t_{LL}$ and $t_{RR}$ correspond to the 4-dimensional part of the Lagrangian and must match their actual physical, 4-dimensional values. In contrast, the prefactors $t_{RL}$ and $t_{LR}$ correspond to evanescent interactions that can be chosen at will. We leave them unspecified at this point.

With these remarks in mind, the generalised generators $\myGenLe^A$ and $\myGenQu^A$ appearing in the lepton/quark--gauge interactions are defined as follows.
First, the physical generators for $\alpha=\beta\in\{L,R\}$ are
\begin{equation}\label{Eq:LeptonQuark-Super-GeneratorPhysical}
    \begin{aligned}
        \myGenLe^A_{{\alpha}\alpha,ab} &=
        \begin{cases}
            g_Y \, \mathcal{Y}^l_{{\alpha},ab}, \quad &A = 1\\[0.75ex]
            g_W \, t^{(A-1)}_{{\alpha},ab}, \quad &A \in \{2,3,4\}\\[0.75ex]
            0, \quad &A \in \{5,\ldots,12\}
        \end{cases}
        \\
        \myGenQu^A_{{\alpha}\alpha,ab,ij} &=
        \begin{cases}
            g_Y \, \mathcal{Y}^q_{{\alpha},ab} \, \delta_{ij}, \quad &A = 1\\[0.75ex]
            g_W \, t^{(A-1)}_{{\alpha},ab} \, \delta_{ij}, \quad &A \in \{2,3,4\}\\[0.75ex]
            g_s \, t^{(A-4)}_{s,ij} \, \delta_{ab}, \quad &A \in \{5,\ldots,12\}.
        \end{cases}
    \end{aligned}
\end{equation}
Here also the gauge couplings $g_Y$, $g_W$ and $g_s$ are introduced. Further, the 
$U(1)_Y$ hypercharges of left- and right-handed leptons and quarks are given (in doublet matrix space) by
\begin{equation}\label{Eq:SM-Hypercharges}
    \begin{alignedat}{2}
        &\mathcal{Y}^l_{L} = 
            \begin{pmatrix}
                \mathcal{Y}_l & 0\\
                0 & \mathcal{Y}_l
            \end{pmatrix} =
            \begin{pmatrix}
                -\frac{1}{2} & 0\\
                0 & -\frac{1}{2}
            \end{pmatrix},
            \qquad
        &&\mathcal{Y}^l_{R} = 
            \begin{pmatrix}
                \mathcal{Y}_{\nu} & 0\\
                0 & \mathcal{Y}_e
            \end{pmatrix} =
            \begin{pmatrix}
                0 & 0\\
                0 & -1
            \end{pmatrix},\\
        &\mathcal{Y}^q_{L} = 
            \begin{pmatrix}
                \mathcal{Y}_q & 0\\
                0 & \mathcal{Y}_q
            \end{pmatrix} =
            \begin{pmatrix}
                \frac{1}{6} & 0\\
                0 & \frac{1}{6}
            \end{pmatrix},
            \qquad
        &&\mathcal{Y}^q_{R} = 
            \begin{pmatrix}
                \mathcal{Y}_u & 0\\
                0 & \mathcal{Y}_d
            \end{pmatrix} =
            \begin{pmatrix}
                \frac{2}{3} & 0\\
                0 & -\frac{1}{3}
            \end{pmatrix},
    \end{alignedat}
\end{equation}
and the weak generators of $SU(2)_L$ and the strong generators of $SU(3)_c$ are given by
\begin{equation}\label{Eq:SM-Generators}
    \begin{aligned}
        t^{(A-1)}_L = \frac{\sigma^{(A-1)}}{2}, 
        \qquad
        &
        t^{(A-1)}_R = 0, 
        \qquad &&A\in\{2,3,4\},\\
        t^{(A-4)}_s = \frac{\lambda^{(A-4)}}{2},
        \qquad &&&A\in\{5,\ldots,12\},
    \end{aligned}
\end{equation}
where $\sigma^A$ denote the Pauli matrices, 
and $\lambda^A$ the Gell-Mann matrices.
The weak and strong generators obey the familiar Lie algebra relations
\begin{equation}\label{Eq:SM-CommutatorRelations}
    \begin{aligned}
        \big[t_L^A,t_L^B\big] = i \varepsilon^{ABC} t_L^C,
        \qquad\quad
        \big[t_s^A,t_s^B\big] = i f^{ABC} t_s^C,
    \end{aligned}
\end{equation}
where $\varepsilon^{ABC}$ denotes the totally antisymmetric Levi-Civita symbol and $f^{ABC}$ the totally antisymmetric structure constants of $SU(3)_c$.
For later use, we define the combinations
\begin{align}\label{Eq:Pauli-Combi-Matrices}
    \sigma_{12}^\pm \coloneqq \frac{\sigma^1 \pm i \sigma^2}{2},
    \qquad\quad
    \sigma_{03}^\pm \coloneqq \frac{\mathbb{1}_{2\times2} \pm  \sigma^3}{2}.
\end{align}

As mentioned before, we do not yet fix the values of the evanescent fermion--gauge interactions. 
We only introduce the following decomposition of the generalised generators appearing in evanescent interactions, i.e.\ with chirality indices $\alpha\ne\beta$,
\begin{equation}\label{Eq:LeptonQuark-Super-GeneratorEvanescent}
    \begin{aligned}
        \myGenLe^A_{LR,ab} &=
        \begin{cases}
            g_Y \, \widehat{\mathcal{Y}}^{l}_{ab}, \quad &A = 1\\[0.75ex]
            g_W \, \widehat{t}^{(A-1)}_{ab}, \quad &A \in \{2,3,4\}\\[0.75ex]
            0, \quad &A \in \{5,\ldots,12\}
        \end{cases} \ ,
        \qquad
        &
        \myGenLe^A_{RL,ab} &=
        \left(\myGenLe^A_{LR}{}^*\right)_{ba} \ ,
        \\
        \myGenQu^A_{LR,ab,ij} &=
        \begin{cases}
            g_Y \, \widehat{\mathcal{Y}}^{q}_{ab} \, \delta_{ij}, \quad &A = 1\\[0.75ex]
            g_W \, \widehat{t}^{(A-1)}_{ab} \, \delta_{ij}, \quad &A \in \{2,3,4\}\\[0.75ex]
            g_s \, \widehat{t}^{(A-4)}_{s,ij} \, \delta_{ab}, \quad &A \in \{5,\ldots,12\}
        \end{cases} \ ,
        \qquad
        &
        \myGenQu^A_{RL,ab,ij} &=
        \left(\myGenQu^A_{LR}{}^*\right)_{ba,ji} \ .
    \end{aligned}
\end{equation}
This decomposition is still general but respects hermiticity of the Lagrangian.
Choices of the matrices $\widehat{\mathcal{Y}}^{l,q}$, $\widehat{t}^A$, $\widehat{t}^A_{s}$ will be discussed in Sec.~\ref{Sec:SymmetryBreaking_in_SM}.

\subsection{Gauge Sector}

The Lagrangian of the gauge field kinetic terms can be written as
\begin{equation}
\begin{aligned}
    \mathcal{L}_\mathrm{gauge} &= - \frac{1}{4} \vgb_{\mu\nu}^A \vgb^{A,\mu\nu}\\
    &= - \frac{1}{4} \big( \partial_\mu \vgb_\nu^A - \partial_\nu \vgb_\mu^A \big)^2
    + \mathscr{C}^{ABC} \partial^\mu \vgb^{A,\nu} \vgb^{B}_{\mu} \vgb^{C}_{\nu}
    - \frac{1}{4} \mathscr{C}^{ABC} \mathscr{C}^{ADE} \vgb_\mu^B \vgb_\nu^C \vgb^{D,\mu} \vgb^{E,\nu},
\end{aligned}
\end{equation}
with the field strength tensor $\vgb_{\mu\nu}^A=\partial_\mu \vgb_\nu^A - \partial_\nu \vgb_\mu^A - \mathscr{C}^{ABC} \vgb_\mu^B \vgb_\nu^C$, expressed in terms of the generalised gauge field twelveplets with $A=1,\ldots,12$ and the generalised structure constants $\mathscr{C}^{ABC}$ defined as 
\begin{equation}\label{Eq:Super-Structure-Constants}
    \begin{aligned}
        \mathscr{C}^{ABC} = 
        \begin{cases}
            g_W \, \varepsilon^{\dot{A} \dot{B} \dot{C}}, \quad &A,B,C \in \{2,3,4\},\\
            g_s \, f^{\ddot{A} \ddot{B} \ddot{C}}, \quad &A,B,C \in \{5,\ldots,12\},\\
            0, \quad &\mathrm{else},
        \end{cases}
    \end{aligned}
\end{equation}
where $\dot{X}=X-1$ and $\ddot{X}=X-4$, for any adjoint colour index $X\in\{A,B,C,\ldots\}$.
This Lagrangian contains the usual field strength tensor terms for the individual $U(1)_Y$,  $SU(2)_L$ and $SU(3)_c$ gauge fields, written in $D$ dimensions.  
Although we work in the unbroken phase, we record here the tree-level relationship between the interaction eigenstate gauge fields and the physical photon field $A_\mu$ and the weak gauge boson fields $Z_{\mu}$ and $W_\mu^\pm$,
\begin{subequations}\label{Eq:EWSM-Gauge-Boson-Flavour-and-Mass-Eigenstates}
    \begin{align}
        W^1_\mu &= \frac{W_\mu^+ + W_\mu^-}{\sqrt{2}},\\
        W^2_\mu &= i \frac{W_\mu^+ - W_\mu^-}{\sqrt{2}},\\
        W^3_\mu &= Z_\mu \cos(\theta_W) + A_\mu \sin(\theta_W),\\
        B_\mu &= - Z_\mu \sin(\theta_W) + A_\mu \cos(\theta_W),\label{Eq:B-Boson}
    \end{align}
\end{subequations}
where $\theta_W$ denotes the weak mixing angle. For future reference we also record 
that the photon couples to leptons via the generalised charge generator
\begin{align}\label{Eq:Electric-Charge-Generator-SMLeptons}
   \mathcal{Q}_{\alpha\beta,ab}^l = \cos(\theta_W)         \myGenLe^1_{\alpha\beta,ab} 
   +\sin(\theta_W)         \myGenLe^4_{\alpha\beta,ab} \ ;
\end{align}
the photon coupling to quarks is similar but involves $\myGenQu^A$ instead of $\myGenLe^A$.
For the physical components with $\alpha\beta=LL,RR$, these expressions reproduce the known electric charges, $\mathcal{Q}_{LL,ab}^l=\mathcal{Q}_{RR,ab}^l=eQ^l_{ab}$ and similar for quarks, with
\begin{align}
    Q^l = \begin{pmatrix}
            0 & 0\\
            0 & -1
        \end{pmatrix},
    \qquad\quad
    Q^q = \begin{pmatrix}
            \frac{2}{3} & 0\\
            0 & -\frac{1}{3}
        \end{pmatrix},
\end{align}
and coupling constant $e = g_Y \cos(\theta_W) = g_W \sin(\theta_W) = \frac{g_Y g_W}{\sqrt{g_Y^2 + g_W^2}}$.

\subsection{Higgs Sector}

The scalar sector of the SM contains two complex fields $\phi_a$, forming the Higgs doublet
\begin{equation}\label{Eq:SM-Higgs-Doublet}
    \begin{aligned}
        \Phi =
        \begin{pmatrix}
            \phi_1\\
            \phi_2
        \end{pmatrix}.
    \end{aligned}
\end{equation}
The Higgs Lagrangian reads
\begin{align}
    \mathcal{L}_\mathrm{Higgs} = (D^\mu \Phi)^\dagger(D_\mu\Phi) - V(\Phi),
\end{align}
with the fully $D$-dimensional covariant derivative
\begin{align}
    D_\mu\Phi = \big( \partial_\mu + i \myGenScl^A \vgb_\mu^A \big)\Phi,
\end{align}
acting on the Higgs doublet $\Phi$ and the scalar generator $\myGenScl^A$ defined as
\begin{equation}\label{Eq:Scalar-Super-Generator}
    \begin{aligned}
        \myGenScl^A_{ab} =
        \begin{cases}
            g_Y \, \mathcal{Y}_{S} \, \delta_{ab}, \quad &A = 1\\[0.75ex]
            g_W \, t^{(A-1)}_{L,ab}, \quad &A \in \{2,3,4\}\\[0.75ex]
            0, \quad &A \in \{5,\ldots,12\}
        \end{cases},
    \end{aligned}
\end{equation}
with $\mathcal{Y}_{S} = 1/2$ and $SU(2)_L$ generators defined in Eq.~(\ref{Eq:SM-Generators}).
The Higgs potential takes the usual form
\begin{align}
    V(\Phi) = \mu^2 \Phi^\dagger \Phi + \lambda (\Phi^\dagger\Phi)^2,
\end{align}
where $\mu^2$ represents the only mass scale of the SM.
In the broken phase, $\mu^2<0$ and the Higgs boson acquires a non-vanishing vacuum expectation value $v=\sqrt{-\mu^2/\lambda}$, and the doublet can be parametrised as
\begin{align}\label{Eq:SM-Higgs-Doublet-broken-phasen}
    \Phi =
        \begin{pmatrix}
            G^{+}\\
            \frac{v+H+iG^{0}}{\sqrt{2}}
        \end{pmatrix},
\end{align}
generating the Higgs mass $M^2_H=2\lambda v^2$ along with the masses of other SM particles.
In the present work, however, we perform the analysis in the unbroken phase and therefore use the parametrisation of Eq.~\eqref{Eq:SM-Higgs-Doublet}.

\subsection{Yukawa Sector}

The Yukawa Lagrangian is given by
\begin{align}\label{Eq:TreeLevel_YukawaLagrangian}
    \mathcal{L}_\mathrm{Yukawa} = - \Big( 
    Y_{l,IJ} {\overline{l}}{}^{a}_I \phi_a \projR l_J^2 + 
    Y_{d,IJ} {\overline{q}}{}^{i,a}_I \phi_a \projR q_J^{i,2} + 
    Y_{u,IJ} {\overline{q}}{}^{i,a}_I \varepsilon_{ab} \phi_b^\dagger \projR q_J^{i,1} + 
    \mathrm{h.c.} \Big),
\end{align}
where $\varepsilon_{ab}$ denotes the totally antisymmetric Levi-Civita symbol in 2 dimensions ($\varepsilon_{12}=-\varepsilon_{21}=1$).
Here we express the Lagrangian consistently in terms of the fermion multiplets introduced in Eq.~\eqref{Eq:SM-Fermion-Doublets}; the right-handed singlets appearing in the Yukawa sector are accessed through the corresponding components of these multiplets.

In general, the Yukawa matrices are complex $3\times3$ matrices $Y'_{l}$, $Y'_{d}$, and $Y'_{u}$.
Using the freedom of field redefinitions, two of them can be diagonalised.
Without loss of generality, we assume a diagonal form of $Y_l$ and $Y_{u}$, leading to
\begin{subequations}
    \begin{align}
        Y_l &=  \mathrm{diag}(y_e,y_\mu,y_\tau),\\
        Y_u &=  \mathrm{diag}(y_u,y_c,y_t),\\
        Y_d &=  V_\mathrm{CKM} Y_d^\mathrm{diag},
    \end{align}
\end{subequations}
where we introduced the CKM-matrix $V_\mathrm{CKM}$ and $Y_d^\mathrm{diag}=\mathrm{diag}(y_d,y_s,y_b)$.

\subsection{Ghosts, Gauge Fixing and BRST Transformations}\label{Sec:Ghosts_GaugeFixing_BRST}

We use $R_\xi$-gauge fixing and introduce Faddeev-Popov ghost and antighost fields and Naka\-nishi-Lautrup auxiliary fields $c^A$, $\overline{c}^A$, $\mathcal{B}^A$ for all generators $A=1,\ldots,12$. In $D$ dimensions, the ghost and gauge-fixing Lagrangian reads
\begin{equation}\label{Eq:SM_Gauge-Fixing_and_Ghost_Lagrangian}
    \begin{aligned}
        \mathcal{L}_\mathrm{ghost+fix} &= \frac{\xi}{2} \mathcal{B}^A\mathcal{B}^A + \mathcal{B}^A\partial^\mu \vgb_\mu^A - {\overline{c}}{}^A \partial^\mu D_\mu^{AC} c^C\\
        &= - \frac{1}{2\xi} (\partial^\mu\vgb^A_\mu)^2 - {\overline{c}}{}^A \partial^\mu D_\mu^{AC} c^C\\
        &= - \frac{1}{2\xi} (\partial^\mu\vgb^A_\mu)^2 - {\overline{c}}{}^A \Box c^A - \mathscr{C}^{ABC} (\partial^\mu {\overline{c}}{}^A) \vgb^B_\mu c^C,
    \end{aligned}
\end{equation}
where $D_\mu^{AC}$ is the covariant derivative in adjoint representation, and where in the second line the Nakanishi–Lautrup auxiliary field has been integrated out according to $\mathcal{B}^A= - \frac{1}{\xi}\partial^\mu \vgb_\mu^A$.
We also introduce the modified gauge parameter $\xi_F \coloneqq 1 - \xi$, such that Feynman gauge corresponds to $\xi_F=0$.
This choice is made for computational convenience, as it reduces the number of terms generated by the gauge boson propagator.

The BRST transformations of the SM in $D$ dimensions are given by
\begin{subequations}\label{Eq:SM-BRST-Transformations}
    \begin{align}
        s_D \vgb_\mu^A &= (D_\mu c)^A = \partial_\mu c^A - \mathscr{C}^{ABC} \vgb_\mu^B c^C,\\
        s_D l^a_{R,I} &= - i c^A \myGenLe_{R,ab}^A l^b_{R,I},\\
        s_D l^a_{L,I} &= - i c^A \myGenLe_{L,ab}^A l^b_{L,I},\\
        s_D q^{i,a}_{R,I} &= - i c^A \myGenQu_{R,ab,ij}^A q^{j,b}_{R,I},\\
        s_D q^{i,a}_{L,I} &= - i c^A \myGenQu_{L,ab,ij}^A q^{j,b}_{L,I},\\
        s_D \phi_a &= - i c^A \myGenScl^A_{ab} \phi_b,\\
        s_D c^A &= \frac{1}{2} \mathscr{C}^{ABC} c^B c^C,\\
        s_D \overline{c}^A &= \mathcal{B}^A = - \frac{1}{\xi} \partial^\mu \vgb_\mu^A,\\
        s_D \mathcal{B}^A &= 0,
    \end{align}
\end{subequations}
where the generators are defined in Eqs.~\eqref{Eq:LeptonQuark-Super-GeneratorPhysical} and \eqref{Eq:Scalar-Super-Generator}, and the structure constants are given in Eq.~\eqref{Eq:Super-Structure-Constants}.

As in general Yang-Mills theories (see Refs.~\cite{Belusca-Maito:2020ala,Kuhler:2025znv}), the BRST transformations of the SM receive nontrivial quantum corrections that are subject to Slavnov-Taylor identities that can be broken and require symmetry-restoring counterterms.
Hence, as is customary (see Refs.~\cite{Piguet:1995er,Belusca-Maito:2023wah}) we couple the BRST transformations in Eq.~\eqref{Eq:SM-BRST-Transformations} to external sources, yielding
\begin{equation}\label{Eq:SM-L_ext-BRST-Trafos-coupled-to-sources}
    \begin{aligned}
        \mathcal{L}_{\mathrm{ext}} &= 
        \rho^{\mu}_A s_D \vgb^A_{\mu} 
        + {\overline{R}}{}^{a}_{l,I} s_D {l}^a_I + (s_D{\overline{l}}{}^a_I) R^{a}_{l,I}
        + {\overline{R}}{}^{i,a}_{q,I} s_D {q}^{i,a}_I + (s_D{\overline{q}}{}^{i,a}_I) R^{i,a}_{q,I}\\
        &+ {\Upsilon^{a}}^{\dagger} s_D \phi_{a} + \Upsilon^{a} s_D \phi_{a}^{\dagger}
        + \zeta_A s_D c^A .        
    \end{aligned}
\end{equation}
Note that we do not couple the BRST variation of the antighost field ${\overline{c}}{}^A$ to a source, i.e.\ we omit a term of the form $\chi_A s_D {\overline{c}}{}^A$, as also discussed in Ref.~\cite{Piguet:1995er}.


\section{Evanescent Interactions and Tree-Level Symmetry Breaking}\label{Sec:SymmetryBreaking_in_SM}

The consistent treatment of $\gamma_5$ in the BMHV scheme leads to a spurious breaking of gauge and BRST symmetry.
This breaking originates solely from the modified $\gamma$-algebra exemplified by Eq.~\eqref{Eq:PLPRrules}, the evanescent part of the fermion kinetic term and the evanescent gauge interactions.
Accordingly, only the fermionic part $\mathcal{L}_\mathrm{fermion}$ (see Eq.~\eqref{Eq:L_fermion-in-the-SM}) contributes to the regularisation-induced symmetry breaking at tree-level, while the rest of the Lagrangian in Eq.~\eqref{Eq:SM-Lagrangian-Complete} remains manifestly BRST invariant in $D$ dimensions.
Here we first provide a more detailed discussion of possible choices of evanescent interactions and then evaluate the resulting tree-level symmetry breaking.

\subsection{Possible Choices of Evanescent Fermion Interactions}\label{Sec:PossibleChoices_of_EvanescentFermions}

As mentioned before, the choice of the evanescent part of the Lagrangian is not unique.
A very extensive discussion of possible choices has been given in Ref.~\cite{Ebert:2024xpy} for a simplified model. 
Our treatment of the kinetic terms corresponds to Option~1 of that reference, where only physical fermion fields are used (except the unavoidable fictitious right-handed neutrino field ${\nu^\mathrm{st}_{R,I}}$ mentioned in Sec.~\ref{Sec:LFermion}). 
For the fermion--gauge interactions, we have expressed the evanescent part through the generalised generators in Eq.~\eqref{Eq:LeptonQuark-Super-GeneratorEvanescent} but not yet fixed their values.
Ref.~\cite{Ebert:2024xpy} found that setting all evanescent interactions to zero yields  the most compact results and thus appears as the preferred choice.

In the Standard Model, however, the situation is more subtle.
Given that electromagnetism and the strong interaction are vector-like, it is natural to ask whether the photon and gluon interactions can be treated in a fully $D$-dimensional way, like in the cases of typical computations in pure QED or pure QCD, and whether such a treatment is advantageous despite the presence of the other gauge bosons with chiral interactions and the mixing between the photon and the $Z$ boson. 
To achieve a fully $D$-dimensional treatment, all four chiral components of the fermionic currents must couple to the corresponding gauge boson through identical generators.

For the strong interaction, this condition is readily satisfied by identifying the corresponding evanescent generator with the strong generator, i.e.\ by setting $\widehat{t}_s^A=t_{s}^A$ in Eq.~\eqref{Eq:LeptonQuark-Super-GeneratorEvanescent}.
The electromagnetic case is less straightforward: here, the electric charge generator must be identical for all chiral covariants, i.e.\ equal to the electric charge itself, leading to the condition
$\mathcal{Q}_{LL,ab}^f  =
    \mathcal{Q}_{RR,ab}^f
    =
    \mathcal{Q}_{LR,ab}^f
    =
    \mathcal{Q}_{RL,ab}^f
    \overset{!}{=}
        eQ^f_{ab}$ for $ f\in\{l,q\}$
in terms of the generalised charge generator of Eq.~\eqref{Eq:Electric-Charge-Generator-SMLeptons}. Consequently, a fully $D$-dimensional photon coupling to all fermions requires the condition
\begin{align}\label{Eq:SM-Condition-on-Electric-Charge-GeneratorLeptonQuark}
\widehat{\mathcal{Y}}^{f}_{ab} + \widehat{t}^3_{ab}
    \overset{!}{=}
        Q^f_{ab},
    \qquad f\in\{l,q\}
\end{align}
for the evanescent hypercharge and weak generators.

Additional insight follows from considering the evanescent interactions of neutrinos.
Using Eq.~\eqref{Eq:LeptonQuark-Super-GeneratorEvanescent} in Eq.~\eqref{Eq:L_fermion-in-the-SM}, the resulting evanescent contribution in the leptonic sector is provided by
\begin{align}\label{Eq:Evanescent-Neutrino-Lepton-Interactions}
          g_W \widehat{t}^{A}_{11} \overline{\nu_{L}}_{,I} {\widehat{\slashed{W}}}{}^A {\nu^\mathrm{st}_{R,I}}  
        + g_W \widehat{t}^{A}_{12} \overline{\nu_{L}}_{,I} {\widehat{\slashed{W}}}{}^A {e_{R,I}} 
        + g_W \widehat{t}^{A}_{21} \overline{e_{L}}_{,I} {\widehat{\slashed{W}}}{}^A {\nu^\mathrm{st}_{R,I}} 
        + g_W \widehat{t}^{A}_{22} \overline{e_{L}}_{,I} {\widehat{\slashed{W}}}{}^A {e_{R,I}}
        + \mathrm{h.c.}
\end{align}
By construction, the sterile right-handed neutrinos ${\nu^\mathrm{st}_{R,I}}$ must not participate in any interaction.
Any interaction term involving them must therefore vanish identically, which fixes the evanescent generator in the lepton sector to the form
\begin{align}\label{Eq:Evanescent-SU2-Generator-Lepton-Sector}
    \widehat{t}^A =
    \begin{pmatrix}
        0 & \widehat{t}^A_{12}\\
        0 & \widehat{t}^A_{22}
    \end{pmatrix}.
\end{align}
Although evanescent generators are not required to form a representation of a Lie algebra, matrices of this form fail to satisfy the $\mathfrak{su}(2)$ commutation relations, and thus unnecessarily complicate computations.
To avoid this, we make the simplifying choice $\widehat{t}^A\equiv0$.
With this choice, Eq.~\eqref{Eq:SM-Condition-on-Electric-Charge-GeneratorLeptonQuark} then requires the evanescent hypercharges to be equal to the right-handed ones, i.e.\ $\widehat{\mathcal{Y}}^{f}_{ab}=\mathcal{Y}^f_{R,ab}$.

In the present work, we do not pursue the most general form of evanescent gauge interactions studied for the Abelian case in Ref.~\cite{Ebert:2024xpy}. 
Instead, in view of the previous discussion, we adopt the following parametrisation:
\begin{equation}\label{Eq:Evanescent-Generators-Choice}
    \begin{aligned}
        \widehat{\mathcal{Y}}^{f}_{ab} &= c_{\mathrm{QED}} \, \mathcal{Y}^f_{R,ab}, \qquad f\in\{l,q\},\\
        \widehat{t}^A_{ab} &\equiv 0,\\
        \widehat{t}^A_{s,ij} &= c_{\mathrm{QCD}} \, t^A_{s,ij},
    \end{aligned}
\end{equation}
with two parameters $c_{\mathrm{QED}}, c_{\mathrm{QCD}} \in \mathbb{R}$.
This allows to explore the effect of treating QED and/or QCD fully 4- or $D$-dimensionally, and also to interpolate between these cases. 
Further, the above choice fulfils the hermiticity relations $\big({\widehat{\mathcal{Y}}^{f}}{}^{\dagger}\big) = \widehat{\mathcal{Y}}^{f}$ and $\big({\widehat{t}^A_{s}}{}^{\dagger}\big)=\widehat{t}^A_{s}$, leading to $\myGenLe^A_{LR}=\myGenLe^A_{RL}=\widehat{\myGenLe}^A$ and $\myGenQu^A_{LR}=\myGenQu^A_{RL}=\widehat{\myGenQu}^A$.

To be more explicit, we provide the following expressions for the gauge interaction terms, obtained by inserting Eq.~\eqref{Eq:Evanescent-Generators-Choice} into Eq.~\eqref{Eq:L_fermion-in-the-SM} and transforming the gauge bosons to mass eigenstates according to Eq.~\eqref{Eq:EWSM-Gauge-Boson-Flavour-and-Mass-Eigenstates}:
\begin{itemize}
    \item photon interaction current
    \begin{equation}\label{Eq:Photon-Interaction-Current-SM}
        \begin{aligned}
            &- e Q^l_{ab} {\overline{l}}{}^{a}_I \slashed{A} l_I^{b} + (1-c_\mathrm{QED}) e Q^l_{ab} {\overline{l}}{}^{a}_I \widehat{\slashed{A}} l_I^{b}\\
            &- \, e Q^q_{ab} {\overline{q}}{}^{i,a}_I \slashed{A} q_I^{i,b} + (1-c_\mathrm{QED}) e Q^q_{ab} {\overline{q}}{}^{i,a}_I \widehat{\slashed{A}} q_I^{i,b},
        \end{aligned}
    \end{equation}
    \item $Z_\mu$ interaction current
    \begin{equation}\label{Eq:Z-Interaction-Current-SM}
        \begin{aligned}
            &\frac{s_W}{c_W} e Q^l_{ab} {\overline{l}}{}^{a}_I \slashed{Z} l_I^{b}
            -\frac{g_W}{c_W} t^3_{L,ab} {\overline{l}}{}^{a}_I \overline{\slashed{Z}} \mathbb{P}_{\mathrm{L}} l_I^{b}
            -(1-c_\mathrm{QED})\frac{s_W}{c_W} e Q^l_{ab} {\overline{l}}{}^{a}_I \widehat{\slashed{Z}} l_I^{b}\\
            +\,&\frac{s_W}{c_W} e Q^q_{ab} {\overline{q}}{}^{i,a}_I \slashed{Z} q_I^{i,b}
            -\frac{g_W}{c_W} t^3_{L,ab} {\overline{q}}{}^{i,a}_I \overline{\slashed{Z}} \mathbb{P}_{\mathrm{L}} q_I^{i,b}
            -(1-c_\mathrm{QED})\frac{s_W}{c_W} e Q^q_{ab} {\overline{q}}^{i,a}_I \widehat{\slashed{Z}} q_I^{i,b},
        \end{aligned}
    \end{equation}
    \item $W_\mu^\pm$ interaction current
    \begin{equation}\label{Eq:W-Interaction-Current-SM}
        \begin{aligned}
            &- \frac{g_W}{\sqrt{2}} (t^1_{L,ab}+it^2_{L,ab}) \, {\overline{l}}{}^{a}_I {\overline{\slashed{W}}}{}^+ \mathbb{P}_{\mathrm{L}} l_I^{b}
            - \frac{g_W}{\sqrt{2}} (t^1_{L,ab}-it^2_{L,ab}) \, {\overline{l}}{}^{a}_I {\overline{\slashed{W}}}{}^- \mathbb{P}_{\mathrm{L}} l_I^{b}\\
            &- \frac{g_W}{\sqrt{2}} (t^1_{L,ab}+it^2_{L,ab})\, {\overline{q}}{}^{i,a}_I {\overline{\slashed{W}}}{}^+ \mathbb{P}_{\mathrm{L}} q_I^{i,b}
            - \frac{g_W}{\sqrt{2}} (t^1_{L,ab}-it^2_{L,ab}) \, {\overline{q}}{}^{i,a}_I {\overline{\slashed{W}}}{}^- \mathbb{P}_{\mathrm{L}} q_I^{i,b},
        \end{aligned}
    \end{equation}
    \item gluon interaction current
    \begin{equation}\label{Eq:Gluon-Interaction-Current-SM}
        \begin{aligned}
            - g_s t^A_{s,ij} {\overline{q}}{}^{i,a}_I {\slashed{G}}{}^A q_I^{j,a} + (1-c_\mathrm{QCD}) g_s t^A_{s,ij} {\overline{q}}{}^{i,a}_I {\widehat{\slashed{G}}}{}^A q_I^{j,a}.
        \end{aligned}
    \end{equation}
\end{itemize}
Two special cases stand out as particularly noteworthy.
Choosing $c_{\mathrm{QED}}=c_{\mathrm{QCD}}=0$ sets all gauge interactions to purely 4 dimensions.
Setting $c_{\mathrm{QED}}=1$ and/or $c_{\mathrm{QCD}}=1$, promotes the photon and/or gluon interaction current, respectively, to fully $D$-dimensional objects, reproducing the standard treatment of QED/QCD in naive DReg.
In contrast, the $Z_\mu$ interaction current cannot be brought into a fully $D$-dimensional form, since it contains a purely 4-dimensional piece $\propto t_{L,ab}^3 \overline{\gamma}^\mu \projL$;
similarly, the $W_\mu^\pm$ interaction currents remain intrinsically 4-dimensional in the parametrisation adopted here.
In fact, no choice of evanescent generators can render the $Z_\mu$ or $W_\mu^\pm$ currents fully $D$-dimensional, because their axial parts admit no uniform continuation to $D$ dimensions (see Ref.~\cite{Ebert:2024xpy}).

\subsection{Regularisation-Induced Breaking of the Slavnov-Taylor Identity}\label{Sec:TreeLevelSymBreaking}

A meaningful physical interpretation of the SM requires that unphysical indefinite-norm states decouple and a unitary, gauge independent time-evolution of physical states can be established. 
The pivotal technical requirement is the Slavnov-Taylor identity. 
It expresses BRST invariance on the level of renormalised Green functions and implies the required physical properties of the theory (see the general reviews Refs.~\cite{Becchi:1975nq,Kugo:1979gm,Piguet:1980nr,Piguet:1995er,Belusca-Maito:2023wah} and references therein, and for discussions specific to the SM see Refs.~\cite{Aoki:1982ed,Kraus:1997bi,Grassi:1999nb,Bohm:2001yx}). 
In the context of dimensional regularisation it can be written as
\begin{align}\label{Eq:UltimateSymmetryRequirement}
    \mathop{\mathrm{LIM}}_{D \, \to \, 4} \, (\mathcal{S}_D(\Gamma_\mathrm{DRen})) = 0.
\end{align}
Here $\Gamma_\mathrm{DRen}$ denotes the effective action, equivalently the generating functional of 1PI Green functions, after renormalisation, i.e.\ evaluated including counterterms, but still in $D$ dimensions, and the limit $\mathop{\text{LIM}}_{D \, \to \, 4}$ sets $D\to4$ and drops all evanescent quantities. 
The Slavnov-Taylor operator acting on a functional $\mathcal{F}$ of the fields and external sources is defined as
\begin{align}
\mathcal{S}_D(\mathcal{F}) = \Dintx \Bigg( \frac{\delta \mathcal{F}}{\delta K_i(x)} \frac{\delta \mathcal{F}}{\delta \phi_i(x)} + \mathcal{B}^A(x) \frac{\delta \mathcal{F}}{\delta \overline{c}^A(x)} \Bigg),
\end{align}
where the sum over all fields $\phi_i$ with non-linear BRST transformations and their external sources $K_i$ is implied (cf.\ Eqs.~\eqref{Eq:SM-BRST-Transformations} and \eqref{Eq:SM-L_ext-BRST-Trafos-coupled-to-sources}). 
The antighost $\overline{c}^A$ transforms linearly, $s_D\overline{c}^A=\mathcal{B}^A$, and its variation is deliberately not coupled to a source, so it enters explicitly through the second term.

By construction of the external-field part of the classical Lagrangian in Eq.~\eqref{Eq:SM-L_ext-BRST-Trafos-coupled-to-sources}, the Slavnov-Taylor operator applied onto the classical action reproduces its BRST transformation, $\mathcal{S}_D(\Dintx\mathcal{L})=s_D\Dintx\mathcal{L}$, with the BRST transformations given in Eq.~\eqref{Eq:SM-BRST-Transformations}. 

As discussed above, the BMHV scheme induces a breaking of gauge and BRST invariance, leading to a violation of the associated Slavnov-Taylor identity.
At tree-level, the breaking of BRST invariance only results from the behaviour of evanescent terms in the fermionic Lagrangian.
Hence, the tree-level breaking of the Slavnov-Taylor identity in $D$ dimensions in the SM is given by
\begin{align}\label{Eq:SM-General-BRST-Breaking-Tree-Level}
        \widehat{\Delta} &= \mathcal{S}_D\left(\Dintx\mathcal{L}\right) = s_D  \Dintx \, \mathcal{L}_\mathrm{fermion} 
        \nonumber\\
        &= - \Dintx \, c^A \,
        \Bigg\{
        {\overline{l}}{}^a_I 
        \bigg[
        \projR 
        \bigg(
        \myGenLe^A_{R,ab} \overset{\leftarrow}{\widehat{\slashed{\partial}}}
        + \myGenLe^A_{L,ab} \overset{\rightarrow}{\widehat{\slashed{\partial}}}
        - \myGenLe^A_{LR,ab} \Big( \overset{\leftarrow}{\widehat{\slashed{\partial}}} 
                            + \overset{\rightarrow}{\widehat{\slashed{\partial}}} \Big)
        \bigg)
        \projR
        \nonumber\\
        &\phantom{
        = \Dintx \, c^A \; \{ {\overline{l}}{}^a_I [
        } 
        + \projL 
        \bigg(
        \myGenLe^A_{R,ab} \overset{\rightarrow}{\widehat{\slashed{\partial}}}
        + \myGenLe^A_{L,ab} \overset{\leftarrow}{\widehat{\slashed{\partial}}}
        - \myGenLe^A_{RL,ab} \Big( \overset{\leftarrow}{\widehat{\slashed{\partial}}} 
                            + \overset{\rightarrow}{\widehat{\slashed{\partial}}} \Big)
        \bigg)
        \projL
        \bigg]
        l_I^b
        \nonumber\\
        &\phantom{
        = \Dintx \, c^A \,\,
        } 
        + {\overline{q}}{}^{i,a}_I 
        \bigg[
        \projR 
        \bigg(
        \myGenQu^A_{R,ab,ij} \overset{\leftarrow}{\widehat{\slashed{\partial}}}
        + \myGenQu^A_{L,ab,ij} \overset{\rightarrow}{\widehat{\slashed{\partial}}}
        - \myGenQu^A_{LR,ab,ij} \Big( \overset{\leftarrow}{\widehat{\slashed{\partial}}} 
                            + \overset{\rightarrow}{\widehat{\slashed{\partial}}} \Big)
        \bigg)
        \projR
        \nonumber\\
        &\phantom{
        = \Dintx \, c^A \; \{ {\overline{l}}{}^a_I [
        } 
        + \projL 
        \bigg(
        \myGenQu^A_{R,ab,ij} \overset{\rightarrow}{\widehat{\slashed{\partial}}}
        + \myGenQu^A_{L,ab,ij} \overset{\leftarrow}{\widehat{\slashed{\partial}}}
        - \myGenQu^A_{RL,ab,ij} \Big( \overset{\leftarrow}{\widehat{\slashed{\partial}}} 
                            + \overset{\rightarrow}{\widehat{\slashed{\partial}}} \Big)
        \bigg)
        \projL
        \bigg]
        q_I^{j,b}
        \Bigg\}
        \nonumber\\
        &\phantom{
        = 
        } 
        - i \Dintx \, c^A \,
        \Bigg\{
        {\overline{l}}{}^a_I
        \bigg[
        \Big(
        \myGenLe^A_{R,ac}\myGenLe^B_{RL,cb} - \myGenLe^B_{RL,ac}\myGenLe^A_{L,cb} - i \mathscr{C}^{ABC} \myGenLe^C_{RL,ab}
        \Big)
        \projL {\widehat{\slashed{\vgb}}}{}^B \projL
        \nonumber\\
        &\phantom{
        = i \Dintx \, c^A \{ {\overline{l}}{}^a_I [
        } 
        + \Big(
        \myGenLe^A_{L,ac}\myGenLe^B_{LR,cb} - \myGenLe^B_{LR,ac}\myGenLe^A_{R,cb} - i \mathscr{C}^{ABC} \myGenLe^C_{LR,ab}
        \Big)
        \projR {\widehat{\slashed{\vgb}}}{}^B \projR
        \bigg]
        l_I^b
        \nonumber\\
        &\phantom{
        i \Dintx \, c^A \,\, \{ 
        } 
        + {\overline{q}}{}^{i,a}_I 
        \bigg[
        \Big(
        \myGenQu^A_{R,ac,ik}\myGenQu^B_{RL,cb,kj} - \myGenQu^B_{RL,ac,ik}\myGenQu^A_{L,cb,kj} - i \mathscr{C}^{ABC} \myGenQu^C_{RL,ab,ij}
        \Big)
        \projL {\widehat{\slashed{\vgb}}}{}^B \projL
        \nonumber\\
        &\phantom{
        = i \Dintx \, c^A \{ {\overline{l}}{}^a_I [
        } 
        + \Big(
        \myGenQu^A_{L,ac,ik}\myGenQu^B_{LR,cb,kj} - \myGenQu^B_{LR,ac,ik}\myGenQu^A_{R,cb,kj} - i \mathscr{C}^{ABC} \myGenQu^C_{LR,ab,ij}
        \Big)
        \projR {\widehat{\slashed{\vgb}}}{}^B \projR
        \bigg]
        q_I^{j,b}
        \Bigg\}
        \nonumber\\
        &\eqqcolon \widehat{\Delta}_1\big[c,\overline{l},l\big] + \widehat{\Delta}_1\big[c,\overline{q},q\big] + \widehat{\Delta}_2\big[c,\vgb,\overline{l},l\big] + \widehat{\Delta}_2\big[c,\vgb,\overline{q},q\big],
\end{align}
where the evanescent generators are still left unspecified, i.e.\ not yet fixed according to Eq.~\eqref{Eq:Evanescent-Generators-Choice}.

No choice of the evanescent generators can make this breaking vanish entirely (see also the related discussion in Ref.~\cite{Ebert:2024xpy}).
However, fixing the evanescent generators according to Eq.~\eqref{Eq:Evanescent-Generators-Choice}, and decomposing Eq.~\eqref{Eq:SM-General-BRST-Breaking-Tree-Level} into its $U(1)_Y$, $SU(2)_L$ and $SU(3)_c$ components --- with gauge bosons $B_\mu$, $W_\mu^A$, and $G_\mu^A$ and associated ghost fields $c_B$, $c_W^A$, and $c_G^A$ --- allows the breaking to be brought into a more tractable form,
\begin{align}\label{Eq:SM-Specified-BRST-Breaking-Tree-Level}
    \widehat{\Delta} = \int \! d^Dx\Big[ 
    \widehat{\Delta}_Y(x) + \widehat{\Delta}_W(x) + \widehat{\Delta}_s(x)
    + \widehat{\Delta}_{YY}(x) + \widehat{\Delta}_{YW}(x) + \widehat{\Delta}_{Ys}(x) + \widehat{\Delta}_{Ws}(x) \Big],
\end{align}
where each $\widehat{\Delta}_X(x)$ collects the contributions proportional to a single gauge coupling $g_X$, while each $\widehat{\Delta}_{XZ}(x)$ collects those proportional to the product of two gauge couplings $g_{X}g_{Z}$, with $X,Z\in\{Y,W,s\}$, also containing a gauge boson.

The individual contributions are given as follows:
\begin{align}
    \begin{split}\label{Eq:SM-Breaking-Y}
        \widehat{\Delta}_Y(x) &= -g_Y c_B \Bigg\{ 
        {\overline{l}}{}^a_I 
        \bigg[
        \projR 
        \bigg( (1-c_\mathrm{QED}) \mathcal{Y}^l_{R,ab} \overset{\leftarrow}{\widehat{\slashed{\partial}}}
        + \big(\mathcal{Y}^l_{L,ab}-c_\mathrm{QED}\mathcal{Y}^l_{R,ab}\big) \overset{\rightarrow}{\widehat{\slashed{\partial}}} \bigg)
        \projR
        \\
        &\phantom{
        = g_Y c_B \{ {\overline{l}}{}^a_I
        } + 
        \projL 
        \bigg( (1-c_\mathrm{QED}) \mathcal{Y}^l_{R,ab} \overset{\rightarrow}{\widehat{\slashed{\partial}}}
        + \big(\mathcal{Y}^l_{L,ab}-c_\mathrm{QED}\mathcal{Y}^l_{R,ab}\big) \overset{\leftarrow}{\widehat{\slashed{\partial}}} \bigg)
        \projL
        \bigg]
        l^b_I
        \Bigg\}
        \\
        &\phantom{= \;} 
        -g_Y c_B \Bigg\{ 
        {\overline{q}}{}^{i,a}_I 
        \bigg[
        \projR 
        \bigg( (1-c_\mathrm{QED}) \mathcal{Y}^q_{R,ab} \overset{\leftarrow}{\widehat{\slashed{\partial}}}
        + \big(\mathcal{Y}^q_{L,ab}-c_\mathrm{QED}\mathcal{Y}^q_{R,ab}\big) \overset{\rightarrow}{\widehat{\slashed{\partial}}} \bigg)
        \projR
        \\
        &\phantom{
        = g_Y c_B \{ {\overline{l}}{}^a_I
        } + 
        \projL 
        \bigg( (1-c_\mathrm{QED}) \mathcal{Y}^q_{R,ab} \overset{\rightarrow}{\widehat{\slashed{\partial}}}
        + \big(\mathcal{Y}^q_{L,ab}-c_\mathrm{QED}\mathcal{Y}^q_{R,ab}\big) \overset{\leftarrow}{\widehat{\slashed{\partial}}} \bigg)
        \projL
        \bigg]
        q^{i,b}_I
        \Bigg\},
    \end{split}\\[1.5ex]
    \begin{split}\label{Eq:SM-Breaking-W}
        \widehat{\Delta}_W(x) &= 
    - g_W t_{L,ab}^A c_W^A \bigg[ 
    {\overline{l}}{}^a_I \Big( \projR \overset{\rightarrow}{\widehat{\slashed{\partial}}} \projR + \projL \overset{\leftarrow}{\widehat{\slashed{\partial}}} \projL \Big) l_I^b
    + {\overline{q}}{}^{i,a}_I \Big( \projR \overset{\rightarrow}{\widehat{\slashed{\partial}}} \projR + \projL \overset{\leftarrow}{\widehat{\slashed{\partial}}} \projL \Big) q_I^{i,b}
    \bigg],
    \end{split}\\[1.5ex]
    \begin{split}\label{Eq:SM-Breaking-s}
    \widehat{\Delta}_s(x) &= - (1-c_\mathrm{QCD}) g_s t_{s,ij}^A c_G^A \widehat{\partial}_\mu \big( 
    {\overline{q}}{}^{i,a}_I \widehat{\gamma}^\mu q_I^{j,a} \big),
    \end{split}\\[1.5ex]
    \begin{split}\label{Eq:SM-Breaking-YY}
    \widehat{\Delta}_{YY}(x) &=
        i c_\mathrm{QED} g_Y^2 \mathcal{Y}^l_{R,ac} \big(\mathcal{Y}^l_{R}-\mathcal{Y}^l_{L}\big)_{cb} c_B {\overline{l}}{}^a_I \Big[ \projR \widehat{\slashed{B}} \projR - \projL \widehat{\slashed{B}} \projL \Big] l_I^b
        \\
        &+ i c_\mathrm{QED} g_Y^2 \mathcal{Y}^q_{R,ac} \big(\mathcal{Y}^q_{R}-\mathcal{Y}^q_{L}\big)_{cb} c_B {\overline{q}}{}^{i,a}_I \Big[ \projR \widehat{\slashed{B}} \projR - \projL \widehat{\slashed{B}} \projL \Big] q_I^{i,b},
    \end{split}\\[1.5ex]
    \begin{split}\label{Eq:SM-Breaking-YW}
    \widehat{\Delta}_{YW}(x) = &- i c_\mathrm{QED} g_Y g_W c_W^A {\overline{l}}{}^a_I \Big[
    \big(t_L^A \mathcal{Y}_R^l\big)_{ab} \projR \widehat{\slashed{B}} \projR
    - \big(\mathcal{Y}_R^l t_L^A\big)_{ab} \projL \widehat{\slashed{B}} \projL
    \Big] l_I^b
    \\
    &- i c_\mathrm{QED} g_Y g_W c_W^A {\overline{q}}{}^{i,a}_I \Big[
    \big(t_L^A \mathcal{Y}_R^q\big)_{ab} \projR \widehat{\slashed{B}} \projR
    - \big(\mathcal{Y}_R^q t_L^A\big)_{ab} \projL \widehat{\slashed{B}} \projL
    \Big] q_I^{i,b},
    \end{split}\\[1.5ex]
    \begin{split}\label{Eq:SM-Breaking-Ys}
    \widehat{\Delta}_{Ys}(x) &= i c_\mathrm{QCD} g_Y g_s \big( \mathcal{Y}_R^q - \mathcal{Y}_L^q \big)_{ab} t_{s,ij}^A c_B {\overline{q}}{}^{i,a}_I \Big[ \projR {\widehat{\slashed{G}}}{}^A \projR - \projL {\widehat{\slashed{G}}}{}^A \projL \Big] q_I^{j,b},
    \end{split}\\[1.5ex]
    \begin{split}\label{Eq:SM-Breaking-Ws}
    \widehat{\Delta}_{Ws}(x) &= - i c_\mathrm{QCD} g_W g_s t_{L,ab}^A t^B_{s,ij} c_W^A {\overline{q}}{}^{i,a}_I \Big[ \projR {\widehat{\slashed{G}}}{}^B \projR - \projL {\widehat{\slashed{G}}}{}^B \projL \Big] q_I^{j,b}.
    \end{split}
\end{align}
Note that we have $\projR {\widehat{\slashed{\vgb}}}{}^A \projR - \projL {\widehat{\slashed{\vgb}}}{}^A \projL = {\widehat{\slashed{\vgb}}}{}^A \gamma_5$, for some gauge boson $\vgb_\mu^A$.

If the evanescent gauge interactions are switched off entirely, i.e.\ $c_\mathrm{QED}=c_\mathrm{QCD}=0$, all $\widehat{\Delta}_{XZ}(x)$-type contributions vanish (Eqs.~\eqref{Eq:SM-Breaking-YY}--\eqref{Eq:SM-Breaking-Ws}), and with them every tree-level breaking term that involves a gauge boson.
The tree-level breaking is thereby reduced considerably.
The breaking term $\widehat{\Delta}_s(x)\propto g_s$ from the $SU(3)_c$ sector (see Eq.~\eqref{Eq:SM-Breaking-s}) persists in this limit because the gluon current is treated purely 4-dimensionally.

This points to a noteworthy and somewhat surprising feature of the QCD sector.
Although the strong interaction is vector-like, and thus allows for a fully $D$-dimensional treatment (see Eq.~\eqref{Eq:Gluon-Interaction-Current-SM}), a tree-level breaking $\propto g_s$ persists regardless. 
Concretely: for $c_\mathrm{QCD}=0$ the mixed contributions in Eqs.~\eqref{Eq:SM-Breaking-Ys}--\eqref{Eq:SM-Breaking-Ws} vanish while $\widehat{\Delta}_s(x)$ in Eq.~\eqref{Eq:SM-Breaking-s} survives, since the gluon current remains purely 4-dimensional; conversely, setting $c_\mathrm{QCD}=1$ removes $\widehat{\Delta}_s(x)$ but reintroduces the mixed breakings in Eqs.~\eqref{Eq:SM-Breaking-Ys}--\eqref{Eq:SM-Breaking-Ws}.
Either way, a regularisation-induced symmetry breaking $\propto g_s$ is unavoidable even at tree-level, regardless of how the evanescent part of the QCD current is chosen.
This shows that the $g_s$-breaking cannot be removed by any choice of $c_\mathrm{QCD}$, so the QCD sector cannot be treated separately or dropped in the symmetry-restoration procedure: the full SM must be treated as a whole.

This conclusion changes in the gaugeless limit of the electroweak sector, $g_Y=0$ and $g_W=0$, combined with a fully $D$-dimensional gluon current, $c_\mathrm{QCD}=1$.
In this configuration, both Eq.~\eqref{Eq:SM-Breaking-s} and Eqs.~\eqref{Eq:SM-Breaking-Ys}--\eqref{Eq:SM-Breaking-Ws} vanish identically, so that all QCD-related breakings disappear.
Since no gauge symmetry is broken in this limit, the BMHV scheme affects only global symmetries.
This scenario is for example studied in Refs.~\cite{DiNoi:2023ygk,DiNoi:2025uhu}.
If instead the same gaugeless limit is taken with a purely 4-dimensional gluon current ($c_\mathrm{QCD}=0$), the breaking $\widehat{\Delta}_s(x)$ in Eq.~\eqref{Eq:SM-Breaking-s} remains, and gauge symmetry-restoring counterterms are still required.


\section{Calculational Procedure}\label{Sec:Procedure}

The essence of the present paper is the determination of the required 1-loop counterterms for the SM in the BMHV scheme, using our general choice of the evanescent interactions with the two parameters $c_{\mathrm{QED,QCD}}$. 
In this section we first define the framework and list conditions on the 1-loop renormalisation; then we describe technical details and cross checks of the calculation.

\subsection{Conditions on the Counterterm Structure}\label{Sec:RenormalisationConditions}

The 1-loop effective action in $D$ dimensions can be decomposed as
\begin{align}\label{eq:GammaDRen1}
    \Gamma_\mathrm{DRen}^{(1)} &= S_0 + \Gamma^{1} + S_\mathrm{sct}^1 + S_\mathrm{fct}^1 \, ,
\end{align}
where $S_0=\Dintx \, \mathcal{L}$ is the tree-level action, $\Gamma^1$ is the 1-loop effective action from genuine 1PI 1-loop diagrams computed using the Feynman rules defined by $S_0$, and where $S_{\text{sct/fct}}^1$ are the singular/finite 1-loop counterterms.
The counterterms are subject to two requirements. 
The first one is UV finiteness, and accordingly the singular counterterms $S_{\mathrm{sct}}^1$ must cancel the $1/\epsilon$ UV divergences appearing in $\Gamma^1$. 
For future multi-loop computations it is necessary that UV divergences are subtracted even if they correspond to evanescent operators. 
Hence, we will provide the full result for $S_{\mathrm{sct}}^1$ below, first in Sec.~\ref{Sec:SM_1-Loop_Renormalisation_Singular_CT} in compact form using the twelveplet $\vgb_\mu^A$, and then expanded by gauge sector with fields $B_\mu$, $W^A_\mu$, and $G^A_\mu$ in Sec.~\ref{Sec:SM-S1_SingularCT}.

The second requirement is the validity of the Slavnov-Taylor identity on the renormalised level after taking $\mathop{\mathrm{LIM}}_{D \, \to \, 4}$, i.e.\ the validity of Eq.~\eqref{Eq:UltimateSymmetryRequirement}. 
This fixes $S_{\mathrm{fct}}^1$ up to the possible addition of evanescent terms and/or symmetric counterterms, the latter of which may be chosen to fulfil certain desired renormalisation conditions. 
We will provide a particularly compact choice of $S_{\mathrm{fct}}^1$ below in Sec.~\ref{Sec:SM_1-Loop_Renormalisation_Finite_CT} (in terms of $\vgb_\mu^A$) and in Sec.~\ref{Sec:SM-S1_FiniteCT} (in terms of $B_\mu$, $W^A_\mu$, and $G^A_\mu$).

In order to determine the form of the symmetry-restoring counterterms $S_{\mathrm{fct}}^1$ we employ the quantum action principle on the regularised level.\footnote{%
For the proof of its validity in DReg see Ref.~\cite{Breitenlohner:1975hg,Breitenlohner:1976te,Breitenlohner:1977hr} and Ref.~\cite{Belusca-Maito:2023wah}.
Earlier applications of this approach to analyses of symmetry violation/preservation can be found in Refs.~\cite{Martin:1999cc,Belusca-Maito:2020ala,Belusca-Maito:2021lnk,Stockinger:2023ndm,vonManteuffel:2025swv,Ebert:2024xpy,Kuhler:2025znv,Solera:2026xnz}, in Ref.~\cite{Cornella:2022hkc} for applications to gauge invariance with background field gauge fixing, and in Refs.~\cite{Stockinger:2005gx,Hollik:2005nn,Stockinger:2018oxe} for applications to supersymmetry.}

The quantum action principle in DReg implies that the potential breaking of the Slavnov-Taylor identity can be rewritten as
\begin{equation}
\label{applyQAP}
  \mathcal{S}_D(\Gamma_\mathrm{DRen})=(\widehat{\Delta}+\Delta_{\mathrm{ct}})\cdot\Gamma_\mathrm{DRen},
\end{equation}
where the right-hand side denotes 1PI Green functions with single operator insertions of the operator
	\begin{align}
	\widehat{\Delta} + \Delta_\mathrm{ct} & =\mathcal{S}_D( S_0+S_\mathrm{ct}) ,
	\label{eq:DelpertEtotal}
	\end{align}
where the evanescent tree-level breaking $\widehat{\Delta}$ was discussed already in Sec.~\ref{Sec:TreeLevelSymBreaking} and the additional term $\Delta_{\mathrm{ct}}$ results from the symmetry-violating counterterm action. 
At the 1-loop level, this quantity can be rewritten using the linearised Slavnov-Taylor operator $b_D$ defined via $\mathcal{S}_D( S_0+\zeta \mathcal{F})=\mathcal{S}_D( S_0)+\zeta b_D\mathcal{F}+\mathcal{O}(\zeta^2)$ for any functional $\mathcal{F}$ and a small parameter $\zeta$. 
Then the requirement that the Slavnov-Taylor identity is restored becomes
\begin{align}\label{Eq:STI_1-loop_RestorationRequirement}
    \mathop{\mathrm{LIM}}_{D \, \to \, 4} \, \big(\widehat{\Delta}\cdot\Gamma^1 +
	b_D S_\mathrm{sct}^1 + b_D S_\mathrm{fct}^1\big) &= 0 .
\end{align}
Hence we first need to evaluate the divergent 1-loop diagrams to obtain $S_{\text{sct}}^1$ that cancels all UV divergences, then evaluate the 1-loop diagrams with a single insertion of $\widehat{\Delta}$ and finally solve the above equation by finding a representative result for $S_{\text{fct}}^1$. 
Clearly, evanescent terms and terms that are invariant (i.e.\ annihilated by $b_D$) drop out of the above equation; these therefore provide the freedom announced before.

\subsection{Implementational Details and Cross Checks}

All results of this work have been obtained with two different and independently developed computational frameworks:
one built around the computer algebra system \texttt{FORM}~\cite{Vermaseren:2000nd,Ruijl:2017dtg,FORM:Manual}, and one based on \texttt{Mathematica}~\cite{mathematica}.
Both have been developed and validated in the context of earlier BMHV calculations, so that their independent implementations provide a strong mutual cross-check.

The \texttt{FORM}-based framework was developed to perform the first 4-loop BMHV calculation, see Ref.~\cite{vonManteuffel:2025swv}, where it is described in detail; an even more comprehensive account of the setup and the underlying methods is given in Ref.~\cite{Weisswange:2025tab}, while first steps towards it were reported in Ref.~\cite{Kuhler:2024fak}.
It consists of a chain of software tools centred on \texttt{FORM}, with \texttt{Mathematica} used solely to interface certain tools and for automated code generation.
Feynman diagrams are generated with \texttt{QGRAF}~\cite{Nogueira:1991ex} and processed by a sequence of \texttt{FORM} routines, which insert the Feynman rules and carry out subsequent algebraic manipulations.
A uniform assignment of loop momenta is achieved with \texttt{Feynson}~\cite{Maheria:2022dsq}, which systematically shifts loop momenta onto a common set of propagator momentum configurations.
The integration-by-parts (IBP) reduction is performed with the \texttt{C++} version of \texttt{FIRE}~\cite{Smirnov:2019qkx,Smirnov:2023yhb}, supplemented by \texttt{Reduze2}~\cite{vonManteuffel:2012np} for the identification and implementation of sector symmetry relations, which was not required for the 1-loop computations considered here.
The setup has been validated against known multi-loop results in ordinary QED up to the 4-loop level and against the chiral gauge theory results of Refs.~\cite{Belusca-Maito:2021lnk,Stockinger:2023ndm}, and has since been applied in Refs.~\cite{vonManteuffel:2025swv,Weisswange:2025tab}.

The \texttt{Mathematica}-based framework was used for the first 2-loop BMHV renormalisation of a non-Abelian theory, see Ref.~\cite{Kuhler:2025znv}; a detailed discussion of the implementation is given therein and in Ref.~\cite{Kuhler:2024fak}.
Here, essentially all steps are performed within \texttt{Mathematica}, employing \texttt{FeynArts}~\cite{Hahn:2000kx} for diagram generation and \texttt{FeynCalc}~\cite{Mertig:1990an,Shtabovenko:2016sxi,Shtabovenko:2020gxv,Shtabovenko:2021hjx} for certain algebraic manipulations, while the IBP reduction is again delegated to \texttt{FIRE}.
The implementation of $\Delta$-vertices, which carry ghost number~1 and fermionic statistics (cf.\ Eq.~\eqref{Eq:SM-General-BRST-Breaking-Tree-Level}), and the vertices involving external sources (see Eq.~\eqref{Eq:SM-L_ext-BRST-Trafos-coupled-to-sources}), whose spin-statistics assignment is reversed, into the \texttt{FeynArts} framework involves several subtleties and required dedicated technical workarounds; these are documented in Refs.~\cite{Kuhler:2025znv,Kuhler:2024fak}.
This setup has been tested against known 1- and 2-loop renormalisation results of QED and QCD, and applied in Ref.~\cite{Kuhler:2025znv}.

In the present work, the divergent parts of all required ordinary 1PI Green functions have been computed using the \texttt{FORM}-based framework, form which the singular 1-loop counterterms $S^1_\mathrm{sct}$ were derived.
The technically more involved $\Delta$-operator inserted 1PI Green functions, associated with $\Delta\cdot\Gamma$, have been evaluated twice and fully independently, once with each of the two frameworks described above; the two computations agree for all of these Green functions and thus provide a strong consistency check of our results.
As in previous works (see Refs.~\cite{Belusca-Maito:2020ala,Belusca-Maito:2021lnk,Belusca-Maito:2023wah,Stockinger:2023ndm,vonManteuffel:2025swv,Ebert:2024xpy,Kuhler:2025znv,Weisswange:2025tab}), both their divergent and their finite contributions have been calculated.
The finite parts determine the finite symmetry-restoring counterterms $S^1_\mathrm{fct}$ through Eq.~\eqref{Eq:STI_1-loop_RestorationRequirement}, whereas the divergent parts cancel exactly against the $b_D$-transformation of $S^1_\mathrm{sct}$, which constitutes a self-consistency check of the singular counterterms.
Both $S^1_\mathrm{sct}$ and $S^1_\mathrm{fct}$ are given below in two equivalent formats: 
in compact form using the twelveplet $\vgb_\mu^A$ in Sec.~\ref{Sec:SM_1-Loop_Renormalisation}, and expanded by gauge sectors in terms of $B_\mu$, $W^A_\mu$, and $G^A_\mu$ in Sec.~\ref{Sec:SM_1-Loop_Renormalisation_Expanded}.
For the finite counterterms, the two representations differ by the addition of finite symmetric counterterms, as mentioned above (see the discussion in Sec.~\ref{Sec:SM-S1_FiniteCT} for details).

As an additional, independent check, the complete $b_D$ variation $b_DS^1_\mathrm{sct}+b_DS^1_\mathrm{fct}$ was evaluated using an in-house \texttt{Mathematica} code once all 1-loop counterterms had been determined. 
This was done for both representations, those of Sec.~\ref{Sec:SM_1-Loop_Renormalisation} and those of Sec.~\ref{Sec:SM_1-Loop_Renormalisation_Expanded}.
In both cases, the variation cancels our result for $\widehat{\Delta}\cdot\Gamma^1$, and thereby confirms that Eq.~\eqref{Eq:STI_1-loop_RestorationRequirement} is indeed satisfied.

Finally, we checked our results in special limiting cases against the literature.
First, in the Abelian limit $g_W=0$ and $g_s=0$, restricted to a single generation of fermions, we recover our earlier results for the Abelian sector of the SM\footnote{%
With fermion kinetic terms treated as described in Sec.~\ref{Sec:PossibleChoices_of_EvanescentFermions} (Option~1 of Ref.~\cite{Ebert:2024xpy}), where physical chiral fermion fields are combined to Dirac spinors whenever possible.}
from Ref.~\cite{Ebert:2024xpy} (see Sec.~6.2 therein) up to finite symmetric counterterms.
Second, staying in the Abelian limit but setting $g_Y=e$ and $\mathcal{Y}^f_L=\mathcal{Y}^f_{R}=Q^f$ together with $c_\mathrm{QED}=1$, and dropping all scalar and Yukawa couplings, our results reduce to those of ordinary QED (see Ref.~\cite{Ebert:2024xpy}, Sec.~6.1).
Third, the QCD limit, obtained for $g_Y=0$ and $g_W=0$ with $c_\mathrm{QCD}=1$ and again switching off all scalar and Yukawa couplings, reproduces the known 1-loop results of ordinary QCD regularised in naive DReg (see Sec.~\ref{Sec:QCD_Sector_in_BMHV}).
In the latter two cases the theory is vector-like, and the resulting counterterm action indeed comes out fully symmetric.

\section{Renormalisation of the SM at the 1-Loop Level}\label{Sec:SM_1-Loop_Renormalisation}

Here we present the complete 1-loop renormalisation of the full SM within the BMHV scheme by constructing a counterterm action $S_\mathrm{ct}^{1} = S^{1}_\mathrm{sct} + S^{1}_\mathrm{fct}$ that satisfies the minimally necessary renormalisation conditions discussed in Sec.~\ref{Sec:RenormalisationConditions}.
It comprises the singular counterterms $S_\mathrm{sct}^1$ in Sec.~\ref{Sec:SM_1-Loop_Renormalisation_Singular_CT}, which cancel all UV divergences, and the finite symmetry-restoring counterterms $S_\mathrm{fct}^1$ in Sec.~\ref{Sec:SM_1-Loop_Renormalisation_Finite_CT}, which re-establish the Slavnov-Taylor identity; together they satisfy Eq.~\eqref{Eq:STI_1-loop_RestorationRequirement}.
Since $S_\mathrm{ct}^1$ is fixed only up to the addition of finite evanescent and/or symmetric counterterms, it is not unique.
This freedom can be used either to cast $S_\mathrm{fct}^1$ into a particular desired form, or to impose further renormalisation conditions --- such as on-shell conditions --- beyond the minimal ones.

In this section, all counterterms are expressed in terms of the twelveplet gauge fields $\vgb_\mu^A$, together with the corresponding generalised generators, the ghost and antighost fields $c^A$ and $\overline{c}^A$, and the external gauge field sources $\rho^A_\mu$, as introduced in Sec.~\ref{Sec:SM-Definition-and-Setup}.
The associated group invariants and generalised coupling structures are collected in App.~\ref{App:SM-GroupInvariants-CouplingStructures}.
As noted in the introduction, the computational advantages of this compact formulation come at the cost of direct accessibility to the individual gauge sectors and of readability.
We therefore restrict ourselves here to a single representative counterterm action and defer the more detailed discussion, in particular concerning the addition of finite symmetric counterterms, to Sec.~\ref{Sec:SM_1-Loop_Renormalisation_Expanded}.

Since we treat fermions by combining physical chiral fermions into Dirac spinors wherever possible (see Sec.~\ref{Sec:PossibleChoices_of_EvanescentFermions} and Option~1 of Ref.~\cite{Ebert:2024xpy}), the evanescent terms mix left- and right-handed fermions carrying different gauge quantum numbers.
As a consequence, global symmetries --- in particular global hypercharge conservation --- are violated as well, so that global hypercharge violating Green functions must be computed in order to obtain the corresponding restoring counterterms (cf.\ Refs.~\cite{Ebert:2024xpy,Weisswange:2025tab}).

All computations were carried out in general $R_\xi$-gauge, with the results expressed throughout in terms of the modified gauge parameter $\xi_F=1-\xi$ rather than $\xi$ itself (see Sec.~\ref{Sec:Ghosts_GaugeFixing_BRST}).
To keep the presentation concise, our results are further written in terms of the counterterm coefficients defined in App.~\ref{App:Results-TheStandardModel}, with the divergent coefficients given in App.~\ref{App:SM-Divergent1LoopCTCoeffs}, and those of the finite symmetry-restoring counterterms in App.~\ref{App:SM-Finite1LoopCTCoeffs}.

\subsection{Singular Counterterm Action at the 1-Loop Level}\label{Sec:SM_1-Loop_Renormalisation_Singular_CT}

In this subsection, we provide the complete singular 1-loop counterterm action, which renders the SM finite at this order.
It is obtained from the divergent parts of all power-counting divergent ordinary 1PI Green functions.
Following the sector decomposition of the SM Lagrangian introduced in Sec.~\ref{Sec:SM-Definition-and-Setup}, we write
\begin{align}\label{Eq:S_sct_1Loop-SectorDecomposition}
    S_\mathrm{sct}^{1} = S^{1}_{\mathrm{sct,fermion}} + S^{1}_{\mathrm{sct,gauge}} + S^{1}_{\mathrm{sct,Higgs}} + S^{1}_{\mathrm{sct,Yukawa}} + S^{1}_{\mathrm{sct,ghost}} + S^{1}_{\mathrm{sct,ext}},
\end{align}
and present the individual contributions in turn.

\paragraph{Fermionic Sector:}
We begin with the fermionic sector, whose singular counterterms read
\begin{equation}\label{Eq:SM-S_sct-fermion}
    \begin{aligned}
        S^{1}_{\mathrm{sct,fermion}} &= \frac{1}{16\pi^2} \Dintx \bigg\{
        \delta {\overline{\mathcal{Z}}}{}^{l,(1)}_{R,ac,IJ} \, {\overline{l}}{}_I^{a} i {\overline{\slashed{\mathcal{D}}}}{}^{l}_{R,cb} \, l_J^b
        + \delta {\overline{\mathcal{Z}}}{}^{l,(1)}_{L,IJ} \, {\overline{l}}{}_I^{a} i {\overline{\slashed{\mathcal{D}}}}{}^{l}_{L,ab} \, l_J^b
        \\
        &
        + \delta {\overline{\mathcal{Z}}}{}^{q,(1)}_{R,ac,IJ} \, {\overline{q}}{}_I^{i,a} i {\overline{\slashed{\mathcal{D}}}}{}^{q}_{R,cb,ij} \, q_J^{j,b}
        + \delta {\overline{\mathcal{Z}}}{}^{q,(1)}_{L,IJ} \, {\overline{q}}{}_I^{i,a} i {\overline{\slashed{\mathcal{D}}}}{}^{q}_{L,ab,ij} \, q_J^{j,b}\\
        &
        - \delta {\overline{\mathcal{X}}}{}^{(1)}_{\overline{f}\vgb f} 
        \Big[ 
        \myGenLe^{A}_{L,ab} {\overline{l}}{}_I^{a} {\overline{\slashed{\vgb}}}{}^{A} \projL l_I^b
        + \myGenQu^{A}_{R,ab,ij} {\overline{q}}{}_I^{i,a} {\overline{\slashed{\vgb}}}{}^{A} \projR q_I^{j,b}
        + \myGenQu^{A}_{L,ab,ij} {\overline{q}}{}_I^{i,a} {\overline{\slashed{\vgb}}}{}^{A} \projL q_I^{j,b}
        \Big]
        \\
        &
        + \delta {\widehat{\mathcal{X}}}{}^{l,(1)}_{ac} \, {\overline{l}}{}_I^{a} i {\widehat{\slashed{\mathcal{D}}}}{}^{l}_{cb} l_I^b
        + \Big[ \delta {\widehat{\mathcal{X}}}{}^{q,(1)}_{ac,IJ} \, {\overline{q}}{}_I^{i,a} i {\widehat{\slashed{\mathcal{D}}}}{}^{q}_{cb,ij} \projR q_J^{j,b} 
        - \delta {\widehat{\mathcal{X}}}{}^{{\overline{q}}{}\vgb q,(1),A}_{ab,ij,IJ} \, {\overline{q}}{}_I^{i,a} {\widehat{\slashed{\vgb}}}{}^{A} \projR q_J^{j,b}
        \mathrlap{ + \mathrm{h.c.} \Big]
        \bigg\},} \;\;
    \end{aligned}
\end{equation}
with covariant derivatives
\begin{equation}
\begin{alignedat}{4}
        &{\overline{\mathcal{D}}}{}^{l,\mu}_{R,ab} &&= 
        \big( \overline{\partial}^{\mu} \delta_{ab} + i \myGenLe_{R,ab}^A \overline{\vgb}^{A,\mu} \big) \projR,
        \qquad 
        &&{\overline{\mathcal{D}}}{}^{q,\mu}_{R,ab,ij} &&= 
        \big( \overline{\partial}^{\mu} \delta_{ab} \delta_{ij} + i \myGenQu_{R,ab,ij}^A \overline{\vgb}^{A,\mu} \big) \projR,\\
        &{\overline{\mathcal{D}}}{}^{l,\mu}_{L,ab} &&= 
        \big( \overline{\partial}^{\mu} \delta_{ab} + i \myGenLe_{L,ab}^A \overline{\vgb}^{A,\mu} \big) \projL,
        \qquad 
        &&{\overline{\mathcal{D}}}{}^{q,\mu}_{L,ab,ij} &&= 
        \big( \overline{\partial}^{\mu} \delta_{ab} \delta_{ij} + i \myGenQu_{L,ab,ij}^A \overline{\vgb}^{A,\mu} \big) \projL,\\
        &{\widehat{\mathcal{D}}}{}^{l,\mu}_{ab} &&= 
        \big( \widehat{\partial}^{\mu} \delta_{ab} + i \widehat{\myGenLe}_{ab}^A \widehat{\vgb}^{A,\mu} \big), 
        \qquad
        &&{\widehat{\mathcal{D}}}{}^{q,\mu}_{ab,ij} &&= 
        \big( \widehat{\partial}^{\mu} \delta_{ab} \delta_{ij} + i \widehat{\myGenQu}_{ab,ij}^A \widehat{\vgb}^{A,\mu} \big),
\end{alignedat}
\end{equation}
and coefficients provided in App.~\ref{App:SM-Divergent1LoopCTCoeffs}.

The individual lines of Eq.~\eqref{Eq:SM-S_sct-fermion} are readily interpreted.
The first two contain the 4-dimensional, BRST-invariant counterterms for leptons and quarks, whose coefficients are independent of the evanescent interaction parameters $c_\mathrm{QED}$ and $c_\mathrm{QCD}$. 
The third line collects the remaining 4-dimensional contributions, which are not BRST-invariant, and the last line the evanescent, BRST-breaking ones.

\paragraph{Gauge Sector:}
Next, we turn to the gauge sector, whose singular counterterms are given by
\begin{align}\label{Eq:SM-S_sct_gauge}
        S^{(1)}_{\mathrm{sct,gauge}} &= \frac{1}{16\pi^2} \Dintx \bigg\{
        - \delta \mathcal{Z}_{\vgb}^{(1)} \frac{1}{4} \big(\partial_{\mu}\vgb^A_{\nu}-\partial_{\nu}\vgb^A_{\mu}\big)^2
        + \delta \mathcal{Z}_{3\vgb}^{(1)} \mathscr{C}^{ABC} \partial_{\mu} \vgb_{\nu}^A \vgb^{B,\mu} \vgb^{C,\nu}
        \nonumber\\
        &\qquad
        - \delta \mathcal{\mathcal{Z}}_{4\vgb}^{(1)} \frac{1}{4} \mathscr{C}^{ABC} \mathscr{C}^{ADE} \vgb_{\mu}^B \vgb_{\nu}^C \vgb^{D,\mu} \vgb^{E,\nu}
        - \delta \overline{\mathcal{Z}}_{\vgb}^{(1)} \frac{1}{4} \overline{\vgb}^A_{\mu\nu} \overline{\vgb}^{A,\mu\nu}
        \nonumber\\
        &\qquad
        - \delta \widehat{\mathcal{Z}}_{\vgb}^{(1)} \frac{1}{4} 
        \Big[ 
        \big(\widehat{\partial}_{\mu}\widehat{\vgb}^A_{\nu}-\widehat{\partial}_{\nu}\widehat{\vgb}^A_{\mu}\big)^2
        - 4 \, c_{\mathrm{QCD}} \mathscr{C}^{ABC} \widehat{\partial}_{\mu} \widehat{\vgb}_{\nu}^A \widehat{\vgb}^{B,\mu} \widehat{\vgb}^{C,\nu}
        \nonumber\\
        &\qquad\qquad\qquad + c_{\mathrm{QCD}}^2 \mathscr{C}^{ABC} \mathscr{C}^{ADE} \widehat{\vgb}_{\mu}^B \widehat{\vgb}_{\nu}^C \widehat{\vgb}^{D,\mu} \widehat{\vgb}^{E,\nu}
        \Big]
        \nonumber\\
        &\qquad
        + \delta \widehat{\mathcal{Z}}_{\vgb}^{(1)} \frac{1}{2} \widehat{\vgb}^A_{\mu} \overline{\partial}^2 \widehat{\vgb}^{A,\mu}
        + \delta \widehat{\mathcal{X}}_{\vgb,12}^{(1),AB} \frac{1}{2} \overline{\vgb}^A_{\mu} \widehat{\partial}^2 \overline{\vgb}^{B,\mu}
        - \delta \widehat{\mathcal{X}}_{\vgb,34}^{(1),AB} \overline{\vgb}^A_{\mu} \overline{\partial}^{\mu} \widehat{\partial}^{\nu} \widehat{\vgb}^{B}_{\nu}
        \nonumber\\
        &\qquad
        + \delta \widehat{\mathcal{Z}}_{\vgb}^{(1)} \mathscr{C}^{ABC} \overline{\partial}_{\mu} \widehat{\vgb}_{\nu}^A \overline{\vgb}^{B,\mu} \widehat{\vgb}^{C,\nu}
        + \delta \widehat{\mathcal{X}}^{(1),ABC}_{3\vgb} \widehat{\partial}_{\mu} \overline{\vgb}_{\nu}^A \widehat{\vgb}^{C,\mu} \overline{\vgb}^{B,\nu}
        \nonumber\\
        &\qquad
        - \delta \widehat{\mathcal{X}}^{(1),ABCD}_{4\vgb} \frac{1}{4} \overline{\vgb}_{\mu}^A \overline{\vgb}^{B,\mu} \widehat{\vgb}^{C}_{\nu} \widehat{\vgb}^{D,\nu}
        \bigg\},
\end{align}
with coefficients listed in App.~\ref{App:SM-Divergent1LoopCTCoeffs}.

The first two lines of Eq.~\eqref{Eq:SM-S_sct_gauge} contain the BRST-invariant counterterms, and the remaining ones the evanescent and generally non-invariant contributions.
Invariance is manifest for the second term of the second line, governed by $\delta \overline{\mathcal{Z}}_{\vgb}^{(1)}$: 
it is a trace over two field strength tensors ${\overline{\vgb}}{}_{\mu\nu}^A$, purely 4-dimensional, and receives only fermionic contributions, in full analogy to the invariant gauge-sector counterterms in Refs.~\cite{vonManteuffel:2025swv,Ebert:2024xpy,Stockinger:2023ndm}.

For the three counterterms proportional to $\delta \mathcal{Z}_{\vgb}^{(1)}$, $\delta \mathcal{Z}_{3\vgb}^{(1)}$ and $\delta \mathcal{Z}_{4\vgb}^{(1)}$, the situation is less transparent.
These terms are fully $D$-dimensional and receive scalar contributions governed by the Dynkin index $\mathcal{S}_2\big(S^{A'}\big)$ (see Eq.~\eqref{Eq:Scalar-Super-Dynkin-Indices}), as well as gauge boson contributions governed by the adjoint quadratic Casimir $\mathcal{C}_2\big(G^{A'}\big)$ (see Eq.~\eqref{Eq:Super-C_A}).
The scalar parts can again be combined into a trace over two field strength tensors, whereas the gauge boson parts differ from one counterterm to the other.
The latter are purely non-Abelian, originating from the gauge boson self-interactions that are absent in Abelian theories.
Taken together, however, the three counterterms are BRST invariant, provided the renormalisation of the BRST transformations in non-Abelian theories themselves is taken into account, which is encoded in the external source contributions $S^{1}_{\mathrm{sct,ext}}$, see Eq.~\eqref{Eq:SM-S_sct_ext}.

A detailed analysis of the evanescent counterterms in the remaining lines, and of the impact of different choices for the evanescent interaction parameters $c_\mathrm{QED,QCD}$, is given in Sec.~\ref{Eq:SM-S1_sct_Lepton_expanded}, where the gauge sectors are discussed individually.

\paragraph{Higgs Sector:}
The divergent counterterms of the Higgs sector read
\begin{align}\label{Eq:SM-S_sct_Higgs}
        S^{1}_{\mathrm{sct,Higgs}} &= \frac{1}{16\pi^2} \Dintx \bigg\{
        \delta \mathcal{Z}_{\phi}^{(1)} \partial^{\mu} \phi_a^{\dagger} \partial_{\mu} \phi_a
        - \delta \mathcal{Z}_{\mu}^{(1)} \mu^2 \phi_a^{\dagger} \phi_a
        + \delta \overline{\mathcal{Z}}_{\phi}^{(1)} \overline{\partial}^{\mu} \phi_a^{\dagger} \overline{\partial}_{\mu} \phi_a
        \nonumber\\
        &+ \delta \widehat{\mathcal{X}}_{\phi}^{(1)} \widehat{\partial}^{\mu} \phi_1^{\dagger} \widehat{\partial}_{\mu} \phi_1
        - \frac{1}{2} \Big[ \delta \widehat{\mathcal{X}}_{\phi\phi}^{(1)} \, \phi_2 \widehat{\Box} \phi_2 + \mathrm{h.c.} \Big]
        \nonumber\\
        &- i \delta \mathcal{Z}^{(1)}_{\phi^{\dagger}\phi\vgb} \myGenScl^{A}_{ab} \vgb_{\mu}^A \big[ \phi_a^{\dagger} (\partial^{\mu}\phi_b) - (\partial^{\mu}\phi^{\dagger}_a) \phi_b \big]
        - i \delta \overline{\mathcal{Z}}^{(1)}_{\phi^{\dagger}\phi\vgb} \myGenScl^{A}_{ab} \overline{\vgb}_{\mu}^A \big[ \phi_a^{\dagger} (\overline{\partial}^{\mu}\phi_b) - (\overline{\partial}^{\mu}\phi^{\dagger}_a) \phi_b \big]
        \nonumber\\
        &- i \delta \widehat{\mathcal{X}}^{(1),A}_{\phi^{\dagger}\phi\vgb} \widehat{\vgb}_{\mu}^A \big[ \phi_1^{\dagger} (\widehat{\partial}^{\mu}\phi_1) - (\widehat{\partial}^{\mu}\phi^{\dagger}_1) \phi_1 \big]
        \nonumber\\
        &+ \delta \mathcal{Z}^{(1)}_{\phi^{\dagger}\phi\vgb\vgb} \myGenScl^{A}_{ac} \myGenScl^{B}_{cb} \vgb_{\mu}^A \vgb^{B,\mu} \phi_a^{\dagger} \phi_b
        + \delta \overline{\mathcal{Z}}^{(1)}_{\phi^{\dagger}\phi\vgb\vgb} \myGenScl^{A}_{ac} \myGenScl^{B}_{cb} \overline{\vgb}_{\mu}^A \overline{\vgb}^{B,\mu} \phi_a^{\dagger} \phi_b
        \nonumber\\
        &+ \delta \widehat{\mathcal{X}}^{(1),AB}_{\phi^{\dagger}\phi\vgb\vgb} \frac{1}{2} \widehat{\vgb}_{\mu}^A \widehat{\vgb}^{B,\mu} \phi_1^{\dagger} \phi_1
        - \delta \mathcal{Z}^{(1)}_{\lambda} \phi_a^{\dagger} \phi_a \phi_b^{\dagger} \phi_b
        \bigg\},
\end{align}
with coefficients collected in App.~\ref{App:SM-Divergent1LoopCTCoeffs}.

Among these, only a single counterterm violates global hypercharge conservation: the purely evanescent term $\phi_2 \widehat{\Box} \phi_2 + \mathrm{h.c.}$ in the second line of Eq.~\eqref{Eq:SM-S_sct_Higgs}.
The associated coefficient $\delta \widehat{\mathcal{X}}_{\phi\phi}^{(1)}$ exhibits no dependence on $c_\mathrm{QED}$ or $c_\mathrm{QCD}$, which identifies the chirality-violating evanescent part of the fermion kinetic term as its only source.
This term would be absent if the fermions were treated by introducing sterile partners for all physical fermions as done e.g.\ in Refs.~\cite{Belusca-Maito:2020ala,Belusca-Maito:2021lnk,Stockinger:2023ndm,Kuhler:2025znv,vonManteuffel:2025swv,Fuentes-Martin:2025meq} (denoted as Option~2 in the general analysis of Ref.~\cite{Ebert:2024xpy}).
In the present treatment that pairs physical chiral fermions wherever possible, the term is unavoidable. 
Ref.~\cite{Ebert:2024xpy} arrives at the analogous conclusion for the model considered there.
Electric and colour charge, unlike global hypercharge, are preserved by every term of Eq.~\eqref{Eq:SM-S_sct_Higgs}, as expected.

In the 1-loop Higgs sector, the dependence on the evanescent generators is restricted to two counterterms:
the term $\propto \delta \widehat{\mathcal{X}}^{(1),A}_{\phi^{\dagger}\phi\vgb}$ in the fourth line and the one $\propto \delta \widehat{\mathcal{X}}^{(1),AB}_{\phi^{\dagger}\phi\vgb\vgb}$ in the first term of the last line.
Both are purely evanescent, contribute only for adjoint colour index $A=1$, i.e.\ in the Abelian $U(1)_Y$ sector, and are controlled by $c_\mathrm{QED}$, vanishing identically for $c_\mathrm{QED}=0$.
We return to these terms in Sec.~\ref{Eq:SM-S1_sct_Lepton_expanded}.

\paragraph{Yukawa Sector:}
Turning to the Yukawa sector, the singular counterterms take the form
\begin{equation}\label{Eq:SM-S_sct_Yukawa}
    \begin{aligned}
        S^{1}_{\mathrm{sct,Yukawa}} &= - \frac{1}{16\pi^2} \Dintx \Big\{ 
        \delta \mathcal{Y}_{l,IJ}^{(1)} \, \overline{l}^{a}_{I} \phi_{a} \projR l_{J}^{2}
        + \delta \mathcal{Y}_{d,IJ}^{(1)} \, \overline{q}^{i,a}_{I} \phi_{a} \projR q_{J}^{i,2}\\
        &\qquad\qquad\qquad\quad
        + \delta \mathcal{Y}_{u,IJ}^{(1)} \, \varepsilon_{ab} \, \overline{q}^{i,a}_{I} \phi_{b}^{\dagger} \projR q_{J}^{i,1}
        + \mathrm{h.c.}
        \Big\},
    \end{aligned}
\end{equation}
with coefficients given in App.~\ref{App:SM-Divergent1LoopCTCoeffs}.

Each term in Eq.~\eqref{Eq:SM-S_sct_Yukawa} is composed of a symmetric field monomial already present in the tree-level Lagrangian (cf.\ Eq.~\eqref{Eq:TreeLevel_YukawaLagrangian}), so that the counterterms amount to a pure multiplicative field and parameter renormalisation of the Yukawa sector and are BRST invariant throughout.
No evanescent structures arise, and accordingly the coefficients are insensitive to the evanescent generators --- in particular they are independent of $c_\mathrm{QED}$ or $c_\mathrm{QCD}$.
The same behaviour was found in Ref.~\cite{Ebert:2024xpy}.

\paragraph{Ghosts and BRST Transformations:}
The remaining singular counterterms concern the ghosts and the renormalisation of the BRST transformations.
No counterterm for the gauge-fixing term is required, since it does not renormalise, see Refs.~\cite{Piguet:1995er,Belusca-Maito:2023wah,Weisswange:2025tab}.
For the ghost sector we obtain
\begin{equation}\label{Eq:SM-S_sct_ghost}
    \begin{aligned}
        S^{1}_{\mathrm{sct,ghost}} &= \frac{1}{16\pi^2} \Dintx \Big\{
        \delta \mathcal{Z}_{c}^{(1)} \partial^{\mu} \overline{c}^{A} \partial_{\mu} c^{A}
        - \delta \mathcal{Z}^{(1)}_{\overline{c}\vgb c} \mathscr{C}^{ABC} \big(\partial^{\mu}\overline{c}^A\big) \vgb^{B}_{\mu} c^{C}
        \Big\},
    \end{aligned}
\end{equation}
with coefficients collected in App.~\ref{App:SM-Divergent1LoopCTCoeffs}.

In a non-Abelian gauge theory the BRST transformations themselves acquire nontrivial quantum corrections, as noted above, and their renormalisation is tracked through the external sources coupled to them, see Eq.~\eqref{Eq:SM-L_ext-BRST-Trafos-coupled-to-sources}.
The associated counterterms are 
\begin{equation}\label{Eq:SM-S_sct_ext}
    \begin{aligned}
        S^{1}_{\mathrm{sct,ext}} &= \frac{1}{16\pi^2} \Dintx \bigg\{
        \delta \mathcal{Z}_{c}^{(1)} \rho_A^{\mu} \partial_{\mu} c^{A}
        - \delta \mathcal{Z}^{(1)}_{\overline{c}\vgb c} \mathscr{C}^{ABC} \rho_A^{\mu} \vgb^{B}_{\mu} c^{C}\\
        &- \Big[ i \delta \mathcal{Z}^{(1)}_{\overline{c}\vgb c} c^A 
        \Big( 
        \overline{R}_{l,I}^a \myGenLe_{L,ab}^A \projL l_I^b
        + \overline{R}_{q,I}^{i,a} \Big( \myGenQu_{R,ab,ij}^A \projR + \myGenQu_{L,ab,ij}^A \projL \Big) q_I^{j,b}
        \Big)
        + \mathrm{h.c.} \Big]\\
        &+ \Big[ i \delta \mathcal{Z}^{(1)}_{\overline{c}\vgb c} \myGenScl^A_{ab} c^A \Upsilon_{a}^{\dagger} \phi_b + \mathrm{h.c.} \Big]
        + \frac{1}{2} \delta \mathcal{Z}^{(1)}_{\overline{c}\vgb c} \mathscr{C}^{ABC} \zeta_A c^B c^C
        \bigg\},
    \end{aligned}
\end{equation}
governed by the same two coefficients as Eq.~\eqref{Eq:SM-S_sct_ghost}.

Both equations share the same characteristics:
all contributions are fully $D$-dimensional, manifestly BRST invariant, and blind to the evanescent generators, so that there is no dependence on $c_\mathrm{QED}$ or $c_\mathrm{QCD}$. 
They amount to a pure multiplicative field and parameter renormalisation, as one would obtain in naive DReg, and thus play no role in the restoration of the broken symmetry.

\subsection{Finite Symmetry-Restoring Counterterm Action at the 1-Loop Level}\label{Sec:SM_1-Loop_Renormalisation_Finite_CT}

This subsection presents the complete finite symmetry-restoring 1-loop counterterm action, which is required to re-establish the Slavnov-Taylor identity at the 1-loop level in the sense of Eq.~\eqref{Eq:UltimateSymmetryRequirement}.
It is determined by the procedure discussed in Sec.~\ref{Sec:RenormalisationConditions}:
for each sector we set up a general Ansatz of admissible local field monomials and fix their coefficients by imposing Eq.~\eqref{Eq:STI_1-loop_RestorationRequirement}, which for the finite contributions amounts to $b_D S_\mathrm{fct}^{1} = - (\widehat{\Delta}\cdot\Gamma)_\mathrm{fin}^{1} + \mathcal{O}(\widehat{.})$, with $\mathcal{O}(\widehat{.})$ denoting finite evanescent terms that drop out under $\mathrm{LIM}_{D\,\to\,4}$.
The resulting $S_\mathrm{fct}^1$ is fixed only up to finite evanescent and finite symmetric counterterms, as stressed above.

The input for this construction is the 1-loop breaking $\widehat{\Delta}\cdot\Gamma^1$ and follows from all power-counting divergent $\Delta$-operator inserted 1PI Green functions, including both their divergent and their finite contributions.
As for the singular counterterms, we organise the result along the sector decomposition of Sec.~\ref{Sec:SM-Definition-and-Setup},
\begin{align}\label{Eq:S_fct_1Loop_SectorDecomposition}
    S_\mathrm{fct}^{1} = S^{1}_{\mathrm{fct,fermion}} + S^{1}_{\mathrm{fct,gauge}} + S^{1}_{\mathrm{fct,Higgs}} + S^{1}_{\mathrm{fct,Yukawa}} + S^{1}_{\mathrm{fct,ext}},
\end{align}
where the ghost sector is absent because it requires no finite symmetry-restoring counterterms at this order.
The individual contributions are discussed in turn.

\paragraph{Fermionic Sector:}
As before, we start with the fermionic sector, whose finite symmetry-restoring counterterms read
\begin{equation}\label{Eq:SM-S_fct_fermion}
\begin{aligned}
    S_\mathrm{fct,fermion}^{1} &= \frac{1}{16\pi^2} \int d^4x \Big\{ 
    \delta \mathcal{F}^{(1)}_{l,ab} {\overline{l}}{}^a_{I} i {\overline{\slashed{\partial}}} l_I^b
    - \delta \mathcal{F}^{(1),A}_{\overline{l}\vgb l,R,ab} {\overline{l}}{}^a_{I} {\overline{\slashed{\vgb}}}{}^A \projR l_I^b
    - \delta \mathcal{F}^{(1),A}_{\overline{l}\vgb l,L,ab,IJ} {\overline{l}}{}^a_{I} {\overline{\slashed{\vgb}}}{}^A \projL l_J^b\\
    &
    + \delta \mathcal{F}^{(1)}_{q,ab} {\overline{q}}{}^{i,a}_{I} i {\overline{\slashed{\partial}}} q_I^{i,b}
    - \delta \mathcal{F}^{(1),A}_{\overline{q}\vgb q,R,ab,ij} {\overline{q}}{}^{i,a}_{I} {\overline{\slashed{\vgb}}}{}^A \projR q_I^{j,b}
    - \delta \mathcal{F}^{(1),A}_{\overline{q}\vgb q,L,ab,ij,IJ} {\overline{q}}{}^{i,a}_{I} {\overline{\slashed{\vgb}}}{}^A \projL q_J^{j,b}
    \Big\},
\end{aligned}
\end{equation}
with coefficients collected in App.~\ref{App:SM-Finite1LoopCTCoeffs}.

A structural difference to the earlier studies of Abelian theories in Refs.~\cite{Belusca-Maito:2021lnk,Stockinger:2023ndm,vonManteuffel:2025swv} is that the finite fermionic breaking can no longer be assigned to either the fermion self-energies or to the gauge interaction vertices alone, but is necessarily distributed over both.
The obstruction comes from the kinetic-term coefficients $\delta \mathcal{F}^{(1)}_{f,ab}$ ($f\in\{l,q\}$), which do not commute with the left-handed generators $\myGenLe_L^A$ or $\myGenQu_L^A$.
They therefore act as a fermion field renormalisation that does not respect the doublet structure, breaking global $SU(2)_L$ invariance, and no finite symmetric counterterm can move this breaking onto the vertices.
The converse --- shifting the vertex breaking entirely into the self-energies --- is obstructed for the same reason.

The obstruction originates in the non-commutativity of $\mathcal{Y}_R^f$ with $t_L^A$, and thus resides entirely in the electroweak sector.
The QCD contributions to the kinetic terms are unaffected and can still be moved to the vertices, provided the gauge sectors are disentangled --- an option we exploit in Sec.~\ref{Sec:SM-S1_FiniteCT}.

Unlike their singular counterparts, all finite fermionic counterterms are sensitive to the evanescent generators, see Eqs.~\eqref{Eq:SM-finite-CT-Leptons-KineticTerm}--\eqref{Eq:SM-finite-CT-Quarks-Left-handed-Interaction}; the resulting dependence on $c_\mathrm{QED}$ and $c_\mathrm{QCD}$ is examined in Sec.~\ref{Sec:SM-S1_FiniteCT}.

\paragraph{Gauge Sector:}
Turning to the gauge sector, the finite symmetry-restoring counterterms are given by
\begin{equation}\label{Eq:SM-S_fct_gauge}
\begin{aligned}
    S_\mathrm{fct,gauge}^{1} &= \frac{1}{16\pi^2} \int d^4x \bigg\{ 
    \delta\mathcal{F}^{(1),AB}_{2\vgb} {\overline{\vgb}}{}_\mu^A \overline{\Box} {\overline{\vgb}}{}^{B,\mu} 
    + \delta\mathcal{F}^{(1),ABC}_{3\vgb\text{-1}} \big({\overline{\partial}}{}^\mu {\overline{\vgb}}{}^{A,\nu}\big) {\overline{\vgb}}{}^B_\mu {\overline{\vgb}}{}^C_\nu \\
    &\qquad\qquad\quad\,\,\,\,\,
    + \delta\mathcal{F}^{(1),ABC}_{3\vgb\text{-2}} {\overline{\vgb}}{}^{A,\nu} \big({\overline{\partial}}{}^\mu {\overline{\vgb}}{}^{B}_{\mu}\big) {\overline{\vgb}}{}^C_\nu 
    + \delta\mathcal{F}^{(1),ABCD}_{4\vgb} {\overline{\vgb}}{}^{A}_{\mu} {\overline{\vgb}}{}^{B,\mu} {\overline{\vgb}}{}^C_\nu {\overline{\vgb}}{}^{D,\nu} \bigg\},
\end{aligned}
\end{equation}
with coefficients listed in App.~\ref{App:SM-Finite1LoopCTCoeffs}.

In contrast to the fermionic sector, none of these coefficients involves the evanescent generators, so that the entire gauge-sector contribution is independent of $c_\mathrm{QED}$ and $c_\mathrm{QCD}$, see Eqs.~\eqref{Eq:SM-finite-CT-gauge-boson-SelfEnergy}--\eqref{Eq:SM-finite-CT-gauge-boson-4Point}.

\paragraph{Higgs Sector:}
Turning to the Higgs sector, the finite symmetry-restoring counterterms read
\begin{equation}\label{Eq:SM-S_fct_Higgs}
\begin{aligned}
    S_\mathrm{fct,Higgs}^{1} &= \frac{1}{16\pi^2} \int d^4x \bigg\{ 
    \delta\mathcal{F}^{(1)}_{\phi^\dagger\phi,ab} \, \phi_a^\dagger \overline{\Box} \phi_b
    + i \delta\mathcal{F}^{(1),A}_{-,ab} \big[ \big({\overline{\partial}}{}^\mu\phi_a^\dagger\big) \phi_b - \phi_a^\dagger \big({\overline{\partial}}{}^\mu\phi_b\big) \big] {\overline{\vgb}}{}^A_\mu\\
    &\qquad\qquad
    + i \delta\mathcal{F}^{(1),A}_{+,ab} \big[ \big({\overline{\partial}}{}^\mu\phi_a^\dagger\big) \phi_b + \phi_a^\dagger \big({\overline{\partial}}{}^\mu\phi_b\big) \big] {\overline{\vgb}}{}^A_\mu
    + \delta\mathcal{F}^{(1),AB}_{\phi^\dagger\phi\vgb\vgb,ab} \phi_a^\dagger \phi_b {\overline{\vgb}}{}^A_\mu {\overline{\vgb}}{}^{B,\mu}\\
    &\qquad\qquad
    + \Big[ \delta\mathcal{F}^{(1)}_{\phi\phi} \, \phi_2 \overline{\Box} \phi_2 
    + i \delta\mathcal{F}_{\phi\phi\vgb,ab}^{(1),A} \big({\overline{\partial}}{}^\mu\phi_a\big) \phi_b {\overline{\vgb}}{}^A_\mu\\
    &\qquad\qquad\quad
    + \delta\mathcal{F}^{(1),AB}_{\phi\phi\vgb\vgb,ab} \phi_a \phi_b {\overline{\vgb}}{}^A_\mu {\overline{\vgb}}{}^{B,\mu} + \mathrm{h.c.} \Big]
    - \delta\mathcal{F}^{(1)}_{\phi^\dagger\phi\phi^\dagger\phi,abcd} \, \phi_a^\dagger\phi_b\phi_c^\dagger\phi_d\\
    &\qquad\qquad
    - \Big[ \delta\mathcal{F}^{(1)}_{\phi^\dagger\phi\phi\phi,abcd} \, \phi_a^\dagger\phi_b\phi_c\phi_d
    + \delta\mathcal{F}^{(1)}_{\phi\phi\phi\phi} \, \phi_2\phi_2\phi_2\phi_2 + \mathrm{h.c.} \Big]
    \bigg\},
\end{aligned}
\end{equation}
with coefficients given in App.~\ref{App:SM-Finite1LoopCTCoeffs}.

As in the gauge sector, the above counterterms are insensitive to evanescent generators, see Eqs.~\eqref{Eq:SM-finite-CT-Scalar-SelfEnergy}--\eqref{Eq:SM-finite-CT-Scalar-DoubleScalar-DoubleGaugeBoson}, Eqs.~\eqref{Eq:SM-finite-CT-Scalar-phi-phi}--\eqref{Eq:SM-finite-CT-Scalar-phi-phi-V-V}, and Eqs.~\eqref{Eq:SM-finite-CT-Scalar-phidagger-phi-phidagger-phi}--\eqref{Eq:SM-finite-CT-Scalar-phi-phi-phi-phi}.

Unlike the singular case in Eq.~\eqref{Eq:SM-S_sct_Higgs}, global hypercharge conservation is now violated by several structures, namely those inside the square bracket spanning the third and fourth lines and those in the last line of Eq.~\eqref{Eq:SM-S_fct_Higgs}.
Their form and properties match the ones encountered in Ref.~\cite{Ebert:2024xpy}.
By contrast, electric and colour charge remain intact throughout.

Two terms have no Abelian counterpart:
those governed by $\delta \mathcal{F}^{(1),A}_{+,ab}$ and $\delta \mathcal{F}^{(1)}_{\phi^\dagger\phi\phi^\dagger\phi,abcd}$.
The former contributes only for twelveplet index $A\in\{2,3\}$ and vanishes for $A=1$. 
The latter amounts to a symmetric counterterm when only $U(1)_Y$ is gauged, as it is hypercharge neutral and only transforms nontrivially under $SU(2)_L$.
Hence neither term appears in the Abelian theory of Ref.~\cite{Ebert:2024xpy}.

\paragraph{Yukawa Sector:}
Next, we turn to the Yukawa sector, whose finite symmetry-restoring counterterms are given by
\begin{equation}\label{Eq:SM-S_fct_Yukawa}
\begin{aligned}
    S_\mathrm{fct,Yukawa}^{1} &= -\frac{1}{16\pi^2} \int d^4x \bigg\{ 
    Y_{l,IJ} \delta \mathcal{F}^{l,(1)}_{Y,abc} {\overline{l}}{}^{a}_I \phi_b \projR l_J^c
    + \delta \mathcal{G}^{l,(1)}_{Y,abc,IJ} {\overline{l}}{}^{a}_I \phi_b^\dagger \projR l_J^c\\
    &\qquad\qquad\qquad
    + Y_{d,IJ} \delta \mathcal{F}^{d,(1)}_{Y,abc} {\overline{q}}{}^{i,a}_I \phi_b \projR q_J^{i,c}
    + \delta \mathcal{G}^{d,(1)}_{Y,abc,IJ} {\overline{q}}{}^{i,a}_I \phi_b^\dagger \projR q_J^{i,c}\\
    &\qquad\qquad\qquad
    + Y_{u,IJ} \delta \mathcal{F}^{u,(1)}_{Y,abc} {\overline{q}}{}^{i,a}_I \phi_b^\dagger \projR q_J^{i,c}
    + \delta \mathcal{G}^{u,(1)}_{Y,abc,IJ} {\overline{q}}{}^{i,a}_I \phi_b \projR q_J^{i,c}
    + \mathrm{h.c.}
    \bigg\},
\end{aligned}
\end{equation}
with coefficients provided in App.~\ref{App:SM-Finite1LoopCTCoeffs}.

They fall into two classes.
Those governed by $\delta \mathcal{F}^{f,(1)}_{Y,abc}$ ($f\in\{l,d,u\}$) break local gauge and BRST invariance while leaving global hypercharge conservation intact, whereas those governed by $\delta \mathcal{G}^{f,(1)}_{Y,abc,IJ}$ ($f\in\{l,d,u\}$) violate global hypercharge conservation as well.
The former counterterms are genuinely non-Abelian:
they are hypercharge neutral but transform nontrivially under $SU(2)_L$ (cf.\ the discussion on finite Yukawa counterterms in Sec.~\ref{Sec:SM-S1_FiniteCT}), and thus amount to symmetric counterterms when only $U(1)_Y$ is gauged.
This is why the Abelian analysis of Ref.~\cite{Ebert:2024xpy} encountered only the $\delta \mathcal{G}_Y$-type counterterms.

Both classes are sensitive to the evanescent generators, see Eqs.~\eqref{Eq:SM-finite-CT-Yukawa-Lepton-F}--\eqref{Eq:SM-finite-CT-Yukawa-Up-Quark-G}.
One consequence is immediate: switching off the evanescent gauge interactions entirely, $c_\mathrm{QED}=c_\mathrm{QCD}=0$, eliminates the hypercharge-conserving counterterms $\propto \delta \mathcal{F}^{f,(1)}_{Y,abc}$ altogether.
We return to the full dependence on $c_\mathrm{QED}$ and $c_\mathrm{QCD}$ in Sec.~\ref{Sec:SM-S1_FiniteCT}, where we also show that disentangling the gauge sectors allows the QCD parts of the $\delta\mathcal{F}_Y$-type counterterms to be removed by adding suitable finite symmetric counterterms, so that only their electroweak contributions remain.

\paragraph{Ghosts and BRST Transformations:}
As mentioned above, the ghost sector needs no finite symmetry-restoring counterterms at the 1-loop level, $S^1_\mathrm{fct,ghost}=0$, so that only the renormalisation of the BRST transformations remains to be addressed. 
Through their coupling to external sources, it manifests itself in
\begin{equation}\label{Eq:SM-S_fct_ext}
\begin{aligned}
    S_\mathrm{fct,ext}^{1} &= \frac{1}{16\pi^2} \int d^4x \bigg\{ 
    - i \delta \mathcal{F}^{(1)}_{R_f} 
    \Big( c^A {\overline{R}}{}^a_{l,I} \big[ \myGenLe^A_{R,ab} \projR + \myGenLe^A_{L,ab} \projL \big] l_I^b\\
    &\qquad\qquad\quad
    + c^A {\overline{R}}{}^{i,a}_{q,I} \big[ \myGenQu^A_{R,ab,ij} \projR + \myGenQu^A_{L,ab,ij} \projL \big]  q_I^{j,b}
    - c^A {\overline{R}}{}^{i,a}_{q,I} {\widehat{\myGenQu}}{}^A_{ab,ij} q_I^{j,b} \Big)
    + \mathrm{h.c.}
    \bigg\},
\end{aligned}
\end{equation}
with coefficients given in App.~\ref{App:SM-Finite1LoopCTCoeffs}.

Evidently, finite contributions are restricted to the fermionic source terms, and leptons and quarks share the single coefficient $\delta \mathcal{F}^{(1)}_{R_f}$, which is proportional to the adjoint quadratic Casimir and independent of the evanescent generators, see Eq.~\eqref{Eq:SM-finite-CT-external-Sources}.
For leptons the minimal Ansatz ${\overline{R}}{}^a_{l,I} (s_D l_I^a)$ already suffices, whereas for the quark case the additional term involving ${\widehat{\myGenQu}}{}^A$ is required --- the only point at which evanescent generators enter this sector at the 1-loop order.
We discuss this dependence in Sec.~\ref{Sec:SM-S1_FiniteCT}.


\section{Discussion of the SM 1-Loop Counterterm Action}\label{Sec:SM_1-Loop_Renormalisation_Expanded}

In this section, we revisit the 1-loop counterterm action of Sec.~\ref{Sec:SM_1-Loop_Renormalisation} in a form that makes the contributions of the individual gauge sectors more explicit.
For this purpose, we disentangle the gauge sectors by decomposing the twelveplets and generalised generators into their $U(1)_Y$, $SU(2)_L$ and $SU(3)_c$ components, proceeding as for the tree-level breaking in Sec.~\ref{Sec:TreeLevelSymBreaking}.
All results are thus expressed in terms of the gauge fields $B_\mu$, $W_\mu^A$ and $G^A_\mu$, the associated ghosts $c_B$, $c^A_W$ and $c^A_G$, and correspondingly for the antighosts and external sources.
What is lost in compactness is gained in transparency:
each gauge sector becomes directly accessible, and the impact of the evanescent gauge interactions --- encoded in the parameters $c_\mathrm{QED}$ and $c_\mathrm{QCD}$ --- can be traced sector by sector.

The singular counterterms $S^1_\mathrm{sct}$ are discussed in Sec.~\ref{Sec:SM-S1_SingularCT}, and the finite symmetry-restoring counterterms $S^1_\mathrm{fct}$ in Sec.~\ref{Sec:SM-S1_FiniteCT}.
In the latter we also exploit the freedom of adding finite symmetric counterterms to cast $S^1_\mathrm{fct}$ into a more convenient form than the representative given in Sec.~\ref{Sec:SM_1-Loop_Renormalisation_Finite_CT}.
The counterterms obtained in this way are the ones we recommend for phenomenological applications.

As before, the results are written in terms of counterterm coefficients defined in App.~\ref{App:CountertermCoeffs_Expanded}, with the divergent ones collected in App.~\ref{App:Divergent_CountertermCoeffs_Expanded} and those of the finite symmetry-restoring counterterms in App.~\ref{App:Finite_CountertermCoeffs_Expanded}.

\subsection{Singular Counterterm Action at the 1-Loop Level}\label{Sec:SM-S1_SingularCT}

As before, we begin with the singular 1-loop counterterms and present them sector by sector, following the decomposition of Eq.~\eqref{Eq:S_sct_1Loop-SectorDecomposition}.

\paragraph{Fermionic Sector:}
Decomposing Eq.~\eqref{Eq:SM-S_sct-fermion} into the individual gauge sectors and splitting the resulting fermionic counterterm action further into a lepton and a quark part as $S^1_\mathrm{sct,fermion}=S^1_\mathrm{sct,lepton}+S^1_\mathrm{sct,quark}$, one obtains for the leptons
\begin{equation}\label{Eq:SM-S1_sct_Lepton_expanded}
    \begin{aligned}
        S^{1}_{\mathrm{sct,lepton}} &= \frac{1}{16\pi^2} \Dintx \Big\{
        \delta \overline{Z}_{R,ac,IJ}^{l,(1)} \, {\overline{l}}{}_I^{a} i {\overline{\slashed{D}}}{}^{l}_{R,cb} \, l_J^b
        + \delta \overline{Z}_{L,IJ}^{l,(1)} \, {\overline{l}}{}_I^{a} i {\overline{\slashed{D}}}{}^{l}_{L,ab} \, l_J^b\\
        &\qquad \qquad \qquad 
        - \delta {\overline{X}}{}^{(1)}_{\overline{l}Wl} \, g_W t_{L,ab}^A \, {\overline{l}}{}_I^{a} {\overline{\slashed{W}}}{}^{A} \projL l_I^b
        + \delta {\widehat{X}}{}^{l,(1)}_{ac} \, {\overline{l}}{}_I^{a} i {\widehat{\slashed{D}}}{}^{l}_{cb} l_I^b
        \Big\},
    \end{aligned}
\end{equation}
and for the quarks
\begin{equation}\label{Eq:SM-S1_sct_Quark_expanded}
    \begin{aligned}
        S^{1}_{\mathrm{sct,quark}} &= \frac{1}{16\pi^2} \Dintx \Big\{
        \delta \overline{Z}^{q,(1)}_{R,ac,IJ} \, {\overline{q}}{}^{i,a}_{I} i {\overline{\slashed{D}}}{}^{q}_{R,cb,ij} \, q^{j,b}_{J}
        + \delta \overline{Z}^{q,(1)}_{L,IJ} \, {\overline{q}}{}^{i,a}_{I} i {\overline{\slashed{D}}}{}^{q}_{L,ab,ij} \, q^{j,b}_{J}\\
        &\quad
        - \delta {\overline{X}}{}^{(1)}_{\overline{q}Wq} \, g_W t^A_{L,ab} \, {\overline{q}}{}^{i,a}_{I} {\overline{\slashed{W}}}{}^{A} \projL q^{i,b}_{I}
        - \delta {\overline{X}}{}^{(1)}_{\overline{q}Gq} \, g_s t^A_{s,ij} \, {\overline{q}}{}^{i,a}_{I} {\overline{\slashed{G}}}{}^{A} q^{j,a}_{I}\\
        &\quad
        + \delta {\widehat{X}}{}^{q,(1)}_{ac} \, {\overline{q}}{}^{i,a}_{I} i {\widehat{\slashed{D}}}{}^{q}_{cb,ij} \, q^{j,b}_{I}
        + \Big[
        \delta {\widehat{X}}{}^{q,(1)}_{Y,ac,IJ} \, {\overline{q}}{}^{i,a}_{I} i {\widehat{\slashed{D}}}{}^{q}_{cb,ij} \projR q^{j,b}_{J}
        \\
        &\quad
        - \delta {\widehat{X}}{}^{(1)}_{\overline{q}Bq,ab,IJ} \, g_Y {\overline{q}}{}^{i,a}_{I} {\widehat{\slashed{B}}} \projR q^{i,b}_{J}
        - \delta {\widehat{X}}{}^{(1),A}_{\overline{q}Wq,ab,IJ} \, g_W {\overline{q}}{}^{i,a}_{I} {\widehat{\slashed{W}}}{}^{A} \projR q^{i,b}_{J}
        + \mathrm{h.c.}
        \Big]
        \\
        &\quad
        - \delta {\widehat{X}}{}^{(1)}_{\overline{q}Gq} g_s t^A_{s,ij} \, {\overline{q}}{}^{i,a}_{I} {\widehat{\slashed{G}}}{}^{A} q^{j,a}_{I}
        \Big\},
    \end{aligned}
\end{equation}
with the respective covariant derivatives
\begin{align}
    {\overline{D}}{}^{l,\mu}_{R,ab} &= 
    \big( {\overline{\partial}}{}^{\mu} \delta_{ab} + i g_Y \mathcal{Y}_{R,ab}^l {\overline{B}}{}^{\mu} \big) \projR,\nonumber\\
    {\overline{D}}{}^{l,\mu}_{L,ab} &= 
    \big( {\overline{\partial}}{}^{\mu} \delta_{ab} + i g_Y \mathcal{Y}_{L,ab}^l {\overline{B}}{}^{\mu} + i g_W t_{L,ab}^A {\overline{W}}{}^{A,\mu} \big) \projL,\nonumber\\
    {\widehat{D}}{}^{l,\mu}_{ab} &= 
    \big( {\widehat{\partial}}{}^{\mu} \delta_{ab} + i c_\mathrm{QED} g_Y \mathcal{Y}_{R,ab}^l {\widehat{B}}{}^{\mu} \big),
\end{align}
and
\begin{align}\label{Eq:Covariant_Derivatives_Quarks}
    {\overline{D}}{}^{q,\mu}_{R,ab,ij} &= 
    \big( {\overline{\partial}}{}^{\mu} \delta_{ab} \delta_{ij} + i g_Y \mathcal{Y}_{R,ab}^q \delta_{ij} {\overline{B}}{}^{\mu} + i g_s t_{s,ij}^A \delta_{ab} {\overline{G}}{}^{A,\mu} \big) \projR, \nonumber\\
    {\overline{D}}{}^{q,\mu}_{L,ab,ij} &= 
    \big( {\overline{\partial}}{}^{\mu} \delta_{ab} \delta_{ij} + i g_Y \mathcal{Y}_{L,ab}^q \delta_{ij} {\overline{B}}{}^{\mu} + i g_W t_{L,ab}^A \delta_{ij} {\overline{W}}{}^{A,\mu} + i g_s t_{s,ij}^A \delta_{ab} {\overline{G}}{}^{A,\mu} \big) \projL, \nonumber\\
    {\widehat{D}}{}^{q,\mu}_{ab,ij} &= 
    \big( {\widehat{\partial}}{}^{\mu} \delta_{ab} \delta_{ij} + i c_\mathrm{QED} g_Y \mathcal{Y}_{R,ab}^q \delta_{ij} {\widehat{B}}{}^{\mu} + i c_\mathrm{QCD} g_s t_{s,ij}^A \delta_{ab} {\widehat{G}}{}^{A,\mu} \big),
\end{align}
and coefficients collected in App.~\ref{App:Divergent_CountertermCoeffs_Expanded}.

As expected, the singular 1-loop fermion counterterms depend on evanescent generators, i.e.\ on the parameters $c_\mathrm{QED}$ and $c_\mathrm{QCD}$, only through their evanescent contributions.
The expressions simplify for any choice $c_\mathrm{QED},c_\mathrm{QCD}\in\{0,1\}$, most notably for ($c_\mathrm{QED}=0$, $c_\mathrm{QCD}=0$) and ($c_\mathrm{QED}=0$, $c_\mathrm{QCD}=1$).

At the 1-loop level, the lepton part, Eq.~\eqref{Eq:SM-S1_sct_Lepton_expanded}, is entirely free of QCD contributions $\propto g_s$, and hence independent of $c_\mathrm{QCD}$.
It depends on $c_\mathrm{QED}$ only via the last term of Eq.~\eqref{Eq:SM-S1_sct_Lepton_expanded}, governed by the coefficient $\delta {\widehat{X}}{}^{l,(1)}_{ab}$; the most economical choice is $c_\mathrm{QED}=0$, which minimises the number of terms in this coefficient and removes the evanescent breaking $\propto {\overline{l}}{}^a_I \widehat{\slashed{B}} l_I^b$.

The quark sector, Eq.~\eqref{Eq:SM-S1_sct_Quark_expanded}, is more intricate.
Here both $c_\mathrm{QED}$ and $c_\mathrm{QCD}$ enter, through the third-line terms via $\delta {\widehat{X}}{}^{q,(1)}_{ab}$ and ${\widehat{\slashed{D}}}{}^{q}_{ab,ij}$, the first term of the fourth line via $\delta {\widehat{X}}{}^{(1)}_{\overline{q}Bq,ab,IJ}$, and the last-line term via $\delta {\widehat{X}}{}^{(1)}_{\overline{q}Gq}$ --- all of them are evanescent.
As for the leptons, $c_\mathrm{QED}=0$ is the most economical choice: 
it removes the $(\mathcal{Y}_R^q\mathcal{Y}_R^q)_{ab}$-term from $\delta {\widehat{X}}{}^{q,(1)}_{ab}$ and the $\widehat{\slashed{B}}$-term from ${\widehat{\slashed{D}}}{}^{q}_{ab,ij}$.
The counterterm coefficient $\delta {\widehat{X}}{}^{(1)}_{\overline{q}Bq,ab,IJ}$ admits a simplification only at $c_\mathrm{QED}=1/2$, which is neither a physically motivated value (see Sec.~\ref{Sec:PossibleChoices_of_EvanescentFermions}) nor one that simplifies any other term; for the motivated values $c_\mathrm{QED}\in\{0,1\}$ this coefficient merely flips its overall sign, with everything else unchanged.

For $c_\mathrm{QCD}$ the situation is less straightforward.
Setting $c_\mathrm{QCD}=1$ renders the purely non-Abelian contribution $\propto g_s^2 C_2(G_s)$ fully $D$-dimensional, since it equates the coefficients $\delta {\overline{X}}{}^{(1)}_{\overline{q}Gq}$ and $\delta {\widehat{X}}{}^{(1)}_{\overline{q}Gq}$ (last term of the second line and last term of the last line of Eq.~\eqref{Eq:SM-S1_sct_Quark_expanded}).
Likewise, the QCD contribution $\propto g_s^2 C_2(F_s)$ to the quark kinetic term and the gluon interaction current (first line and the first term of the third line) becomes also fully $D$-dimensional.
However, this choice comes at a price:
it generates the non-invariant, evanescent QCD--Yukawa contribution $-\delta {\widehat{X}}{}^{q,(1)}_{Y,ab,IJ}g_st^A_{s,ij}{\overline{q}}{}^{i,a}_{I}{\widehat{\slashed{G}}}{}^{A}\projR q^{j,b}_{J}+\mathrm{h.c.}$, which stems from the gluon term in the evanescent covariant derivative ${\widehat{\slashed{D}}}{}^{q}_{ab,ij}$ and is absent for $c_\mathrm{QCD}=0$.
Hence, from the singular quark counterterms alone, neither value of $c_\mathrm{QCD}$ is unambiguously preferable.
We return to this choice, and to the QCD sector in general, in Sec.~\ref{Sec:QCD_Sector_in_BMHV}.

\paragraph{Gauge Sector:}
Proceeding in the same way for the gauge sector, we decompose Eq.~\eqref{Eq:SM-S_sct_gauge} and split the resulting counterterm action into an electroweak and a strong contribution, $S^1_\mathrm{sct,gauge}=S^{1}_{\mathrm{sct,\text{EW-gauge}}}+S^{1}_{\mathrm{sct,\text{QCD-gauge}}}$.
The electroweak part reads
\begin{align}\label{Eq:SM-S1_sct_EW-gauge_expanded}
        &S^{1}_{\mathrm{sct,\text{EW-gauge}}} = \frac{1}{16\pi^2} \Dintx \bigg\{
        - \delta Z^{(1)}_{B} \frac{1}{4} B_{\mu\nu}B^{\mu\nu}
        - \delta \overline{Z}^{(1)}_{B} \frac{1}{4} {\overline{B}}{}_{\mu\nu}{\overline{B}}{}^{\mu\nu}
        - c_{\mathrm{QED}}^2 \, \delta \widehat{Z}^{(1)}_{B} \frac{1}{4} {\widehat{B}}{}_{\mu\nu}{\widehat{B}}{}^{\mu\nu}\nonumber\\
        &\qquad\quad
        + c_{\mathrm{QED}}^2 \, \delta \widehat{Z}^{(1)}_{B} \frac{1}{2} {\widehat{B}}{}_\mu \overline{\Box}\, {\widehat{B}}{}^{\mu}
        + \delta \widehat{X}^{(1)}_{B,12} \, \frac{1}{2} {\overline{B}}{}_\mu \widehat{\Box}\, {\overline{B}}{}^{\mu}
        - c_{\mathrm{QED}} \, \delta \widehat{X}^{(1)}_{B,34} \, {\overline{B}}{}_\mu {\overline{\partial}}{}^\mu {\widehat{\partial}}{}^\nu {\widehat{B}}{}_\nu\nonumber\\
        &\qquad\quad
        + \delta \widehat{X}^{(1),A}_{BW,12} \, \frac{1}{2} {\overline{B}}{}^\mu \widehat{\Box}\, {\overline{W}}{}^{A}_{\mu}
        - c_{\mathrm{QED}} \, \delta \widehat{X}^{(1),A}_{BW,12} \, {\widehat{B}}{}_\mu {\widehat{\partial}}{}^\mu {\overline{\partial}}{}^\nu {\overline{W}}{}^{A}_{\nu}\nonumber\\
        &\qquad\quad
        - c_{\mathrm{QED}} \frac{g_W}{2} \delta \widehat{X}^{(1),D}_{BW,12} \, \varepsilon^{ABD} \big( {\widehat{\partial}}{}^\mu {\overline{W}}{}^{A,\nu} \big) {\widehat{B}}{}_{\mu} {\overline{W}}{}^{B}_{\nu}
        - c_{\mathrm{QED}}^2 \frac{g_Y g_W}{4} \delta \widehat{X}^{(1),AB}_{BBWW} \, {\overline{W}}{}^{A}_{\mu} {\overline{W}}{}^{B,\mu} {\widehat{B}}{}_{\nu} {\widehat{B}}{}^{\nu}\nonumber\\
        &\qquad\quad
        - \delta Z^{(1)}_{W} \frac{1}{4} \big( \partial_\mu W^A_\nu - \partial_\nu W^A_\mu \big)^2
        + \delta Z^{(1)}_{3W} \, g_W \varepsilon^{ABC} \big( \partial^\mu W^{A,\nu} \big) W^{B}_{\mu} W^{C}_{\nu}\nonumber\\
        &\qquad\quad
        - \delta Z^{(1)}_{4W} \frac{g_W^2}{4} \varepsilon^{ABC} \varepsilon^{ADE} W^{B}_{\mu} W^{C}_{\nu} W^{D,\mu} W^{E,\nu}\nonumber\\
        &\qquad\quad
        - \delta \overline{Z}^{(1)}_{W} \frac{1}{4} {\overline{W}}{}^A_{\mu\nu}{\overline{W}}{}^{A,\mu\nu}
        + \delta \overline{Z}^{(1)}_{W} \frac{1}{4} {\overline{W}}{}^A_\mu \widehat{\Box}\, {\overline{W}}{}^{A,\mu}
        \bigg\},
\end{align}
and the strong part
\begin{align}\label{Eq:SM-S1_sct_QCD-gauge_expanded}
        &S^{1}_{\mathrm{sct,\text{QCD-gauge}}} = \frac{1}{16\pi^2} \Dintx \bigg\{
        - \delta Z^{(1)}_{G} \frac{1}{4} \big( \partial_\mu G^A_\nu - \partial_\nu G^A_\mu \big)^2
        + \delta Z^{(1)}_{3G} \, g_s f^{ABC} \big( \partial^\mu G^{A,\nu} \big) G^{B}_{\mu} G^{C}_{\nu}
        \nonumber\\
        &\qquad
        - \delta Z^{(1)}_{4G} \frac{g_s^2}{4} f^{ABC} f^{ADE} G^{B}_{\mu} G^{C}_{\nu} G^{D,\mu} G^{E,\nu}
        - \delta \overline{Z}^{(1)}_{G} \frac{1}{4} G^A_{\mu\nu} G^{A,\mu\nu} 
        \nonumber\\
        &\qquad
        + \big( 1 - c_{\mathrm{QCD}}^2 \big) \delta \overline{Z}^{(1)}_{G} \frac{1}{4} \big( {\widehat{\partial}}{}_\mu {\widehat{G}}{}^A_\nu - {\widehat{\partial}}{}_\nu {\widehat{G}}{}^A_\mu \big)^2
        - \big( 1 - c_{\mathrm{QCD}}^2 \big) \delta \overline{Z}^{(1)}_{G} \frac{1}{2} {\widehat{G}}{}^A_\mu \overline{\Box}\, {\widehat{G}}{}^{A,\mu}
        \nonumber\\
        &\qquad
        + \big( 1 - c_{\mathrm{QCD}} \big) \delta \overline{Z}^{(1)}_{G} \, {\overline{G}}{}^A_\mu {\overline{\partial}}{}^\mu {\widehat{\partial}}{}^\nu {\widehat{G}}{}^A_\nu
        - \big( 1 - c_{\mathrm{QCD}} \big) \delta \overline{Z}^{(1)}_{G} g_s f^{ABC} \big( {\widehat{\partial}}{}^\mu {\overline{G}}{}^{A,\nu} \big) {\widehat{G}}{}^{B}_{\mu} {\overline{G}}{}^{C}_{\nu}
        \nonumber\\
        &\qquad
        - \big( 1 - c_{\mathrm{QCD}}^2 \big) \delta \overline{Z}^{(1)}_{G} g_s f^{ABC} \big( {\overline{\partial}}{}^\mu {\widehat{G}}{}^{A,\nu} \big) {\overline{G}}{}^{B}_{\mu} {\widehat{G}}{}^{C}_{\nu}
        - \big( 1 - c_{\mathrm{QCD}}^3 \big) \delta \overline{Z}^{(1)}_{G} g_s f^{ABC} \big( {\widehat{\partial}}{}^\mu {\widehat{G}}{}^{A,\nu} \big) {\widehat{G}}{}^{B}_{\mu} {\widehat{G}}{}^{C}_{\nu}
        \nonumber\\
        &\qquad
        + \big( 1 - c_{\mathrm{QCD}}^2 \big) \delta \overline{Z}^{(1)}_{G} \frac{g_s^2}{4} f^{ABC} f^{ADE}
          \Big[ {\overline{G}}{}^{B}_{\mu} {\widehat{G}}{}^{C}_{\nu} {\overline{G}}{}^{D,\mu} {\widehat{G}}{}^{E,\nu}
              + {\widehat{G}}{}^{B}_{\mu} {\overline{G}}{}^{C}_{\nu} {\widehat{G}}{}^{D,\mu} {\overline{G}}{}^{E,\nu} \Big]
        \nonumber\\
        &\qquad
        + \big( 1 - c_{\mathrm{QCD}}^4 \big) \delta \overline{Z}^{(1)}_{G} \frac{g_s^2}{4} f^{ABC} f^{ADE} {\widehat{G}}{}^{B}_{\mu} {\widehat{G}}{}^{C}_{\nu} {\widehat{G}}{}^{D,\mu} {\widehat{G}}{}^{E,\nu}
        \bigg\},
\end{align}
with field strength tensors 
\begin{equation}
    \begin{aligned}
        B_{\mu\nu} &= \partial_\mu B_\nu - \partial_\nu B_\mu \ ,\\
        W^A_{\mu\nu} &= \partial_\mu W^A_\nu - \partial_\nu W^A_\mu - g_W \varepsilon^{ABC} W^B_\mu W^C_\nu \ ,\\
        G^A_{\mu\nu} &= \partial_\mu G^A_\nu - \partial_\nu G^A_\mu - g_s f^{ABC} G^B_\mu G^C_\nu \ ,
    \end{aligned}
\end{equation}
and coefficients listed in App.~\ref{App:Divergent_CountertermCoeffs_Expanded}.

At the 1-loop level, the two parts do not mix: the electroweak part contains neither gluons nor QCD corrections to its coefficients, and the strong part neither electroweak bosons nor electroweak corrections.
Consequently, $c_\mathrm{QED}$ enters only the electroweak part, Eq.~\eqref{Eq:SM-S1_sct_EW-gauge_expanded}, and $c_\mathrm{QCD}$ only the strong part, Eq.~\eqref{Eq:SM-S1_sct_QCD-gauge_expanded}, so that their impact can be analysed separately.

\emph{Electroweak part:} 
The first two lines of Eq.~\eqref{Eq:SM-S1_sct_EW-gauge_expanded} are pure $B_\mu$ terms from the $U(1)_Y$ sector, the third and fourth lines are mixed $B_\mu$--$W^A_\mu$ terms reflecting the intermeshing of the electroweak sector, and the last three lines are pure $W^A_\mu$ terms from the $SU(2)_L$ sector.
No value of $c_\mathrm{QED}$ allows the $4$- and  $(D-4)$-dimensional $B_\mu$ terms to be combined into a single $D$-dimensional, possibly BRST-invariant term:
for any $c_\mathrm{QED}$, the coefficients $c_\mathrm{QED}^2\delta \widehat{Z}^{(1)}_{B}$, $\delta \widehat{X}^{(1)}_{B,12}$ and $c_{\mathrm{QED}} \, \delta \widehat{X}^{(1)}_{B,34}$ differ from $\delta \overline{Z}^{(1)}_{B}$ and from one another.
Equality of these coefficients would require $\mathcal{Y}_R^f=\mathcal{Y}_L^f$ together with $c_\mathrm{QED}=1$, which is not realised in the SM.
The choice $c_\mathrm{QED}=0$, however, eliminates most of the evanescent breaking terms in the $B_\mu$ and mixed $B_\mu$--$W^A_\mu$ sectors, leaving only those with field monomials ${\overline{B}}{}^\mu \widehat{\Box}\, {\overline{B}}{}_{\mu}$ and ${\overline{B}}{}^\mu \widehat{\Box}\, {\overline{W}}{}^{A}_{\mu}$.
It is therefore the most economical choice.
The pure $W_\mu^A$ terms are independent of any evanescent generators at this order, since we set $\widehat{t}_{ab}^A\equiv0$, see Eq.~\eqref{Eq:Evanescent-Generators-Choice}.

\emph{Strong part:}
To arrive at the form of Eq.~\eqref{Eq:SM-S1_sct_QCD-gauge_expanded}, we added and subtracted the $D$-dimensional term $\delta \overline{Z}^{(1)}_{G} G^A_{\mu\nu} G^{A,\mu\nu} / 4$, which leaves the action unchanged.
One copy is kept as is (second term of the second line), while the other is decomposed into $4$- and $(D-4)$-dimensional pieces as
\begin{subequations}\label{Eq:Dimensional_Decomposition_of_Gauge_Terms}
    \begin{align}
        \begin{split}
\frac{1}{4}\big( \partial_\mu G^A_\nu - \partial_\nu G^A_\mu \big)^2 
    &= 
    \frac{1}{4} \big( {\overline{\partial}}{}_\mu {\overline{G}}{}^A_\nu - {\overline{\partial}}{}_\nu {\overline{G}}{}^A_\mu \big)^2 
    + \frac{1}{4} \big( {\widehat{\partial}}{}_\mu {\widehat{G}}{}^A_\nu - {\widehat{\partial}}{}_\nu {\widehat{G}}{}^A_\mu \big)^2
    \\
    &\quad 
    - \frac{1}{2} {\overline{G}}{}^A_\mu \widehat{\Box} {\overline{G}}{}^{A,\mu} 
    - \frac{1}{2} {\widehat{G}}{}^A_\mu \overline{\Box} {\widehat{G}}{}^{A,\mu} 
    + {\overline{G}}{}^A_\mu {\overline{\partial}}{}^\mu {\widehat{\partial}}{}^\nu {\widehat{G}}{}^A_\nu ,
        \end{split}\\[0.75ex]
        \begin{split}
\big( \partial^\mu G^{A,\nu} \big) G^{B}_{\mu} G^{C}_{\nu}
    &= 
    \big( {\overline{\partial}}{}^\mu {\overline{G}}{}^{A,\nu} \big) {\overline{G}}{}^{B}_{\mu} {\overline{G}}{}^{C}_{\nu} 
    + \big( {\overline{\partial}}{}^\mu {\widehat{G}}{}^{A,\nu} \big) {\overline{G}}{}^{B}_{\mu} {\widehat{G}}{}^{C}_{\nu}
    \\
    &\quad 
    + \big( {\widehat{\partial}}{}^\mu {\overline{G}}{}^{A,\nu} \big) {\widehat{G}}{}^{B}_{\mu} {\overline{G}}{}^{C}_{\nu} 
    + \big( {\widehat{\partial}}{}^\mu {\widehat{G}}{}^{A,\nu} \big) {\widehat{G}}{}^{B}_{\mu} {\widehat{G}}{}^{C}_{\nu} ,
        \end{split}\\[0.75ex]
        \begin{split}
G^{B}_{\mu} G^{C}_{\nu} G^{D,\mu} G^{E,\nu} 
    &= 
    {\overline{G}}{}^{B}_{\mu} {\overline{G}}{}^{C}_{\nu} {\overline{G}}{}^{D,\mu} {\overline{G}}{}^{E,\nu} 
    + {\overline{G}}{}^{B}_{\mu} {\widehat{G}}{}^{C}_{\nu} {\overline{G}}{}^{D,\mu} {\widehat{G}}{}^{E,\nu} 
    \\
    &\quad
    + {\widehat{G}}{}^{B}_{\mu} {\overline{G}}{}^{C}_{\nu} {\widehat{G}}{}^{D,\mu} {\overline{G}}{}^{E,\nu} 
    + {\widehat{G}}{}^{B}_{\mu} {\widehat{G}}{}^{C}_{\nu} {\widehat{G}}{}^{D,\mu} {\widehat{G}}{}^{E,\nu}.
        \end{split}
    \end{align}
\end{subequations}
These combine with the other $4$- and $(D-4)$-dimensional terms such that all of them carry a prefactor $(1-c_\mathrm{QCD}^n)$ for some integer $n$ (see Eq.~\eqref{Eq:SM-S1_sct_QCD-gauge_expanded}).
This arrangement points to $c_\mathrm{QCD}=1$, which indeed eliminates every $4$- and $(D-4)$-dimensional term and leaves only the $D$-dimensional terms of the first two lines.
These are BRST invariant, so the gluonic gauge terms become fully BRST invariant.
Setting $c_\mathrm{QCD}=0$ instead and decomposing $G^A_{\mu\nu} G^{A,\mu\nu}$ according to Eq.~\eqref{Eq:Dimensional_Decomposition_of_Gauge_Terms} leaves the first three terms of Eq.~\eqref{Eq:SM-S1_sct_QCD-gauge_expanded} unchanged and retains only the 4-dimensional piece $-\delta {\overline{Z}}{}^{(1)}_{G} {\overline{G}}{}^A_{\mu\nu} {\overline{G}}{}^{A,\mu\nu} / 4$ and an evanescent piece $+\delta{\overline{Z}}{}_G^{(1)}{\overline{G}}{}^A_\mu\widehat{\Box}{\overline{G}}{}^{A,\mu}/2$ of the decomposed term.
The first three terms and the 4-dimensional piece are BRST invariant, whereas the evanescent term is not and thus constitutes a breaking absent for $c_\mathrm{QCD}=1$.
Within the gauge sector, $c_\mathrm{QCD}=1$ is therefore preferable, since it avoids this and any other breaking.
Unlike in the quark sector, no counterbalancing cost arises here.

\paragraph{Higgs Sector:}
Writing out the gauge twelveplets by gauge sector, the singular 1-loop counterterms of the Higgs sector in Eq.~\eqref{Eq:SM-S_sct_Higgs} can be brought into the form
\begin{align}\label{Eq:SM-S1_sct_Higgs_expanded}
        S^{1}_{\mathrm{sct,Higgs}} &= \frac{1}{16\pi^2} \Dintx \bigg\{
        \delta Z^{(1)}_{\phi} \big( \partial^\mu \phi^\dagger_a \big) \big( \partial_\mu \phi_a \big)
        - i \delta Z^{(1)}_{\phi} g_Y \mathcal{Y}_{S} B_\mu 
        \Big[ 
        \phi^\dagger_a \big( \partial^\mu \phi_a \big) - \big( \partial^\mu \phi^\dagger_a \big) \phi_a 
        \Big]
        \nonumber\\
        &
        - i \delta Z^{(1)}_{\phi^\dagger\phi W} g_W t^A_{L,ab} W^A_\mu 
        \Big[ 
        \phi^\dagger_a \big( \partial^\mu \phi_b \big) - \big( \partial^\mu \phi^\dagger_a \big) \phi_b 
        \Big]
        + \delta Z^{(1)}_{\phi} g_Y^2 \mathcal{Y}_S^2 \, B^\mu B_\mu \phi^\dagger_a \phi_a
        \nonumber\\
        &
        + 2 \delta Z^{(1)}_{\phi^\dagger\phi W} g_Y g_W \mathcal{Y}_S \, t^A_{L,ab} B^\mu W^{A}_{\mu} \phi^\dagger_a \phi_b
        + \delta Z^{(1)}_{\phi^\dagger\phi WW} g_W^2 \big(t^A_{L} t^B_{L}\big)_{ab} \, W^A_\mu W^{B,\mu} \phi^\dagger_a \phi_b
        \nonumber\\
        &
        + \delta \overline{Z}^{(1)}_{\phi} \big( {\overline{D}}{}^\mu \phi_a \big)^\dagger \big( {\overline{D}}{}_\mu \phi_a \big)
        + \delta {\widehat{X}}{}^{(1)}_{\phi,ab} \big( {\widehat{\partial}}{}^\mu \phi^\dagger_a \big) \big( {\widehat{\partial}}{}_\mu \phi_b \big)
        - \frac{1}{2} \Big[ 
        \delta \widehat{X}^{(1)}_{\phi\phi} \, \phi_2 \widehat{\Box} \phi_2 
        + \mathrm{h.c.}
        \Big]
        \nonumber\\
        &
        - c_\mathrm{QED} \delta \widehat{X}^{(1)}_{\phi^\dagger\phi B}
        \Big(
        i g_Y \mathcal{Y}_S {\widehat{B}}{}_\mu \Big[ \phi^\dagger_1 \big( {\widehat{\partial}}{}^\mu \phi_1 \big) - \big( {\widehat{\partial}}{}^\mu \phi^\dagger_1 \big) \phi_1 \Big]
        - 2 c_\mathrm{QED} g_Y^2 \mathcal{Y}_S^2 {\widehat{B}}{}_\mu {\widehat{B}}{}^\mu \phi^\dagger_1 \phi_1
        \Big)
        \nonumber\\
        &
        - \delta Z^{(1)}_{\mu} \mu^2 \, \phi^\dagger_a \phi_a
        - \delta Z^{(1)}_{\lambda} \, \phi^\dagger_a \phi_a \phi^\dagger_b \phi_b
        \bigg\},
\end{align}
with covariant derivative
\begin{align}
    {\overline{D}}{}_\mu \phi_a = \big( {\overline{\partial}}{}_\mu \delta_{ab} + i g_Y \mathcal{Y}_S \delta_{ab} {\overline{B}}{}_\mu + i g_W t^A_{L,ab} {\overline{W}}{}^A_\mu \big) \phi_b \ ,
\end{align}
and coefficients provided in App.~\ref{App:Divergent_CountertermCoeffs_Expanded}.
Compared to Eq.~\eqref{Eq:SM-S_sct_Higgs}, the scalar kinetic terms have been reorganised so that the purely 4-dimensional terms combine into covariant form (first term of the fourth line).

The properties established in Sec.~\ref{Sec:SM_1-Loop_Renormalisation_Singular_CT} carry over:
the 1-loop Higgs sector contains no QCD contributions $\propto g_s$ and is therefore independent of $c_\mathrm{QCD}$, and the hypercharge-violating term $\phi_2 \widehat{\Box} \phi_2 + \mathrm{h.c.}$ remains unaffected by any evanescent generator, as it is caused by the evanescent part of the fermion kinetic term.
The only dependence on $c_\mathrm{QED}$ resides in the terms of the penultimate line of Eq.~\eqref{Eq:SM-S1_sct_Higgs_expanded} governed by $\delta \widehat{X}^{(1)}_{\phi^\dagger\phi B}$, which vanish identically for $c_\mathrm{QED}=0$.
This is hence the most economical choice.

\paragraph{Yukawa Sector:}
The Yukawa sector contains no gauge or ghost fields, so it is unaffected by the decomposition of the twelveplets and keeps the form of Eq.~\eqref{Eq:SM-S_sct_Yukawa}, i.e.\ 
\begin{equation}\label{Eq:SM-S1_sct_Yukawa_expanded}
    \begin{aligned}
        S^{1}_{\mathrm{sct,Yukawa}} &= - \frac{1}{16\pi^2} \Dintx \Big\{ 
        \delta Y_{l,IJ}^{(1)} \, \overline{l}^{a}_{I} \phi_{a} \projR l_{J}^{2}
        + \delta Y_{d,IJ}^{(1)} \, \overline{q}^{i,a}_{I} \phi_{a} \projR q_{J}^{i,2}\\
        &\qquad\qquad\qquad\quad
        + \delta Y_{u,IJ}^{(1)} \, \varepsilon_{ab} \, \overline{q}^{i,a}_{I} \phi_{b}^{\dagger} \projR q_{J}^{i,1}
        + \mathrm{h.c.}
        \Big\}.
    \end{aligned}
\end{equation}
The counterterm coefficients are unchanged but now expressed in terms of the quadratic Casimirs of the individual gauge sectors, and are listed in App.~\ref{App:Divergent_CountertermCoeffs_Expanded}.
As before, it is BRST invariant and independent of $c_\mathrm{QED}$ and $c_\mathrm{QCD}$.

\paragraph{Ghosts and BRST Transformations:}
Decomposing the ghost and antighost twelveplets in Eq.~\eqref{Eq:SM-S_sct_ghost} yields
\begin{equation}\label{Eq:SM-S1_sct_ghost_expanded}
    \begin{aligned}
        S^{1}_{\mathrm{sct,ghost}} &= 
        \frac{1}{16\pi^2} \Dintx \Big\{
        \delta Z^{(1)}_{c_W} \, \partial^{\mu} {\overline{c}}{}^{A}_{W} \partial_{\mu} c^A_W
        - \delta Z^{(1)}_{\overline{c} W c} \, g_W \varepsilon^{ABC} \big( \partial^\mu {\overline{c}}{}^{A}_{W} \big) W^B_\mu c^C_W\\
        &\qquad\qquad\qquad\,
        + \delta Z^{(1)}_{c_G} \, \partial^{\mu} {\overline{c}}{}^{A}_{G} \partial_{\mu} c^A_G
        - \delta Z^{(1)}_{\overline{c} G c} \, g_s f^{ABC} \big( \partial^\mu {\overline{c}}{}^{A}_{G} \big) G^B_\mu c^C_G
        \Big\},
    \end{aligned}
\end{equation}
and similarly for the BRST transformations coupled to external sources in Eq.~\eqref{Eq:SM-S_sct_ext}, we find
\begin{equation}\label{Eq:SM-S1_sct_ext_expanded}
    \begin{aligned}
        S^{1}_{\mathrm{sct,ext}} &= \frac{1}{16\pi^2} \Dintx \bigg\{
        \delta Z^{(1)}_{c_W} \, \rho^{A}_{W,\mu} \partial^\mu c^A_W
        - \delta Z^{(1)}_{\overline{c} W c} \, g_W \varepsilon^{ABC} \rho^{A}_{W,\mu} W^{B,\mu} c^C_W\\
        &\qquad
        + \delta Z^{(1)}_{c_G} \, \rho^{A}_{G,\mu} \partial^\mu c^A_G
        - \delta Z^{(1)}_{\overline{c} G c} \, g_s f^{ABC} \rho^{A}_{G,\mu} G^{B,\mu} c^C_G\\
        &\qquad
        - \Big[ 
        i \delta Z^{(1)}_{\overline{c} W c} \, g_W t^A_{L,ab} \, c^A_W {\overline{R}}{}^{a}_{l,I} \projL l^b_I
        + i \delta Z^{(1)}_{\overline{c} W c} \, g_W t^A_{L,ab} \, c^A_W {\overline{R}}{}^{i,a}_{q,I} \projL q^{i,b}_I\\
        &\qquad\quad
        + i \delta Z^{(1)}_{\overline{c} G c} \, g_s t^A_{s,ij} \, c^A_G {\overline{R}}{}^{i,a}_{q,I} q^{j,a}_I
        - i \delta Z^{(1)}_{\overline{c} W c} \, g_W t^A_{L,ab} \, c^A_W \Upsilon^\dagger_a \phi_b
        + \mathrm{h.c.}
        \Big]\\
        &\qquad
        + \frac{1}{2} \delta Z^{(1)}_{\overline{c} W c} \, g_W \varepsilon^{ABC} \, \zeta_{W}^A c^B_W c^C_W
        + \frac{1}{2} \delta Z^{(1)}_{\overline{c} G c} \, g_s f^{ABC} \, \zeta_{G}^A c^B_G c^C_G
        \bigg\}.
    \end{aligned}
\end{equation}
Both are governed by the same four coefficients, collected in App.~\ref{App:Divergent_CountertermCoeffs_Expanded}.
As stated in Sec.~\ref{Sec:SM_1-Loop_Renormalisation_Singular_CT}, these counterterms are fully $D$-dimensional and independent of $c_\mathrm{QED,QCD}$ at the 1-loop level.

\subsection{Finite Symmetry-Restoring Counterterm Action at the 1-Loop Level}\label{Sec:SM-S1_FiniteCT}

We now turn to the finite symmetry-restoring 1-loop counterterms, proceeding as for the singular ones.
Besides decomposing the gauge twelveplets, we exploit the freedom of adding finite symmetric counterterms to simplify the fermionic and the Yukawa sector.

\paragraph{Fermionic Sector:}
Starting from Eq.~\eqref{Eq:SM-S_fct_fermion}, we split the action into a lepton and a quark part, $S^{1}_{\mathrm{fct,fermion}}=S^{1}_{\mathrm{fct,lepton}}+S^{1}_{\mathrm{fct,quark}}$, and decompose the gauge field twelveplet $\vgb_\mu^A$ into $B_\mu$, $W_\mu^A$, and $G_\mu^A$.
Since the gauge sectors are now individually accessible, we can add the finite symmetric counterterm
\begin{align}\label{Eq:Finite_Symmetric_Quark_Counterterm}
    + \, g_s^2 c_\mathrm{QCD} \bigg(\frac{\xi_F}{3} - c_\mathrm{QCD} \frac{3-\xi_F}{3}\bigg) C_2(F_s) \bigg[
    {\overline{q}}{}^{i,a}_{I} i {\overline{\slashed{D}}}{}^{q}_{R,ab,ij} \, q^{j,b}_{I}
    + {\overline{q}}{}^{i,a}_{I} i {\overline{\slashed{D}}}{}^{q}_{L,ab,ij} \, q^{j,b}_{I}
    \bigg],
\end{align}
with covariant derivatives as defined in Eq.~\eqref{Eq:Covariant_Derivatives_Quarks}, which removes the QCD contribution from the quark kinetic counterterm in $S^{1}_{\mathrm{fct,quark}}$ entirely.
The electroweak contributions to the quark kinetic counterterm, and the lepton kinetic counterterm (which is purely electroweak at this order), cannot be removed in this way because $[t_L^A,\mathcal{Y}^f_R]\neq0$.
Recall that this non-commutation was the obstruction that prevented removing the kinetic counterterms with coefficients $\delta\mathcal{F}_{l,ab}^{(1)}$ and $\delta\mathcal{F}_{q,ab}^{(1)}$ in Eq.~\eqref{Eq:SM-S_fct_fermion}.
The QCD contribution is not subject to it and can thus be eliminated once the sectors are separated.

The finite fermionic counterterms then read, for the leptons,
\begin{equation}\label{Eq:SM-S1_fct_Lepton_expanded}
    \begin{aligned}
        S^{1}_{\mathrm{fct,lepton}} &= \frac{1}{16\pi^2} \int d^4x \Big\{
        \delta F^{(1)}_{l,ab} \, {\overline{l}}{}_I^{a} i {\overline{\slashed{\partial}}} \, l_I^b
        - \delta F^{(1)}_{\overline{l}Bl,R,ac} \, g_Y \mathcal{Y}^l_{R,cb} \, {\overline{l}}{}_I^{a} {\overline{\slashed{B}}} \projR l_I^b \\
        &\qquad\qquad\qquad
        - \delta F^{(1)}_{\overline{l}Wl,R} \, g_W \big( \mathcal{Y}^l_R t^A_L \mathcal{Y}^l_R \big)_{ab} \, {\overline{l}}{}_I^{a} {\overline{\slashed{W}}}{}^{A} \projR l_I^b\\
        &\qquad\qquad\qquad
        - \delta F^{(1)}_{\overline{l}Bl,L,ab,IJ} \, g_Y \, {\overline{l}}{}_I^{a} {\overline{\slashed{B}}} \projL l_J^b
        - \delta F^{(1),A}_{\overline{l}Wl,L,ab,IJ} \, g_W \, {\overline{l}}{}_I^{a} {\overline{\slashed{W}}}{}^{A} \projL l_J^b
        \Big\},
    \end{aligned}
\end{equation}
and for the quarks,
\begin{equation}\label{Eq:SM-S1_fct_Quark_expanded}
    \begin{aligned}
        S^{1}_{\mathrm{fct,quark}} &= \frac{1}{16\pi^2} \int d^4x \Big\{
        \delta F^{(1)}_{q,ab} \, {\overline{q}}{}^{i,a}_{I} i {\overline{\slashed{\partial}}} \, q^{i,b}_{I} 
        - \delta F^{(1)}_{\overline{q}Gq,ab} \, g_s t^A_{s,ij} \, {\overline{q}}{}^{i,a}_{I} {\overline{\slashed{G}}}{}^{A} q^{j,b}_{I}\\
        &\qquad\qquad
        - \delta F^{(1)}_{\overline{q}Bq,R,ab} \, g_Y \, {\overline{q}}{}^{i,a}_{I} {\overline{\slashed{B}}} \projR q^{i,b}_{I}
        - \delta F^{(1),A}_{\overline{q}Wq,R,ab} \, g_W \, {\overline{q}}{}^{i,a}_{I} {\overline{\slashed{W}}}{}^{A} \projR q^{i,b}_{I}\\
        &\qquad\qquad
        - \delta F^{(1)}_{\overline{q}Bq,L,ab,IJ} \, g_Y \, {\overline{q}}{}^{i,a}_{I} {\overline{\slashed{B}}} \projL q^{i,b}_{J}
        - \delta F^{(1),A}_{\overline{q}Wq,L,ab,IJ} \, g_W \, {\overline{q}}{}^{i,a}_{I} {\overline{\slashed{W}}}{}^{A} \projL q^{i,b}_{J}\Big\},
    \end{aligned}
\end{equation}
with coefficients given in App.~\ref{App:Finite_CountertermCoeffs_Expanded}.

The lepton counterterms depend only on $c_\mathrm{QED}$, being free of QCD contributions.
The quark counterterms generally depend on both parameters, without mixing within a coefficient at 1-loop order: the $g_Y$-contributions depend on $c_\mathrm{QED}$ and the $g_s$-contributions on $c_\mathrm{QCD}$.
The only exception is the quark kinetic counterterm $\delta F^{(1)}_{q}$, which after adding Eq.~\eqref{Eq:Finite_Symmetric_Quark_Counterterm} depends only on $c_\mathrm{QED}$.
All expressions simplify considerably for any of the physically motivated values $c_\mathrm{QED},c_\mathrm{QCD}\in\{0,1\}$.

For the leptons, $c_\mathrm{QED}=0$ is clearly preferable.
It makes the kinetic counterterm coefficient $\delta F^{(1)}_{l}$ vanish, simplifies the coefficients of the left-handed interactions visibly (without making them vanish), and leaves the right-handed ones structurally unchanged compared to $c_\mathrm{QED}=1$.

For the quarks, $(c_\mathrm{QED},c_\mathrm{QCD})=(0,1)$ gives the most economical result.
For this choice, the quark kinetic counterterm and the quark--gluon interaction counterterm of the first line of Eq.~\eqref{Eq:SM-S1_fct_Quark_expanded} vanish entirely.
In the left-handed interactions of the last line, $c_\mathrm{QED}=0$ reduces the number of electroweak terms in the coefficients, whereas both values $c_\mathrm{QCD}\in\{0,1\}$ differ only in the prefactors of the QCD terms.
For the right-handed interactions of the second line, no choice is preferable, since all $c_\mathrm{QED},c_\mathrm{QCD}\in\{0,1\}$ yield similar results that differ only in prefactors.

In summary, the finite fermionic sector favours $(c_\mathrm{QED},c_\mathrm{QCD})=(0,1)$, which minimises the number of finite fermion counterterms.

\paragraph{Gauge Sector:}
Decomposing Eq.~\eqref{Eq:SM-S_fct_gauge} into gauge sectors $U(1)_Y$, $SU(2)_L$ and $SU(3)_c$, with respective gauge fields $B_\mu$, $W_\mu^A$ and $G_\mu^A$, yields
\begin{align}\label{Eq:SM-S1_fct_gauge_expanded}
    S_\mathrm{fct,gauge}^{1} &= \frac{1}{16\pi^2} \int d^4x \bigg\{ 
    \frac{1}{2} \delta F^{(1)}_{B} {\overline{B}}{}^\mu \overline{\Box} \, {\overline{B}}{}_{\mu} 
    + \delta F^{(1),A}_{BW} {\overline{B}}{}^\mu \overline{\Box} \, {\overline{W}}{}^{A}_{\mu}
    + \frac{1}{2} \delta F^{(1)}_{W} {\overline{W}}{}^{A,\mu} \overline{\Box} \, {\overline{W}}{}^{A}_{\mu}
    \nonumber\\
    &
    + \delta F^{(1),C}_{BW} g_W \varepsilon^{ABC} \Big[ 
    \big({\overline{\partial}}{}^\mu {\overline{W}}{}^{A,\nu}\big) {\overline{W}}{}^{B}_{\mu} {\overline{B}}{}_{\nu}
    - \big({\overline{\partial}}{}^\mu {\overline{W}}{}^{A,\nu}\big) {\overline{B}}{}_{\mu} {\overline{W}}{}^{B}_{\nu}
    + \big({\overline{\partial}}{}^\mu {\overline{W}}{}^{A}_\mu\big) {\overline{W}}{}^{B}_{\nu} {\overline{B}}{}^{\nu}
    \Big]
    \nonumber\\
    &
    + 2 \delta F^{(1)}_{W} g_W \varepsilon^{ABC} \big({\overline{\partial}}{}^\mu {\overline{W}}{}^{A,\nu}\big) {\overline{W}}{}^{B}_{\mu} {\overline{W}}{}^C_{\nu}
    + \frac{g_Y^2}{8} \delta F^{(1)}_{4B} {\overline{B}}{}^{\mu} {\overline{B}}{}_{\mu} {\overline{B}}{}^{\nu} {\overline{B}}{}_{\nu}
    \nonumber\\
    &
    + \delta F^{(1),A}_{BBBW} {\overline{B}}{}^{\mu} {\overline{B}}{}_{\mu} {\overline{B}}{}^{\nu} {\overline{W}}{}^A_{\nu}
    + \delta F^{(1),AB}_{BBWW\text{-1}} {\overline{B}}{}^{\mu} {\overline{B}}{}_{\mu} {\overline{W}}{}^A_{\nu} {\overline{W}}{}^{B,\nu}
    + \delta F^{(1),AB}_{BBWW\text{-2}} {\overline{B}}{}^{\mu} {\overline{B}}{}^{\nu} {\overline{W}}{}^A_{\mu} {\overline{W}}{}^{B}_{\nu}
    \nonumber\\
    &
    + \delta F^{(1),ABC}_{BWWW} {\overline{B}}{}^{\mu} {\overline{W}}{}^A_{\nu} {\overline{W}}{}^B_{\mu} {\overline{W}}{}^{C,\nu}
    + g_W^2 \delta F^{(1),ABCD}_{4W} {\overline{W}}{}^A_{\mu} {\overline{W}}{}^B_{\nu} {\overline{W}}{}^{C,\mu} {\overline{W}}{}^{D,\nu}
    \bigg\},
\end{align}
with coefficients collected in App.~\ref{App:Finite_CountertermCoeffs_Expanded}.

This form of $S_\mathrm{fct,gauge}^{1}$ makes manifest that there are no gluonic finite symmetry-restoring counterterms at the 1-loop order.
Additionally, none of the coefficients receives QCD contributions.
The result is purely electroweak and, as stated in Sec.~\ref{Sec:SM_1-Loop_Renormalisation_Finite_CT}, independent of evanescent gauge interaction parameters $c_\mathrm{QED,QCD}$.

\paragraph{Higgs Sector:}
Proceeding in the same way for the Higgs sector, Eq.~\eqref{Eq:SM-S_fct_Higgs}, we obtain
\begin{align}\label{Eq:SM-S1_fct_Higgs_expanded}
        &S^{1}_{\mathrm{fct,Higgs}} = \frac{1}{16\pi^2} \int d^4x \bigg\{
        \delta F^{(1)}_{\phi,ab} \, \phi^\dagger_a \overline{\Box} \phi_b
        + i \delta F^{(1)}_{B,-,ab} \, {\overline{B}}{}_\mu \Big[ \big( {\overline{\partial}}{}^\mu \phi^\dagger_a \big) \phi_b - \phi^\dagger_a \big( {\overline{\partial}}{}^\mu \phi_b \big) \Big] 
        \nonumber\\
        &\,\,
        + i \delta F^{(1),A}_{W,-,ab} \, {\overline{W}}{}^A_\mu \Big[ \big( {\overline{\partial}}{}^\mu \phi^\dagger_a \big) \phi_b - \phi^\dagger_a \big( {\overline{\partial}}{}^\mu \phi_b \big) \Big]
        + i \delta F^{(1),A}_{W,+,ab} \, {\overline{W}}{}^A_\mu \Big[ \big( {\overline{\partial}}{}^\mu \phi^\dagger_a \big) \phi_b + \phi^\dagger_a \big( {\overline{\partial}}{}^\mu \phi_b \big) \Big]
        \nonumber\\
        &\,\,
        + \delta F^{(1)}_{\phi^\dagger\phi BB,ab} \, {\overline{B}}{}_\mu {\overline{B}}{}^\mu \phi^\dagger_a \phi_b
        + \delta F^{(1),A}_{\phi^\dagger\phi BW,ab} \, {\overline{B}}{}^\mu {\overline{W}}{}^{A}_{\mu} \phi^\dagger_a \phi_b
        + \delta F^{(1),AB}_{\phi^\dagger\phi WW,ab} \, {\overline{W}}{}^A_\mu {\overline{W}}{}^{B,\mu} \phi^\dagger_a \phi_b
        \nonumber\\
        &\,\,
        + \Big[
        \delta F^{(1)}_{\phi\phi} \phi_2 \overline{\Box} \phi_2
        + i \delta F^{(1)}_{\phi\phi B} \, {\overline{B}}{}_\mu \big( {\overline{\partial}}{}^\mu \phi_2 \big) \phi_2
        + i \delta F^{(1),A}_{\phi\phi W,ab} \, {\overline{W}}{}^A_\mu \big( {\overline{\partial}}{}^\mu \phi_a \big) \phi_b
        \nonumber\\
        &\,\,\quad
        + \delta F^{(1)}_{\phi\phi BB}  {\overline{B}}{}_\mu {\overline{B}}{}^\mu \phi_2 \phi_2
        + \delta F^{(1),A}_{\phi\phi BW,ab}  {\overline{B}}{}^\mu {\overline{W}}{}^{A}_{\mu} \phi_a \phi_b
        + \delta F^{(1),AB}_{\phi\phi WW,ab}  {\overline{W}}{}^A_\mu {\overline{W}}{}^{B,\mu} \phi_a \phi_b
        + \mathrm{h.c.}
        \Big]
        \nonumber\\
        &\,\,
        - \delta F^{(1)}_{\phi^\dagger\phi\phi^\dagger\phi,abcd} \, \phi^\dagger_a \phi_b \phi^\dagger_c \phi_d
        - \Big[
        \delta F^{(1)}_{\phi^\dagger\phi\phi\phi,abcd} \, \phi^\dagger_a \phi_b \phi_c \phi_d
        + \delta F^{(1)}_{\phi\phi\phi\phi} \, \phi_2 \phi_2 \phi_2 \phi_2
        + \mathrm{h.c.}
        \Big] \!
        \bigg\},
\end{align}
with coefficients provided in App.~\ref{App:Finite_CountertermCoeffs_Expanded}.

As for the gauge sector, no finite symmetry-restoring counterterms involving gluons arise at the 1-loop level, and all coefficients are free of QCD corrections and independent of $c_\mathrm{QED}$ and $c_\mathrm{QCD}$.
The decomposition also makes explicit, as stated in Sec.~\ref{Sec:SM_1-Loop_Renormalisation_Finite_CT}, that only the ${\overline{W}}{}^A_\mu$-part of the $\delta\mathcal{F}^{(1),A}_{+}$-counterterm in Eq.~\eqref{Eq:SM-S_fct_Higgs} contributes (here governed by $\delta F^{(1),A}_{W,+,ab}$) and that the $b_D$-variation of the $\delta F^{(1)}_{\phi^\dagger\phi\phi^\dagger\phi,abcd}$-counterterm is purely $SU(2)_L$, proportional to $g_W c_W^A$.
The terms violating global hypercharge conservation, identified in Sec.~\ref{Sec:SM_1-Loop_Renormalisation_Finite_CT}, appear here in the square brackets spanning the fourth and fifth lines and the last two terms of the final line, while electric and colour charge remain conserved throughout.

\paragraph{Yukawa Sector:}
The Yukawa sector contains no gauge or ghost fields and is thus unaffected by the decomposition of the twelveplets.
The counterterm coefficients of Eq.~\eqref{Eq:SM-S_fct_Yukawa}, however, can be rewritten by decomposing the quadratic Casimirs used in Eqs.~\eqref{Eq:SM-finite-CT-Yukawa-Lepton-F}--\eqref{Eq:SM-finite-CT-Yukawa-Up-Quark-G}, hence disentangling the contributions of the three gauge sectors.
This allows us to simplify the global hypercharge-conserving $\delta\mathcal{F}_Y$-type counterterms by adding finite symmetric counterterms.

For the down-type quarks, adding 
\begin{align}\label{Eq:Finite_Symmetric_DownTypeYukawa_CT}
    + \, c_\mathrm{QCD} \, g_s^2 \bigg(\frac{15-11\xi_F}{12}+c_\mathrm{QCD}\frac{\xi_F}{6}\bigg) C_2(F_s) \, Y_{d,IJ} \, {\overline{q}}{}^{i,a}_{I} \phi_a \projR q^{i,2}_{J} + \mathrm{h.c.},
\end{align}
removes the entire QCD contribution to the counterterm governed by $\delta \mathcal{F}^{d,(1)}_{Y,abc}$.
For the up-type quarks, adding 
\begin{align}\label{Eq:Finite_Symmetric_UpTypeYukawa_CT}
    + \, \bigg[ \frac{g_Y^2}{9}  - g_s^2 \bigg(2+ c_\mathrm{QCD} \frac{9-\xi_F}{6}+c_\mathrm{QCD}^2\frac{\xi_F}{3}\bigg) C_2(F_s) \bigg] Y_{u,IJ} \,  {\overline{q}}{}^{i,a}_{I} \varepsilon_{ab} \phi_b^\dagger \projR q^{i,1}_{J} + \mathrm{h.c.},
\end{align}
removes the QCD contribution to $\delta \mathcal{F}^{u,(1)}_{Y,abc}$ together with a symmetric electroweak part that carries no $c_\mathrm{QED}$ dependence.
The remaining electroweak parts of both counterterms, as well as the leptonic counterterm governed by $\delta \mathcal{F}^{l,(1)}_{Y,abc}$, cannot be removed in this way: as for the fermion kinetic terms above, $[t_L^A,\mathcal{Y}_R^f]\neq0$ rules out a suitable symmetric counterterm.
The $\delta\mathcal{G}_Y$-type counterterms violate global hypercharge conservation and are therefore not removable by symmetric counterterms at all.

The resulting finite 1-loop Yukawa counterterms read
\begin{equation}\label{Eq:SM-S1_fct_Yukawa_expanded}
    \begin{aligned}
        S^{1}_{\mathrm{fct,Yukawa}} &= - \frac{1}{16\pi^2} \int d^4x \bigg\{
        c_{\mathrm{QED}} \, Y_{l,IJ} \delta F^{l,(1)}_{Y,ab} \, {\overline{l}}{}^{a}_{I} \phi_b \projR l^{2}_{J}
        + \delta G^{l,(1)}_{Y,abc,IJ} \, {\overline{l}}{}^{a}_{I} \phi^\dagger_b \projR l^{c}_{J}\\
        &\qquad\quad
        + c_{\mathrm{QED}} \, Y_{d,IJ} \delta F^{d,(1)}_{Y,ab} \, {\overline{q}}{}^{i,a}_{I} \phi_b \projR q^{i,2}_{J}
        + \delta G^{d,(1)}_{Y,abc,IJ} \, {\overline{q}}{}^{i,a}_{I} \phi^\dagger_b \projR q^{i,c}_{J}\\
        &\qquad\quad
        + c_{\mathrm{QED}} \, Y_{u,IJ} \delta F^{u,(1)}_{Y,ab} \, {\overline{q}}{}^{i,a}_{I} \phi^\dagger_b \projR q^{i,1}_{J}
        + \delta G^{u,(1)}_{Y,abc,IJ} \, {\overline{q}}{}^{i,a}_{I} \phi_b \projR q^{i,c}_{J}
        + \mathrm{h.c.}
        \bigg\},
    \end{aligned}
\end{equation}
with coefficients provided in App.~\ref{App:Finite_CountertermCoeffs_Expanded}.

After adding the two terms in Eqs.~\eqref{Eq:Finite_Symmetric_DownTypeYukawa_CT} and \eqref{Eq:Finite_Symmetric_UpTypeYukawa_CT}, the $\delta\mathcal{F}_Y$-type counterterms contain only electroweak contributions (independent of $c_\mathrm{QCD}$), with a global factor $c_\mathrm{QED}$ that can be pulled out.
They are thus sourced entirely by the evanescent ${\widehat{B}}{}_\mu$--fermion interactions and vanish for $c_\mathrm{QED}=0$, which is therefore the preferable choice.

The $\delta\mathcal{G}_Y$-type quark counterterms contain both electroweak and QCD contributions.
At 1-loop order, as in the fermionic sector, $c_\mathrm{QED}$ appears only in the $g_Y$-contributions and $c_\mathrm{QCD}$ only in the $g_s$-contributions, without mixing.
The leptonic counterterm is purely electroweak and independent of $c_\mathrm{QCD}$.
None of the $\delta\mathcal{G}_Y$-type counterterms vanishes for any choice of $c_\mathrm{QED}$ and $c_\mathrm{QCD}$, but all simplify for $c_\mathrm{QED},c_\mathrm{QCD}\in\{0,1\}$:
\begin{itemize}
    \item Leptons: neither value of $c_\mathrm{QED}$ is better, as they differ only in the prefactor of the $(\mathcal{Y}^l_R)^2$ term;
    \item Down-type quarks: $c_\mathrm{QED}=0$ reduces the number of electroweak terms in the coefficient, while the two values of $c_\mathrm{QCD}$ differ only in the prefactors of the QCD terms;
    \item Up-type quarks: $c_\mathrm{QED}=0$ again reduces the number of electroweak terms, and $c_\mathrm{QCD}=0$ is slightly preferable, leaving one QCD term in $\delta G^{u,(1)}_{Y,abc,IJ}$ instead of two.
\end{itemize}

In summary, the Yukawa sector favours $(c_\mathrm{QED},c_\mathrm{QCD})=(0,0)$, which minimises the number of finite counterterms, closely followed by $(c_\mathrm{QED},c_\mathrm{QCD})=(0,1)$.

\paragraph{Ghosts and BRST Transformations:}
As stated in Sec.~\eqref{Sec:SM_1-Loop_Renormalisation_Finite_CT}, the ghost sector needs no finite symmetry-restoring counterterms at the 1-loop level, $S^1_\mathrm{fct,ghost}=0$.
For the BRST transformations, decomposing the ghost twelveplet $c^A$ into $c_B$, $c_W^A$ and $c_G^A$ in Eq.~\eqref{Eq:SM-S_fct_ext} yields
\begin{equation}\label{Eq:SM-S1_fct_ext_expanded}
\begin{aligned}
    S_\mathrm{fct,ext}^{1} &= \frac{1}{16\pi^2} \int d^4x \bigg\{ 
    i \delta F^{(1)}_{R,W} \, g_W t_{L,ab}^A c_W^A \Big[
    {\overline{R}}{}^a_{l,I} \projL l_I^b + {\overline{R}}{}^{i,a}_{q,I} \projL q_I^{i,b} \Big]\\
    &\qquad\qquad\qquad
    + i (1-c_\mathrm{QCD}) \delta F^{(1)}_{R,G} \, g_s t_{s,ij}^A c_G^A {\overline{R}}{}^{i,a}_{q,I} q_I^{j,a}
    + \mathrm{h.c.}
    \bigg\},
\end{aligned}
\end{equation}
with coefficients collected in App.~\ref{App:Finite_CountertermCoeffs_Expanded}.

These counterterms are independent of $c_\mathrm{QED}$ and depend on $c_\mathrm{QCD}$ only through the prefactor $(1-c_\mathrm{QCD})$ of the last line of Eq.~\eqref{Eq:SM-S1_fct_ext_expanded}.
Choosing $c_\mathrm{QCD}=1$ removes the QCD counterterm $g_s t_{s,ij}^A c_G^A {\overline{R}}{}^{i,a}_{q,I} q_I^{j,a} + \mathrm{h.c.}$ entirely, leaving only the weak one $\propto g_W t_{L,ab}^Ac_W^A$, and is therefore preferable.

\subsection{The 1-Loop QCD Sector in the BMHV Framework}\label{Sec:QCD_Sector_in_BMHV}

The QCD sector deserves separate attention in the BMHV framework:
although vector-like, it always contributes to the regularisation-induced breaking at tree-level, independently of the evanescent gluon--quark interaction (see Sec.~\ref{Sec:SymmetryBreaking_in_SM}).
Recall that a purely 4-dimensional treatment ($c_\mathrm{QCD}=0$) causes a pure QCD breaking, Eq.~\eqref{Eq:SM-Breaking-s}, while a fully $D$-dimensional one ($c_\mathrm{QCD}=1$) causes mixed breakings, Eqs.~\eqref{Eq:SM-Breaking-Ys}--\eqref{Eq:SM-Breaking-Ws}.
The preceding sections showed that $c_\mathrm{QCD}=1$ is mostly, but not always, preferable (see the discussions for the finite Yukawa and singular fermion sectors).
Here we study the pure QCD sector at the 1-loop level, in particular how the tree-level breaking of Eq.~\eqref{Eq:SM-Breaking-s}, which is then the only breaking, propagates and depends on $c_\mathrm{QCD}$.

For this purpose, we specialise the counterterms of Secs.~\ref{Sec:SM-S1_SingularCT} and \ref{Sec:SM-S1_FiniteCT} to pure QCD by dropping all lepton, scalar and Yukawa contributions and setting all couplings except $g_s$ to zero.
The resulting counterterm action $S_\mathrm{ct,QCD}^1=S_\mathrm{sct,QCD}^1+S_\mathrm{fct,QCD}^1$ then produces the 1-loop breaking
\begin{align}
    \Delta_\mathrm{ct,QCD}^{1} = \Delta_\mathrm{sct,QCD}^{1} + \Delta_\mathrm{fct,QCD}^{1} = b_D ( S_\mathrm{sct,QCD}^1 ) + b_D ( S_\mathrm{fct,QCD}^1 ),
\end{align}
where $b_D$ is the linearised Slavnov-Taylor operator as introduced above Eq.~\eqref{Eq:STI_1-loop_RestorationRequirement}.
The singular part is given by
\begin{align}\label{Eq:Delta1_sct_QCD}
        &\Delta^{1}_{\mathrm{sct,QCD}} = \frac{( 1 - c_{\mathrm{QCD}} )}{16\pi^2} \Dintx \bigg\{
        \delta {\widehat{X}}{}^{(1)}_{c\overline{q}q} \, g_s t^A_{s,ij} \, c^A_G \, {\widehat{\partial}}{}_\mu \big( {\overline{q}}{}^{i,a}_{I} {\widehat{\gamma}}{}^{\mu} q^{j,a}_{I} \big)
        \nonumber\\
        &\qquad\quad
        - \delta \overline{Z}^{(1)}_{G} \bigg[ 
        c^A_G \, \widehat{\Box} \, {\overline{\partial}}{}^{\mu} {\overline{G}}{}^{A}_{\mu}
        - c_{\mathrm{QCD}} \, c^A_G \, \overline{\Box} \, {\widehat{\partial}}{}^{\mu} {\widehat{G}}{}^{A}_{\mu}
        - g_s f^{ABC} c^A_G \, {\overline{G}}{}^{B}_{\mu} \, \widehat{\Box} \, {\overline{\eta}}{}^{\mu\nu} {\overline{G}}{}^{C}_{\nu}
        \nonumber\\
        &\qquad\qquad\quad
        + c_{\mathrm{QCD}} \, g_s f^{ABC} \Big[
        c_{\mathrm{QCD}} \, c^A_G {\widehat{G}}{}^{B}_{\mu} \big( {\widehat{\partial}}{}^{\mu} {\widehat{\partial}}{}^{\nu} - \widehat{\Box} \, {\widehat{\eta}}{}^{\mu\nu} \big) {\widehat{G}}{}^{C}_{\nu}
        + 2 \, c^A_G {\overline{G}}{}^{B}_{\mu}  {\overline{\partial}}{}^{\mu} {\widehat{\partial}}{}^{\nu} {\widehat{G}}{}^{C}_{\nu}
        \nonumber\\
        &\qquad\qquad\qquad\qquad\qquad
        - c^A_G {\widehat{G}}{}^{B}_{\mu} {\widehat{\partial}}{}^{\mu} {\overline{\partial}}{}^{\nu} {\overline{G}}{}^{C}_{\nu}
        - 2 \, c^A_G \big( {\overline{\partial}}{}^{\mu} {\widehat{G}}{}^{B}_{\nu} \big) \big( {\widehat{\partial}}{}^{\nu} {\overline{G}}{}^{C}_{\mu} \big)
        + c^A_G \big( {\overline{\partial}}{}^{\mu} {\overline{G}}{}^{B}_{\mu} \big) \big( {\widehat{\partial}}{}^{\nu} {\widehat{G}}{}^{C}_{\nu} \big)
        \Big]
        \nonumber\\
        &\qquad\qquad\quad
        + c_{\mathrm{QCD}} \, g_s^2 f^{ABE} f^{CDE} c^A_G \Big[
        {\widehat{\partial}}{}^{\mu} \big( {\overline{G}}{}^{B}_{\nu} {\widehat{G}}{}^{C}_{\mu} {\overline{G}}{}^{D,\nu} \big)
        + c_{\mathrm{QCD}}^2 \, {\widehat{\partial}}{}^{\mu} \big( {\widehat{G}}{}^{B}_{\nu} {\widehat{G}}{}^{C}_{\mu} {\widehat{G}}{}^{D,\nu} \big)
        \Big]
        \bigg]
        \bigg\},
\end{align}
and the finite part by
\begin{align}\label{Eq:Delta1_fct_QCD}
        \Delta^{1}_{\mathrm{fct,QCD}} &= \frac{( 1 - c_{\mathrm{QCD}} )}{16\pi^2} \Dintx \bigg\{
        \delta F^{(1)}_{c\overline{q}q} \, g_s t^A_{s,ij} \, c^A_G \, {\overline{\partial}}{}_\mu \big( {\overline{q}}{}^{i,a}_{I} {\overline{\gamma}}{}^{\mu} q^{j,a}_{I} \big)
        \nonumber\\
        &
        + \delta F^{(1)}_{R,G} \bigg[ g_s t^A_{s,ij} \, c^A_G \, \partial_\mu \big( {\overline{q}}{}^{i,a}_{I} {\widehat{\gamma}}{}^{\mu} q^{j,a}_{I} \big)
        - g_s^2 f^{ABC} t^C_{s,ij} \, c^A_G \, {\overline{q}}{}^{i,a}_{I} \Big[ \slashed{G}^{B} - ( 1 - c_{\mathrm{QCD}} ) {\widehat{\slashed{G}}}{}^{B} \Big] q^{j,a}_{I}
        \nonumber\\
        &\qquad\qquad
        - \frac{i}{2} g_s^2 f^{ABC} t^C_{s,ij} \Big(
        c^A_G c^B_G \, {\overline{R}}{}^{i,a}_{q,I} q^{j,a}_{I}
        - c^A_G c^B_G \, {\overline{q}}{}^{i,a}_{I} R^{j,a}_{q,I}
        \Big)
        \bigg]
        \bigg\}.
\end{align}
Here $\delta \overline{Z}^{(1)}_{G}$ and $\delta F^{(1)}_{R,G}$ are given in Eqs.~\eqref{Eq:SCT-dZGbar1} and \eqref{Eq:FCT-dFRG1-ExternalFermionSource}, and the newly introduced coefficients are
\begin{subequations}\label{Eq:QCD_Only_Breaking_Coefficients}
    \begin{align}\label{Eq:QCD_Only_SCT_dXcQq}
        \begin{split}
            \delta {\widehat{X}}{}^{(1)}_{c\overline{q}q} &= \frac{g_s^2}{\epsilon} \bigg\{
            \bigg[ \frac{6-2\xi_F}{3}-c_\mathrm{QCD}\bigg( \frac{2\xi_F}{3} + c_\mathrm{QCD} \frac{3-\xi_F}{3} \bigg) \bigg] C_2(F_s)
            \\
            &\quad\,\,
            - \frac{1}{4} \bigg[ \frac{6+4\xi_F}{3}-c_\mathrm{QCD}\bigg( \xi_F + c_\mathrm{QCD} \frac{6-2\xi_F}{3} \bigg) \bigg] C_2(G_s)
            \bigg\},
        \end{split}\\[0.75ex]
        \begin{split}\label{Eq:QCD_Only_FCT_dFcQq}
            \delta F^{(1)}_{c\overline{q}q} &= - \frac{3 + ( 2 + c_\mathrm{QCD} ) \xi_F}{12} g_s^2 \, C_2(G_s) \ .
        \end{split}
    \end{align}
\end{subequations}
Both Eq.~\eqref{Eq:Delta1_sct_QCD} and Eq.~\eqref{Eq:Delta1_fct_QCD} carry a global factor $(1-c_\mathrm{QCD})$.
For $c_\mathrm{QCD}=1$ the breaking therefore vanishes entirely, as expected because the present scenario then corresponds to ordinary QCD in naive DReg.
For $c_\mathrm{QCD}=0$ it is nonvanishing, although some terms of Eq.~\eqref{Eq:Delta1_sct_QCD} (the second term of the second line and the three last lines) vanish for both $c_\mathrm{QCD}\in\{0,1\}$ due to an additional prefactor $c_\mathrm{QCD}$.
The breaking $\Delta_\mathrm{ct,QCD}^{1}$ constitutes the 1-loop version of the tree-level breaking of Eq.~\eqref{Eq:SM-Breaking-s} and includes a finite 4-dimensional part.
The finite corrections to the non-Abelian BRST transformation of the quarks, see Eq.~\eqref{Eq:SM-S1_fct_ext_expanded}, do not compensate it but add further finite breakings (last two lines of Eq.~\eqref{Eq:Delta1_fct_QCD}).
Hence, even vector-like QCD requires divergent and finite symmetry-restoring counterterms if the gluon--quark interaction is not fully $D$-dimensional, owing to the mismatch with the $D$-dimensional quark kinetic term.

In summary, $c_\mathrm{QCD}=1$ is the best choice for the pure QCD sector.
However, this does not remove all 1-loop breakings $\propto g_s$:
some mixed breakings, the 1-loop analogue of Eqs.~\eqref{Eq:SM-Breaking-Ys}--\eqref{Eq:SM-Breaking-Ws} and not considered here, favour $c_\mathrm{QCD}=0$.
This was already hinted at in Sec.~\ref{Sec:SM-S1_SingularCT} for the singular fermion counterterms and is also addressed in the next subsection.

\subsection{Concluding Comments on Evanescent Gauge Interactions}\label{Sec:Concluding_Comments_on_Evanescent_Gauge_Interactions}

Throughout the electroweak sector, $c_\mathrm{QED}=0$ yields the most economical counterterms:
many of them vanish or simplify considerably for this choice.
Introducing evanescent $U(1)_Y$ gauge interactions ($c_\mathrm{QED}\neq0$) is thus disadvantageous, in agreement with Ref.~\cite{Ebert:2024xpy}.

Since our analysis is performed in the symmetric phase of the SM, the question arises whether this conclusion persists in the broken phase, expressed in terms of mass eigenstates $A_\mu$, $Z_\mu$ and $W_\mu^{\pm}$.
After all, $c_\mathrm{QED}$ was introduced solely to render the photon current fully $D$-dimensional, see Eq.~\eqref{Eq:Photon-Interaction-Current-SM}, not the $B_\mu$ current.
A complete broken-phase analysis is left for future work, but we expect the conclusion to hold.
Because $B_\mu$ contains both $A_\mu$ and $Z_\mu$, see Eq.~\eqref{Eq:B-Boson}, a nonzero $c_\mathrm{QED}$ also generates additional evanescent $Z$ interactions, see Eq.~\eqref{Eq:Z-Interaction-Current-SM}.
Unlike for the vector-like gluon current, these cannot combine into fully $D$-dimensional or symmetric structures, since the axial part of a chiral gauge interaction admits no uniform continuation to $D$ dimensions, see Ref.~\cite{Ebert:2024xpy}.
This remains true for any nonzero evanescent generator $\widehat{t}^A$, which we avoided because of the complications in the lepton--neutrino sector, see the discussion around Eqs.~\eqref{Eq:Evanescent-Neutrino-Lepton-Interactions} and \eqref{Eq:Evanescent-SU2-Generator-Lepton-Sector}.
The $Z$ interaction current therefore remains neither fully $D$-dimensional nor vector-like, and the additional terms only source further counterterms.
The same argument applies to the $W_\mu^\pm$ current, which cannot be rendered $D$-dimensional either.
 
For the QCD sector, the following considerations apply equally in the broken phase, since $SU(3)_c$ is unbroken.
In most sectors $c_\mathrm{QCD}=1$ (fully $D$-dimensional gluon interaction current) yields the most economical result.
In particular, it renders the 1-loop gluon counterterm action in Eq.~\eqref{Eq:SM-S1_sct_QCD-gauge_expanded} fully $D$-dimensional and BRST invariant, and, as shown in Sec.~\ref{Sec:QCD_Sector_in_BMHV}, it is the only choice for which pure QCD corrections preserve BRST invariance.
However, this preference is less straightforward than for $c_\mathrm{QED}$.
Already at tree-level, $c_\mathrm{QCD}=1$ trades the pure QCD breaking of Eq.~\eqref{Eq:SM-Breaking-s} for the mixed breakings of Eqs.~\eqref{Eq:SM-Breaking-Ys}--\eqref{Eq:SM-Breaking-Ws}.
The evanescent QCD--Yukawa counterterm $-\delta {\widehat{X}}{}^{q,(1)}_{Y,ab,IJ}g_st^A_{s,ij}{\overline{q}}{}^{i,a}_{I}{\widehat{\slashed{G}}}{}^{A}\projR q^{j,b}_{J}+\mathrm{h.c.}$ of Eq.~\eqref{Eq:SM-S1_sct_Quark_expanded} is one example that this pattern persists at 1-loop order.
This becomes more explicit in the $b_D$-variation of the singular quark counterterms, Eq.~\eqref{Eq:SM-S1_sct_Quark_expanded}, of which we display a relevant excerpt,
\begin{align}
    b_D &S^1_\mathrm{sct,quark} \supset \frac{1}{16\pi^2} \Dintx \bigg\{
    \delta {\overline{X}}{}^{(1)}_{\overline{q}Gq} \, g_s t^A_{s,ij} \, c^A_G \, \partial_\mu \big( {\overline{q}}{}^{i,a}_{I} \gamma^{\mu} q^{j,a}_{I} \big)
    \nonumber\\
    &\qquad
    - (1-c_\mathrm{QCD}) 
    \frac{g_s^2}{4\epsilon} \bigg[ \frac{\xi_F}{3} - c_\mathrm{QCD} \Big( \xi_F + c_\mathrm{QCD} \frac{6-2\xi_F}{3} \Big) \bigg] C_2(G_s)
    g_s t^A_{s,ij} \, c^A_G \, {\widehat{\partial}}{}_\mu \big( {\overline{q}}{}^{i,a}_{I} {\widehat{\gamma}}{}^{\mu} q^{j,a}_{I} \big)
    \nonumber\\
    &\qquad
    - (1-c_\mathrm{QCD}) \delta{\widehat{X}}{}^{q,(1)}_{ab} g_s t^A_{s,ij} \, c^A_G \, {\widehat{\partial}}{}_\mu \big( {\overline{q}}{}^{i,a}_{I} {\widehat{\gamma}}{}^{\mu} q^{j,b}_{I} \big)
    \nonumber\\
    &\qquad
    - \Big[
    ( 1 - c_{\mathrm{QCD}} ) \delta {\widehat{X}}{}^{q,(1)}_{Y,ab,IJ} \, g_s t^A_{s,ij} \, c^A_G \, {\widehat{\partial}}{}_\mu \big( {\overline{q}}{}^{i,a}_{I} {\widehat{\gamma}}{}^{\mu} \projR q^{j,b}_{J} \big)
    \nonumber\\
    &\qquad\quad
    + i g_Y g_s \, \delta {\widehat{X}}{}^{(1)}_{\overline{q}Gq} \big( \mathcal{Y}^q_L - \mathcal{Y}^q_R \big)_{ab} t^A_{s,ij} \, c_B {\overline{q}}{}^{i,a}_{I} {\widehat{\slashed{G}}}{}^{A} \projR q^{j,b}_{I}
    \nonumber\\
    &\qquad\quad
    + i c_{\mathrm{QCD}} g_Y g_s \big( \mathcal{Y}^q_L - \mathcal{Y}^q_R \big)_{ac} \delta {\widehat{X}}{}^{q,(1)}_{cb} t^A_{s,ij} \, c_B  {\overline{q}}{}^{i,a}_{I} {\widehat{\slashed{G}}}{}^{A} \projR q^{j,b}_{I}
    \nonumber\\
    &\qquad\quad
    + i c_{\mathrm{QCD}} g_Y g_s \big( \mathcal{Y}^q_L - \mathcal{Y}^q_R \big)_{ac} \delta {\widehat{X}}{}^{q,(1)}_{Y,cb,IJ} t^A_{s,ij} \, c_B {\overline{q}}{}^{i,a}_{I} {\widehat{\slashed{G}}}{}^{A} \projR q^{j,b}_{J}
    \nonumber\\
    &\qquad\quad
    + i g_W g_s \, \delta {\widehat{X}}{}^{(1)}_{\overline{q}Gq} t^A_{L,ab} t^B_{s,ij} \, c^A_W {\overline{q}}{}^{i,a}_{I} {\widehat{\slashed{G}}}{}^{B} \projR q^{j,b}_{I}
    \nonumber\\
    &\qquad\quad
    + i c_{\mathrm{QCD}} g_W g_s t^A_{L,ac} \delta {\widehat{X}}{}^{q,(1)}_{cb} t^B_{s,ij} \, c^A_W {\overline{q}}{}^{i,a}_{I} {\widehat{\slashed{G}}}{}^{B} \projR q^{j,b}_{I}
    \nonumber\\
    &\qquad\quad
    + i c_{\mathrm{QCD}} g_W g_s t^A_{L,ac} \delta {\widehat{X}}{}^{q,(1)}_{Y,cb,IJ} t^B_{s,ij} \, c^A_W {\overline{q}}{}^{i,a}_{I}{\widehat{\slashed{G}}}{}^{B} \projR q^{j,b}_{J}
    + \mathrm{h.c.} 
    \Big]
    \bigg\},
\end{align}
with coefficients collected in App.~\ref{App:Divergent_CountertermCoeffs_Expanded}.
The second to fourth lines are pure QCD breakings and vanish for $c_\mathrm{QCD}=1$.
The sixth, seventh and last two lines are mixed breakings, which arise from evanescent quark--gluon terms that are not invariant under the electroweak part of their BRST transformations, and vanish for $c_\mathrm{QCD}=0$. 
The remaining lines survive for either choice.
The 1-loop breaking thus mirrors the tree-level dilemma (see Sec.~\ref{Sec:TreeLevelSymBreaking}), with neither value removing all breakings $\propto g_s$.

We nevertheless can identify the following four advantages of the choice $c_\mathrm{QCD}=1$.
First, it eliminates all pure QCD breakings, so that every remaining breaking $\propto g_s$ is of mixed electroweak--QCD nature.
Second, it reduces the number of counterterms in most sectors and makes the gluon 1-loop counterterms of Eq.~\eqref{Eq:SM-S1_sct_QCD-gauge_expanded} fully BRST invariant, which in turn reduces the number of counterterm and $\Delta$ vertices required in higher-order computations.
Third, pure QCD radiative corrections take the familiar form of naive DReg calculations.
Fourth, in the electroweak gaugeless limit ($g_Y=0$, $g_W=0$), employed when higher-order QCD corrections dominate phenomenologically over electroweak ones, only $SU(3)_c$ is gauged and the BMHV scheme then affects only global symmetries, so that no local gauge and BRST symmetry-restoring counterterms are needed.

For this reason, we recommend treating the electroweak gauge interactions purely 4-dimen\-sion\-ally and the gluon interaction fully $D$-dimensionally, i.e.\ $(c_\mathrm{QED},c_\mathrm{QCD})=(0,1)$.
The choice $(c_\mathrm{QED},c_\mathrm{QCD})=(0,0)$, where all gauge interaction currents are treated purely 4-dimensionally, is slightly less advantageous but still practical, while all other values are discouraged.


\section{Conclusions}\label{Sec:Conclusion}

In this work, we have presented the complete 1-loop renormalisation of the full Standard Model in dimensional regularisation with non-anticommuting $\gamma_5$ in the BMHV scheme.
The calculation was performed in general $R_\xi$-gauge and analysed within the BRST formalism.
The regularisation-induced breaking of gauge and BRST invariance was determined via the regularised quantum action principle, where the breaking is rewritten in terms of a local operator insertion.
We have provided the full 1-loop counterterm action, including both singular and finite symmetry-restoring counterterms, which render the theory UV finite and re-establish the Slavnov-Taylor identity for Green functions with arbitrary off-shell momenta.
To our knowledge, this constitutes the first complete 1-loop renormalisation of the full Standard Model in $R_\xi$-gauge within the self-consistent BMHV scheme that includes both types of counterterms, providing a finite theory and the explicit restoration of the spuriously broken symmetry.

All results were obtained employing automated computational frameworks.
The $\Delta$-operator inserted Green functions, which are technically more challenging than ordinary ones, were computed twice using two independently developed computational setups, one built around \texttt{FORM} and the other based on \texttt{Mathematica}, and agree across both.
Further nontrivial consistency checks include the explicit verification that the complete set of counterterms cancels the 1-loop breaking $\widehat{\Delta}\cdot\Gamma^1$, satisfying the 1-loop restoration requirement, and the reproduction of known results in specific limits: the Abelian sector of the SM, as well as ordinary QED and ordinary QCD.
We presented the counterterms in two complementary forms: a compact representation in terms of gauge-field twelveplets and generalised generators (see Sec.~\ref{Sec:SM_1-Loop_Renormalisation}), well suited for automated implementations, and a sector-wise representation in terms of $B_\mu$, $W^A_\mu$ and $G^A_\mu$ (see Sec.~\ref{Sec:SM_1-Loop_Renormalisation_Expanded}), which makes the structure of the individual gauge sectors transparent and allows for a better discussion of the results.

A central aspect of this work is the comprehensive analysis of evanescent gauge interactions in the full SM, parametrised by $c_\mathrm{QED}$ and $c_\mathrm{QCD}$. 
For suitable choices, the vector-like photon and gluon interaction currents can be rendered fully $D$-dimensional.
In contrast, this is impossible for the $Z$ and $W^\pm$ currents, because the axial part of a chiral gauge interaction admits no uniform continuation to $D$ dimensions. 
Irrespective of the treatment of the gluon interaction, we showed that QCD always contributes to the regularisation-induced symmetry breaking, both at tree level and at the 1-loop level.
A purely 4-dimensional gluon current causes pure QCD breakings, while a fully $D$-dimensional one removes these but induces mixed electroweak--QCD breakings.
As a consequence, despite its vector-like nature, QCD cannot be omitted from the symmetry-restoration procedure, and the SM must be treated as a whole. 
Only in the electroweak gaugeless limit, combined with a fully $D$-dimensional gluon current, local gauge and BRST symmetry remain unbroken; in this case the BMHV scheme affects only global symmetries. As an example, the global hypercharge-violating counterterm $\phi_2 \overline{\Box} \phi_2$ in Eq.~(\ref{Eq:SM-S1_fct_Higgs_expanded}) persists even in the gaugeless limit.

We recommend treating the electroweak gauge interactions including the photon purely 4-dimensionally, i.e.\ setting $c_\mathrm{QED}=0$. This choice significantly simplifies the structure of symmetry-restoring counterterms. For the gluon interaction, both options $c_\mathrm{QCD}=0$ or  $c_\mathrm{QCD}=1$ have their respective advantages and are practical. In total, we assess the   fully $D$-dimensional treatment of the gluon, i.e.\ $c_\mathrm{QCD}=1$, as superior because this choice eliminates all pure QCD breakings, minimises the number of counterterms in most sectors, renders the purely gluonic 1-loop counterterms fully $D$-dimensional and BRST invariant, and reproduces the familiar form of pure QCD corrections. 

Although our analysis is performed in the symmetric phase of the SM, it already determines all 1-loop counterterms of dimension-4 operators, and thus those required for the 1-loop $\beta$-functions.
Only the counterterms of operators with mass dimension $\leq 3$ remain to be adjusted in the broken phase, because for $\xi\ne0$ the gauge-fixing functional for $Z$ and $W^\pm$ then involves dimensionful terms, the Goldstone bosons and the vacuum expectation value.
Nonetheless, we expect our recommendation for the evanescent gauge interactions to persist in the broken phase, as nonzero evanescent hypercharge interactions also affect the $Z$ boson, whose current cannot be made fully $D$-dimensional.

The results presented here are an essential stepping stone towards a rigorous and fully self-consistent multi-loop renormalisation of the SM in the BMHV scheme, which requires the complete 1-loop counterterm action. 
Such a framework will be indispensable for higher-order predictions of electroweak precision observables, needed to match the steadily increasing precision of current and future experiments.


\section*{Acknowledgments}

We would like to thank Andreas von Manteuffel, Andrea Quadri and Javier Fuentes-Mart{\'\i}n for valuable discussions. 
M.W.\ acknowledges support by the Deutsche Forschungsgemeinschaft (DFG, German Research Foundation) under Germany's Excellence Strategy --- EXC 2121 ``Quantum Universe'' --- 390833306.
This work has been partially funded by the DFG project~491245950, and we acknowledge financial support by the DFG, grant STO 876/8-2.


\begin{appendices}
\addappheadtotoc
\clearpage

\section{Group Invariants, Coupling Structures and Abbreviations}\label{App:SM-GroupInvariants-CouplingStructures}

Here, we collect a number of structures constructed from the generators and coupling matrices of the Standard Model, such as group invariants, which will be used in the presentation of the counterterm results.

We begin with the group invariants associated with the fundamental generators of the SM gauge groups.
In particular, the Dynkin indices are defined as
\begin{equation}\label{Eq:Dynkin-Indices}
    \begin{alignedat}{2}
        S_{2}\big(F_L\big) \delta^{AB} &= \myTrbig{ t_L^A t_L^B } &&= \frac{1}{2} \, \delta^{AB},\\
        S_{2}\big(F_s\big) \delta^{AB} &= \myTrbig{ t_s^A t_s^B } &&= \frac{1}{2} \, \delta^{AB},
    \end{alignedat}
\end{equation}
and the quadratic Casimirs by
\begin{equation}\label{Eq:Quadratic-Casimirs}
    \begin{alignedat}{3}
        C_{2}\big(F_L\big) \delta_{ab} &= t^A_{L,ac} t^A_{L,cb} &&= \frac{3}{4} \, \delta_{ab}, &&\\
        C_{2}\big(F_s\big) \delta_{ij} &= t^A_{s,ik} t^A_{s,kj} &&= \frac{4}{3} \, \delta_{ij}, &&\\
        C_{2}\big(G_L\big) \delta^{AB} &= \varepsilon^{ACD}\varepsilon^{BCD} &&= N_L \, \delta^{AB} &&= 2 \, \delta^{AB},\\
        C_{2}\big(G_s\big) \delta^{AB} &= f^{ACD}f^{BCD} &&= N_c \, \delta^{AB} &&= 3 \, \delta^{AB}.
    \end{alignedat}
\end{equation}

Next, we extend these definitions to the generalised generators introduced in Eqs.~\eqref{Eq:LeptonQuark-Super-GeneratorPhysical}, \eqref{Eq:LeptonQuark-Super-GeneratorEvanescent} and \eqref{Eq:Scalar-Super-Generator}, and the structure constants in Eq.~\eqref{Eq:Super-Structure-Constants}. 
For readability we do not explicitly use the shifted index notation introduced in Eq.~\eqref{Eq:LeptonQuark-Super-GeneratorPhysical} for the fundamental generators $t_L^A$ and $t_s^A$ (cf.\ Eq.~\eqref{Eq:SM-Generators}), or the $\dot{X}$ and $\ddot{X}$ notation ($X\in\{A,B,C,\ldots\}$) from Eq.~\eqref{Eq:Super-Structure-Constants}, as the index ranges are obvious from the context.
Analogous to the above constructions, we define generalised Dynkin indices, which for the scalar generators read
\begin{equation}\label{Eq:Scalar-Super-Dynkin-Indices}
    \begin{alignedat}{3}
        \mathcal{S}_2\big(S^{A'}\big) \delta^{AB} 
        &= \myTrbig{\Theta^A\Theta^B} &&= 
        \begin{cases}
            g_Y^2 \, \mathcal{Y}_S^2 \, N_L
            , \quad &A,B = 1\\
            g_W^2 \, S_{2}\big(F_L\big) \, \delta^{AB}, \quad &A,B\in\{2,3,4\}\\
            0, \quad &\mathrm{else}.
        \end{cases}
    \end{alignedat}
\end{equation}
For the lepton generators, we define
\begin{equation}\label{Eq:Lepton-Super-Dynkin-Indices}
    \begin{alignedat}{2}
        \mathcal{S}_2\big(F^{l,A'}_R\big) \delta^{AB} 
        &= \myTrbig{\myGenLe_R^A\myGenLe_R^B} &&= 
        \begin{cases}
            g_Y^2 \, \myTrbig{(\mathcal{Y}^l_R)^2}
            , \quad &A,B = 1\\
            0, \quad &\mathrm{else},
        \end{cases}\\
        \mathcal{S}_2\big(F^{l,A'}_L\big) \delta^{AB} 
        &= \myTrbig{\myGenLe_L^A\myGenLe_L^B} &&= 
        \begin{cases}
            g_Y^2 \, \myTrbig{(\mathcal{Y}^l_L)^2}
            , \quad &A,B = 1\\
            g_W^2 \, S_{2}\big(F_L\big) \delta^{AB}, \quad &A,B\in\{2,3,4\}\\
            0, \quad &\mathrm{else}.
        \end{cases}
    \end{alignedat}
\end{equation}
Similarly, for the quark generators, we find
\begin{equation}\label{Eq:Quark-Super-Dynkin-Indices}
    \begin{alignedat}{2}
        \mathcal{S}_2\big(F^{q,A'}_R\big) \delta^{AB} 
        &= \myTrbig{\myGenQu_R^A\myGenQu_R^B} &&= 
        \begin{cases}
            g_Y^2 \, N_c \, \myTrbig{(\mathcal{Y}^q_R)^2} 
            , \quad &A,B = 1\\
            g_s^2 \, N_L \, S_{2}\big(F_s\big) \delta^{AB}, \quad &A,B\in\{5,\ldots,12\}\\
            0, \quad &\mathrm{else},
        \end{cases}\\
        \mathcal{S}_2\big(F^{q,A'}_L\big) \delta^{AB} 
        &= \myTrbig{\myGenQu_L^A\myGenQu_L^B} &&= 
        \begin{cases}
            g_Y^2 \, N_c \, \myTrbig{(\mathcal{Y}^q_L)^2} 
            , \quad &A,B = 1\\
            g_W^2 \, N_c \, S_{2}\big(F_L\big) \delta^{AB}, \quad &A,B\in\{2,3,4\}\\
            g_s^2 \, N_L \, S_{2}\big(F_s\big) \delta^{AB}, \quad &A,B\in\{5,\ldots,12\}\\
            0, \quad &\mathrm{else}.
        \end{cases}
    \end{alignedat}
\end{equation}

At the 1-loop level and beyond, one also encounters mixed traces involving both left- and right-handed generators.
For the leptons, we define
\begin{equation}\label{Eq:Lepton-Super-Dynkin-Indices-Mixed}
    \begin{alignedat}{2}
        \mathcal{S}^{AB}_2\big(F^{l}_R,F^{l}_L\big) 
        &= \myTrbig{\myGenLe_R^A\myGenLe_L^B} &&= 
        \begin{cases}
            g_Y^2 \, \myTrbig{\mathcal{Y}^l_R\mathcal{Y}^l_L} 
            , \quad &A,B = 1\\
            g_Y g_W \, \myTrbig{ \mathcal{Y}^l_R t_L^B }, \quad &A=1,B\in\{2,3,4\}\\
            0, \quad &\mathrm{else},
        \end{cases}\\
        \mathcal{S}^{AB}_2\big(F^{l}_L,F^{l}_R\big) 
        &= \myTrbig{\myGenLe_L^A\myGenLe_R^B} &&= \mathcal{S}^{BA}_2\big(F^{l}_R,F^{l}_L\big).
    \end{alignedat}
\end{equation}
Analogously, for the quarks we obtain
\begin{equation}\label{Eq:Quark-Super-Dynkin-Indices-Mixed}
    \begin{alignedat}{2}
        \mathcal{S}^{AB}_2\big(F^{q}_R,F^{q}_L\big) 
        &= \myTrbig{\myGenQu_R^A\myGenQu_L^B} &&= 
        \begin{cases}
            g_Y^2 \, N_c \, \myTrbig{\mathcal{Y}^q_R\mathcal{Y}^q_L} 
            , \quad &A,B = 1\\
            g_Y g_W \, N_c \, \myTrbig{ \mathcal{Y}^q_R t_L^B }, \quad &A=1,B\in\{2,3,4\}\\
            g_s^2 \, N_L \, S_{2}\big(F_s\big) \delta^{AB}, \quad &A,B\in\{5,\ldots,12\}\\
            0, \quad &\mathrm{else},
        \end{cases}\\
        \mathcal{S}^{AB}_2\big(F^{q}_L,F^{q}_R\big) 
        &= \myTrbig{\myGenQu_L^A\myGenQu_R^B} &&= \mathcal{S}^{BA}_2\big(F^{q}_R,F^{q}_L\big).
    \end{alignedat}
\end{equation}

Since the analysis also includes evanescent generators (see Eqs.~\eqref{Eq:LeptonQuark-Super-GeneratorEvanescent} and \eqref{Eq:Evanescent-Generators-Choice}), traces involving these must likewise be considered.
For the leptons, we find
\begin{equation}\label{Eq:Lepton-Super-Dynkin-Indices-Evanescent}
    \begin{alignedat}{2}
        \mathcal{S}_2\big(F^{l,A'}_{R},F^{l,B'}_{LR}\big) \delta^{AB} 
        &= \myTrbig{\myGenLe^A_R\widehat{\myGenLe}^B} &&= 
        \begin{cases}
            g_Y^2 \, c_{\mathrm{QED}} \myTrbig{(\mathcal{Y}^l_R)^2}, \quad &A,B = 1\\
            0, \quad &\mathrm{else},
        \end{cases}\\
        \mathcal{S}_2\big(F^{l,A'}_{LR},F^{l,B'}_{R}\big) \delta^{AB}
        &= \myTrbig{\widehat{\myGenLe}^A\myGenLe^B_R} &&= \mathcal{S}_2\big(F^{l,A'}_{R},F^{l,B'}_{LR}\big) \delta^{AB},\\
        \mathcal{S}^{AB}_2\big(F^{l}_{L},F^{l}_{LR}\big) 
        &= \myTrbig{\myGenLe^A_L\widehat{\myGenLe}^B} &&= 
        \begin{cases}
            g_Y^2 \, c_{\mathrm{QED}} \myTrbig{\mathcal{Y}^l_L\mathcal{Y}^l_R}, \quad &A,B = 1\\
            g_Y g_W \, c_{\mathrm{QED}} \myTrbig{t^A_L\mathcal{Y}^l_R}, \quad &A \in \{2,3,4\}, B = 1\\
            0, \quad &\mathrm{else},
        \end{cases}\\
        \mathcal{S}^{AB}_2\big(F^{l}_{LR},F^{l}_{L}\big)
        &= \myTrbig{\widehat{\myGenLe}^A\myGenLe^B_L} &&= \mathcal{S}^{BA}_2\big(F^{l}_{L},F^{l}_{LR}\big),\\
        \mathcal{S}_2\big(F^{l,A'}_{LR}\big) \delta^{AB}
        &= \myTrbig{\widehat{\myGenLe}^A\widehat{\myGenLe}^B} &&= 
        \begin{cases}
            g_Y^2 \, c_{\mathrm{QED}}^2 \myTrbig{(\mathcal{Y}^l_R)^2}, \quad &A,B = 1\\
            0, \quad &\mathrm{else}.
        \end{cases}
    \end{alignedat}
\end{equation}
For the quarks, we analogously obtain
{\allowdisplaybreaks[4]
\begin{alignat}{2}
    \mathcal{S}_2\big(F^{q,A'}_{R},F^{q,B'}_{LR}\big) \delta^{AB} 
    &= \myTrbig{\myGenQu^A_R\widehat{\myGenQu}^B} &&= 
    \begin{cases}
        g_Y^2 \, c_{\mathrm{QED}} N_c \, \myTrbig{(\mathcal{Y}^q_R)^2}, \quad &A,B = 1\\
        g_s^2 \, c_{\mathrm{QCD}} N_L \, S_{2}\big(F_s\big) \delta^{AB}, \quad &A,B \in \{5,\ldots,12\}\\
        0, \quad &\mathrm{else},
    \end{cases}\notag\\
    \mathcal{S}_2\big(F^{q,A'}_{LR},F^{q,B'}_{R}\big) \delta^{AB}
    &= \myTrbig{\widehat{\myGenQu}^A\myGenQu^B_R} &&= \mathcal{S}_2\big(F^{q,A'}_{R},F^{q,B'}_{LR}\big) \delta^{AB},\notag\\
    \mathcal{S}^{AB}_2\big(F^{q}_{L},F^{q}_{LR}\big) 
    &= \myTrbig{\myGenQu^A_L\widehat{\myGenQu}^B} &&= 
    \begin{cases}
        g_Y^2 \, c_{\mathrm{QED}} N_c \, \myTrbig{\mathcal{Y}^q_L\mathcal{Y}^q_R}, \quad &A,B = 1\\
        g_Y g_W \, c_{\mathrm{QED}} N_c \, \myTrbig{t^A_L\mathcal{Y}^q_R}, \quad &A \in \{2,3,4\}, B = 1\\
        g_s^2 \, c_{\mathrm{QCD}} N_L \, S_{2}\big(F_s\big) \delta^{AB}, \quad &A,B \in \{5,\ldots,12\}\\
        0, \quad &\mathrm{else},
    \end{cases}\notag\\
    \mathcal{S}^{AB}_2\big(F^{q}_{LR},F^{q}_{L}\big)
    &= \myTrbig{\widehat{\myGenQu}^A\myGenQu^B_L} &&= \mathcal{S}^{BA}_2\big(F^{q}_{L},F^{q}_{LR}\big),\notag\\
    \mathcal{S}_2\big(F^{q,A'}_{LR}\big) \delta^{AB}
    &= \myTrbig{\widehat{\myGenQu}^A\widehat{\myGenQu}^B} &&= 
    \begin{cases}
        g_Y^2 \, c_{\mathrm{QED}}^2 N_c \, \myTrbig{(\mathcal{Y}^q_R)^2}, 
        \quad &A,B = 1\\
        g_s^2 \, c_{\mathrm{QCD}}^2 N_L \, S_{2}\big(F_s\big) \delta^{AB}, 
        \quad &A,B\in\{5,\ldots,12\}\\
        0, \quad &\mathrm{else}.
    \end{cases}
    \label{Eq:Quarks-Super-Dynkin-Indices-Evanescent}
\end{alignat}
}

Furthermore, for traces involving four generators, we define
\begin{align}
    \mathcal{S}_{4}^{ABCD}\big(X_1,X_2,X_3,X_4\big) &= \myTrbig{X_1^A X_2^B X_3^C X_4^D} + \myTrbig{X_2^B X_1^A X_3^C X_4^D} + \myTrbig{X_1^A X_2^B X_4^D X_3^C}\nonumber\\
    &+ \myTrbig{X_2^B X_1^A X_4^D X_3^C} - \myTrbig{X_1^A X_3^C X_2^B X_4^D} - \myTrbig{X_2^B X_3^C X_1^A X_4^D}\nonumber\\
    &- \myTrbig{X_1^A X_4^D X_2^B X_3^C} - \myTrbig{X_2^B X_4^D X_1^A X_3^C}.
\end{align}
Here, we use $F^{f}_Y$, with $Y\in\{L,R\}$, to indicate the corresponding generator.
For instance,
\begin{equation}
    \begin{aligned}
        \mathcal{S}_{4}^{ABCD}&\big(F^{q}_R+F_L^q,F^{q}_R+F_L^q,F_{LR}^q,F_{LR}^q\big)
        = \myTrbig{\big(\myGenQu_R^A + \myGenQu_L^A\big) \big(\myGenQu_R^B + \myGenQu_L^B\big) \widehat{\myGenQu}^C \widehat{\myGenQu}^D} + \ldots 
    \end{aligned}
\end{equation}

Likewise, we also define generalised quadratic Casimirs.
For the scalar generators,
\begin{equation}\label{Eq:Quadratic-Super-Casimir-Scalar}
    \begin{aligned}
        \mathcal{C}_2\big(S\big) \delta_{ab} &= \myGenScl^A_{ac}\myGenScl^A_{cb} 
        = \Big[ g_Y^2 \mathcal{Y}_S^2 + g_W^2 C_{2}\big(F_L\big) \Big] \delta_{ab}
        = \frac{g_Y^2 + 3 g_W^2}{4} \, \delta_{ab}.
    \end{aligned}
\end{equation}
For purely left- or right-handed generators, the quadratic lepton Casimirs are
\begin{equation}\label{Eq:Quadratic-Super-Casimir-Leptons}
    \begin{alignedat}{2}
        \mathcal{C}^{ab}_2\big(F^l_R\big) &= \myGenLe^A_{R,ac}\myGenLe^A_{R,cb} &&= g_Y^2 \big((\mathcal{Y}^l_R)^2\big)_{ab},\\
        \mathcal{C}_2\big(F^l_L\big) \delta_{ab} &= \myGenLe^A_{L,ac}\myGenLe^A_{L,cb} &&= \Big[ g_Y^2 \mathcal{Y}_l^2 + g_W^2 C_{2}(F_L) \Big] \delta_{ab},
    \end{alignedat}
\end{equation}
and for quarks
\begin{equation}\label{Eq:Quadratic-Super-Casimir-Quarks}
    \begin{alignedat}{2}
        \mathcal{C}^{ab}_2\big(F^q_R\big) \delta_{ij} &= \myGenQu^A_{R,ac,ik}\myGenQu^A_{R,cb,kj} &&= \Big[ g_Y^2 \big((\mathcal{Y}^q_R)^2\big)_{ab} + g_s^2 C_{2}(F_s) \delta_{ab} \Big] \delta_{ij},\\
        \mathcal{C}_2\big(F^q_L\big) \delta_{ab} \, \delta_{ij} &= \myGenQu^A_{L,ac,ik}\myGenQu^A_{L,cb,kj} &&= \Big[ g_Y^2 \mathcal{Y}_q^2 + g_W^2 C_{2}(F_L) + g_s^2 C_{2}(F_s) \Big] \delta_{ab} \, \delta_{ij}.
    \end{alignedat}
\end{equation}
For mixed chiralities, the quadratic lepton and quark Casimirs read
\begin{equation}\label{Eq:Super-C_2-Leptons-Quarks-mixed}
    \begin{alignedat}{3}
        \mathcal{C}^{ab}_2\big(F^l_R,F^l_L\big) &= \myGenLe^A_{R,ac}\myGenLe^A_{L,cb} &&= g_Y^2 \big(\mathcal{Y}^l_R \mathcal{Y}^l_L\big)_{ab} &&= \mathcal{C}^{ab}_2\big(F^l_L,F^l_R\big),\\
        \mathcal{C}^{ab}_2\big(F^q_R,F^q_L\big) \delta_{ij} &= \myGenQu^A_{R,ac,ik}\myGenQu^A_{L,cb,kj} &&= \Big[ g_Y^2 \big(\mathcal{Y}^q_R \mathcal{Y}^q_L\big)_{ab} + g_s^2 C_{2}(F_s) \delta_{ab} \Big] \delta_{ij} &&= \mathcal{C}^{ab}_2\big(F^q_L,F^q_R\big) \delta_{ij}.
    \end{alignedat}
\end{equation}
Including evanescent generators (see Eqs.~\eqref{Eq:LeptonQuark-Super-GeneratorEvanescent} and \eqref{Eq:Evanescent-Generators-Choice}), we obtain for leptons
\begin{equation}\label{Eq:Super-C_2-Leptons-Evanescent}
    \begin{alignedat}{3}
        \mathcal{C}^{ab}_2\big(F^l_R,F^l_{LR}\big) &=
        \myGenLe^A_{R,ac}\widehat{\myGenLe}^A_{cb} &&= 
        c_{\mathrm{QED}} g_Y^2 \big((\mathcal{Y}^l_R)^2\big)_{ab} &&= 
        \mathcal{C}^{ab}_2\big(F^l_{LR},F^l_{R}\big),\\
        \mathcal{C}^{ab}_2\big(F^l_L,F^l_{LR}\big) &=
        \myGenLe^A_{L,ac}\widehat{\myGenLe}^A_{cb} &&= 
        c_{\mathrm{QED}} g_Y^2 \big(\mathcal{Y}^l_L \mathcal{Y}^l_R\big)_{ab} &&= 
        \mathcal{C}^{ab}_2\big(F^l_{LR},F^l_{L}\big),\\
        \mathcal{C}^{ab}_2\big(F^l_{LR}\big) &=
        \widehat{\myGenLe}^A_{ac}\widehat{\myGenLe}^A_{cb} &&= 
        c_{\mathrm{QED}}^2 g_Y^2 \big((\mathcal{Y}^l_R)^2\big)_{ab} &&,
    \end{alignedat}
\end{equation}
and for quarks
\begin{equation}\label{Eq:Super-C_2-Quarks-Evanescent}
    \begin{alignedat}{2}
        \mathcal{C}^{ab}_2\big(F^q_R,F^q_{LR}\big) \delta_{ij} &=
        \mathcal{C}^{ab}_2\big(F^q_{LR},F^q_{R}\big) \delta_{ij} &&=
        \myGenQu^A_{R,ac,ik}\widehat{\myGenQu}^A_{cb,kj}\\
        & &&= \Big[
        c_{\mathrm{QED}} g_Y^2 \big((\mathcal{Y}^q_R)^2\big)_{ab} +
        c_{\mathrm{QCD}} g_s^2 C_{2}(F_s) \delta_{ab} \Big] \delta_{ij},\\
        \mathcal{C}^{ab}_2\big(F^q_L,F^q_{LR}\big) \delta_{ij} &= 
        \mathcal{C}^{ab}_2\big(F^q_{LR},F^q_{L}\big) \delta_{ij} &&=
        \myGenQu^A_{L,ac,ik}\widehat{\myGenQu}^A_{cb,kj}\\
        & &&= \Big[
        c_{\mathrm{QED}} g_Y^2 \big(\mathcal{Y}^q_L \mathcal{Y}^q_R\big)_{ab} +
        c_{\mathrm{QCD}} g_s^2 C_{2}(F_s) \delta_{ab} \Big] \delta_{ij},\\
        \mathcal{C}^{ab}_2\big(F^q_{LR}\big) \delta_{ij} &=
        \widehat{\myGenQu}^A_{ac,ik}\widehat{\myGenQu}^A_{cb,kj} &&= \Big[
        c_{\mathrm{QED}}^2 g_Y^2 \big((\mathcal{Y}^q_R)^2\big)_{ab} +
        c_{\mathrm{QCD}}^2 g_s^2 C_{2}(F_s) \delta_{ab} \Big] \delta_{ij}.
    \end{alignedat}
\end{equation}
The generalised quadratic Casimir in the adjoint representation is
\begin{equation}\label{Eq:Super-C_A}
    \begin{aligned}
        \mathcal{C}_2\big(G^{A'}\big) \delta^{AB} = \mathscr{C}^{ACD}\mathscr{C}^{BCD} =
        \begin{cases}
            g_W^2 \, C_{2}\big(G_L\big) \delta^{AB}, \quad &A,B,\ldots\in\{2,3,4\}\\
            g_s^2 \, C_{2}\big(G_s\big) \delta^{AB}, \quad &A,B,\ldots\in\{5,\ldots,12\}\\
            0, \quad &\mathrm{else}.
        \end{cases}
    \end{aligned}
\end{equation}

For traces of Yukawa matrices, we define
\begin{equation}\label{Eq:Yukawa-Traces}
    \begin{aligned}
        H_2(L) &= \myTrbig{Y_l^{\dagger}Y_l},\\
        H_2(D) &= \myTrbig{Y_d^{\dagger}Y_d},\\
        H_2(U) &= \myTrbig{Y_u^{\dagger}Y_u},
    \end{aligned}
\end{equation}
\begin{equation}\label{Eq:Yukawa-4-Traces}
    \begin{aligned}
        H_4(L) &= \myTrbig{Y_l^{\dagger}Y_l Y_l^{\dagger}Y_l},\\
        H_4(D) &= \myTrbig{Y_d^{\dagger}Y_d Y_d^{\dagger}Y_d},\\
        H_4(U) &= \myTrbig{Y_u^{\dagger}Y_u Y_u^{\dagger}Y_u},
    \end{aligned}
\end{equation}
together with their sums over all fermion species,
\begin{equation}\label{Eq:Yukawa-Trace-Combi}
    \begin{aligned}
        \mathcal{H}_2(Y) &= H_2(L) + N_c \big[ H_2(D) + H_2(U) \big],\\
        \mathcal{H}_4(Y) &= H_4(L) + N_c \big[ H_4(D) + H_4(U) \big].
    \end{aligned}
\end{equation}
Traces in which Yukawa matrices appear without a Hermitian-conjugate partner arise as well. We denote them by
\begin{equation}\label{Eq:Yukawa-Trace-Combi-2}
    \begin{aligned}
        \mathcal{H}_2(D,U) &= N_c \Big[ \myTrbig{Y_d Y_u} + \myTrbig{Y_d^\dagger Y_u^\dagger} \Big],\\
        \mathcal{H}_2(L,D,\overline{U}) &= \myTrbig{Y_lY_l} + N_c \Big[ \myTrbig{Y_d Y_d} + \myTrbig{Y_u^\dagger Y_u^\dagger} \Big],\\
        \mathcal{H}_4(L,D,\overline{U}) &= \myTrbig{Y_lY_lY_lY_l} + N_c \Big[ \myTrbig{Y_d Y_dY_d Y_d} + \myTrbig{Y_u^\dagger Y_u^\dagger Y_u^\dagger Y_u^\dagger} \Big].
    \end{aligned}
\end{equation}
For later use we further introduce the combinations
\begin{subequations}\label{Eq:Def-YukawaCoeffs-h_1--h_4}
\begin{align}
    h_1 &= - \frac{19}{48} \mathcal{H}_2(Y) + \frac{1}{12} \mathcal{H}_2(D,U),\\
    h_2 &=   \frac{7}{48}  \mathcal{H}_2(Y) + \frac{1}{12} \mathcal{H}_2(D,U),\\
    h_3 &=   \frac{5}{48}  \mathcal{H}_2(Y) - \frac{1}{12} \mathcal{H}_2(D,U),\\
    h_4 &=   \frac{3}{48}  \mathcal{H}_2(Y),
\end{align}
\end{subequations}
and
\begin{subequations}\label{Eq:Def-YukawaCoeffs-varrho_0--varrho_3}
\begin{align}
    \varrho_0 &= \frac{1}{6} \Big[\mathcal{H}_2(L,D,\overline{U})  - 4 N_c \mathrm{Tr}\big(Y_u^\dagger Y_d\big)\Big],\\
    \varrho_1 &= \frac{1}{6} \Big[\mathcal{H}_2(L,D,\overline{U})  -   N_c \mathrm{Tr}\big(Y_u^\dagger Y_d\big)\Big],\\
    \varrho_2 &= \frac{1}{4} \mathcal{H}_2(L,D,\overline{U}) ,\\
    \varrho_3 &= \frac{1}{12} \Big[\mathcal{H}_2(L,D,\overline{U}) + 2 N_c \mathrm{Tr}\big(Y_u^\dagger Y_d\big)\Big].
\end{align}
\end{subequations}
Furthermore, the following traces over four Yukawa matrices appear:
\begin{align}
    \begin{split}\label{Eq:Def-YukawaCoeffs-H4_1}
        \mathcal{H}_{4}^{(1)} &= 
        2\,\mathrm{Tr}\big( Y_l^\dagger Y_l^\dagger Y_l Y_l \big)
        + N_c \Big[
        \mathrm{Tr}\big( Y_d Y_d Y_u Y_u \big)
        + \mathrm{Tr}\big( Y_d Y_u Y_d Y_u \big)
        - 2\,\mathrm{Tr}\big( Y_d Y_u Y_u^\dagger Y_u \big)\\
        &
        - 2\,\mathrm{Tr}\big( Y_d^\dagger Y_d Y_d^\dagger Y_u^\dagger \big)
        - 2\,\mathrm{Tr}\big( Y_d^\dagger Y_d Y_u Y_d \big)
        - 2\,\mathrm{Tr}\big( Y_d^\dagger Y_d Y_u Y_u^\dagger \big)
        + 2\,\mathrm{Tr}\big( Y_d^\dagger Y_d^\dagger Y_d Y_d \big)\\
        &
        + \mathrm{Tr}\big( Y_d^\dagger Y_d^\dagger Y_u^\dagger Y_u^\dagger \big)
        + \mathrm{Tr}\big( Y_d^\dagger Y_u^\dagger Y_d^\dagger Y_u^\dagger \big)
        - 2\,\mathrm{Tr}\big( Y_d^\dagger Y_u^\dagger Y_u Y_d \big)
        - 2\,\mathrm{Tr}\big( Y_d^\dagger Y_u^\dagger Y_u Y_u^\dagger \big)\\
        &
        + 2\,\mathrm{Tr}\big( Y_u^\dagger Y_u^\dagger Y_u Y_u \big)
        \Big],
    \end{split}\\
    \begin{split}\label{Eq:Def-YukawaCoeffs-H4_2}
        \mathcal{H}_{4}^{(2)} &= 
        - 2\,\mathrm{Tr}\big( Y_l^\dagger Y_l^\dagger Y_l Y_l \big)
        + N_c \Big[
        \mathrm{Tr}\big( Y_d Y_d Y_u Y_u \big)
        + 2\,\mathrm{Tr}\big( Y_d Y_u Y_u^\dagger Y_u \big)
        + 2\,\mathrm{Tr}\big( Y_d^\dagger Y_d Y_d^\dagger Y_u^\dagger \big)\\
        &
        + 2\,\mathrm{Tr}\big( Y_d^\dagger Y_d Y_u Y_d \big)
        + 2\,\mathrm{Tr}\big( Y_d^\dagger Y_d Y_u Y_u^\dagger \big)
        - 2\,\mathrm{Tr}\big( Y_d^\dagger Y_d^\dagger Y_d Y_d \big)
        + \mathrm{Tr}\big( Y_d^\dagger Y_d^\dagger Y_u^\dagger Y_u^\dagger \big)\\
        &
        + 2\,\mathrm{Tr}\big( Y_d^\dagger Y_u^\dagger Y_u Y_d \big)
        + 2\,\mathrm{Tr}\big( Y_d^\dagger Y_u^\dagger Y_u Y_u^\dagger \big)
        - 2\,\mathrm{Tr}\big( Y_u^\dagger Y_u^\dagger Y_u Y_u \big)
        \Big],
    \end{split}\\
    \begin{split}\label{Eq:Def-YukawaCoeffs-H4_3}
        \mathcal{H}_{4}^{(3)} &= 
        2 \Big(
        \mathrm{Tr}\big( Y_l^\dagger Y_l Y_l Y_l \big)
        + N_c \Big[
        \mathrm{Tr}\big( Y_d^\dagger Y_d Y_d Y_d \big)
        + \mathrm{Tr}\big( Y_u^\dagger Y_u^\dagger Y_u^\dagger Y_u \big)
        \Big] \Big),
    \end{split}\\
    \begin{split}\label{Eq:Def-YukawaCoeffs-H4_4}
        \mathcal{H}_{4}^{(4)} &= 
        N_c \Big[
        \mathrm{Tr}\big( Y_d Y_d Y_d Y_u \big)
        + \mathrm{Tr}\big( Y_d^\dagger Y_u^\dagger Y_u^\dagger Y_u^\dagger \big)
        \Big].
    \end{split}
\end{align}

For matrix products, we introduce
\begin{equation}\label{Eq:Yukawa-Products}
    \begin{alignedat}{2}
        \mathbf{Y}_2(L) &= Y_l^{\dagger}Y_l, 
        \qquad 
        \mathbf{Y}_2(\overline{L}) &&= Y_l Y_l^{\dagger},\\
        \mathbf{Y}_2(D) &= Y_d^{\dagger}Y_d, 
        \qquad 
        \mathbf{Y}_2(\overline{D}) &&= Y_d Y_d^{\dagger},\\
        \mathbf{Y}_2(U) &= Y_u^{\dagger}Y_u, 
        \qquad 
        \mathbf{Y}_2(\overline{U}) &&= Y_u Y_u^{\dagger}.
    \end{alignedat}
\end{equation}

Finally, we introduce the following abbreviations for recurring combinations of the parameters $c_\mathrm{QED}$ and $c_\mathrm{QCD}$ (which parametrise the evanescent gauge interactions via Eq.~\eqref{Eq:Evanescent-Generators-Choice}):
\begin{alignat}{2}
    a_{\mathrm{QED}} &= 1 - \frac{1 + 4 c_{\mathrm{QED}}}{6} \xi_F,
    &\qquad
    b_{\mathrm{QED}} &= 1 - \frac{1 - c_{\mathrm{QED}} ( 2 + c_{\mathrm{QED}} )}{6} \xi_F,
    \label{Eq:SM-a_QED-b_QED}
    \\[1ex]
    u_{\mathrm{QED}} &= 1 + c_{\mathrm{QED}}^2 - ( 1 - c_{\mathrm{QED}} )^2 \, \frac{\xi_F}{6},
    &\qquad
    u_{\mathrm{QCD}} &= 1 + c_{\mathrm{QCD}}^2 - ( 1 - c_{\mathrm{QCD}} )^2 \, \frac{\xi_F}{6},
    \label{Eq:SM-u_QED-u_QCD}
    \\[1ex]
    w_{\mathrm{QCD}} &= \mathrlap{( 1 - c_{\mathrm{QCD}} ) \frac{6 - ( 1 - c_{\mathrm{QCD}} ) \xi_F}{12},}
    \label{Eq:SM-w_QCD}
    \\[1ex]
    k_{\mathrm{QCD}} &= \mathrlap{\frac{1}{12} \Big[ 12 - 2\xi_F - 4\xi_F c_{\mathrm{QCD}} - \big( 6 - 5\xi_F \big) c_{\mathrm{QCD}}^2 + \big( 6 - 2\xi_F \big) c_{\mathrm{QCD}}^3 \Big].}
    \label{Eq:SM-k_QCD}
\end{alignat}

\section{Counterterm Coefficients in Compact Notation} \label{App:Results-TheStandardModel}

Here, we list the explicit 1-loop results for the coefficients entering the compact-notation counterterm action of Sec.~\ref{Sec:SM_1-Loop_Renormalisation}, expressed in terms of the generalised gauge field twelveplets $\vgb^A_\mu$ (see Sec.~\ref{Sec:SM-Definition-and-Setup}).
The coefficients of the divergent counterterms in Sec.~\ref{Sec:SM_1-Loop_Renormalisation_Singular_CT} are given in App.~\ref{App:SM-Divergent1LoopCTCoeffs}, and those of the finite symmetry-restoring counterterms in Sec.~\ref{Sec:SM_1-Loop_Renormalisation_Finite_CT} in App.~\ref{App:SM-Finite1LoopCTCoeffs}.
The corresponding coefficients of the decomposed, sector-wise representation are collected in App.~\ref{App:CountertermCoeffs_Expanded}.

All results were obtained as described in Sec.~\ref{Sec:Procedure}.
They are expressed in terms of the modified gauge parameter $\xi_F=1-\xi$, and the group invariants and coupling structures defined in App.~\ref{App:SM-GroupInvariants-CouplingStructures}.

The following examples illustrate how the compact twelveplet notation applies to the generalised generators and coupling structures:
\begin{subequations}
\begin{align}
    \mathscr{C}^{ABC} \myGenQu_{L,ab,ij}^C &= 
    \begin{cases}
        g_W^2 \varepsilon^{ABC} t_{L,ab}^C \delta_{ij}, \quad &A,B,C\in\{2,3,4\},\\
        g_s^2 f^{ABC} t_{s,ij}^C \delta_{ab}, \quad &A,B,C\in\{5,\ldots,12\},\\
        0, \quad &\mathrm{else},
    \end{cases} \label{Eq:SM-Notation-Example-1} \\
    \mathcal{C}_2\big(G^{A'}\big) \myGenLe_{L,ab}^A &= 
    \begin{cases}
        g_W^3 C_2(G_L) t_{L,ab}^A, \quad &A\in\{2,3,4\},\\
        0, \quad &\mathrm{else},
    \end{cases} \label{Eq:SM-Notation-Example-2} \\
    \mathcal{C}_2\big(G^{A'}\big) \myGenQu_{L,ab,ij}^A &= 
    \begin{cases}
        g_W^3 C_2(G_L) t_{L,ab}^A \delta_{ij}, \quad &A\in\{2,3,4\},\\
        g_s^3 C_2(G_s) t_{s,ij}^A \delta_{ab}, \quad &A\in\{5,\ldots,12\},\\
        0, \quad &\mathrm{else},
    \end{cases} \label{Eq:SM-Notation-Example-3}
\end{align}
\end{subequations}
where we again suppressed the $\dot{X}$ and $\ddot{X}$ notation ($X\in\{A,B,C,\ldots\}$) for the index shifts, introduced below Eq.~\eqref{Eq:Super-Structure-Constants} in Sec.~\ref{Sec:SM-Definition-and-Setup}, as the index range is always clear from the context.
Importantly, adjoint colour indices marked with a prime (such as $A'$ in Eqs.~\eqref{Eq:SM-Notation-Example-2} and \eqref{Eq:SM-Notation-Example-3}) are not implicitly summed over.
They only label which specific component of the adjoint quadratic Casimir (see Eq.~\eqref{Eq:Super-C_A}), or of any other object carrying such an index, contributes.

\subsection{Counterterm Coefficients of Divergent 1-Loop Contributions}\label{App:SM-Divergent1LoopCTCoeffs}

We begin with the coefficients of the divergent 1-loop contributions.

\paragraph{Fermionic Sector:}
The divergent fermionic counterterm action is provided in Eq.~\eqref{Eq:SM-S_sct-fermion}.
For the leptons, we obtain
\begin{align}
    \begin{split}\label{Eq:SM-Lepton-SE-R-CTs}
        \delta\overline{\mathcal{Z}}^{l,(1)}_{R,ab,IJ} = &- \frac{1}{\epsilon} \Big\{ (1-\myxi) \mathcal{C}_2^{ab}\big(F^l_R\big) \delta_{IJ} + \frac{\myNL}{2} \delta_{2a} \delta_{2b} \big(\mathbf{Y}_2(L)\big)_{IJ} \Big\},
    \end{split}\\[1.25ex]
    \begin{split}\label{Eq:SM-Lepton-SE-L-CTs}
        \delta\overline{\mathcal{Z}}^{l,(1)}_{L,IJ} = &- \frac{1}{\epsilon} \Big\{ (1-\myxi) \mathcal{C}_2\big(F_L^l\big) \delta_{IJ} + \frac{1}{2} \big(\mathbf{Y}_2(\overline{L})\big)_{IJ} \Big\},
    \end{split}\\[1.25ex]
    \begin{split}\label{Eq:SM-Lepton-SE-Evanescent-CTs}
        \delta\widehat{\mathcal{X}}^{l,(1)}_{ab} = &- \frac{1}{\epsilon} \bigg\{ \frac{6-2\myxi}{3} \mathcal{C}_2^{ab}\big(F_L^l,F_R^l\big)
        - \frac{\myxi}{3} \Big[ \mathcal{C}_2^{ab}\big(F_R^l,F_{LR}^l\big) + \mathcal{C}_2^{ab}\big(F_L^l,F_{LR}^l\big) \Big]\\
        &- \frac{3-\myxi}{3} \mathcal{C}_2^{ab}\big(F_{LR}^l\big) \bigg\},
    \end{split}
\end{align}
while for the quarks the coefficients are given by
\begin{align}
    \begin{split}\label{Eq:SM-Quark-SE-R-CTs}
        \delta\overline{\mathcal{Z}}^{q,(1)}_{R,ab,IJ} = &- \frac{1}{\epsilon} \Big\{ (1-\myxi) \mathcal{C}_2^{ab}\big(F_R^q\big) \delta_{IJ} + \frac{\myNL}{2} \delta_{2a} \delta_{2b} \big(\mathbf{Y}_2(D)\big)_{IJ} + \delta_{1a} \delta_{1b} \big(\mathbf{Y}_2(U)\big)_{IJ} \Big\},
    \end{split}\\[1.25ex]
    \begin{split}\label{Eq:SM-Quark-SE-L-CTs}
        \delta\overline{\mathcal{Z}}^{q,(1)}_{L,IJ}  = &- \frac{1}{\epsilon} \Big\{ (1-\myxi) \mathcal{C}_2\big(F_L^q\big) \delta_{IJ} + \frac{1}{2} \Big[ \big(\mathbf{Y}_2(\overline{D})\big)_{IJ} + \big(\mathbf{Y}_2(\overline{U})\big)_{IJ} \Big] \Big\} ,
    \end{split}\\[1.25ex]
    \begin{split}\label{Eq:SM-Quark-SE-Evanescent-CTs}
        \delta\widehat{\mathcal{X}}^{q,(1)}_{ab,IJ} = &- \frac{1}{\epsilon} 
        \bigg\{ 
        \frac{6-2\myxi}{3} \mathcal{C}_2^{ab}\big(F_L^q,F_R^q\big) \delta_{IJ}
        - \frac{\myxi}{3} \Big[ 
        \mathcal{C}_2^{ab}\big(F_R^q,F_{LR}^q\big) + \mathcal{C}_2^{ab}\big(F_L^q,F_{LR}^q\big) 
        \Big] \delta_{IJ}\\
        &- \frac{3-\myxi}{3} \mathcal{C}_2^{ab}\big(F_{LR}^q\big) \delta_{IJ}
        - \frac{1}{2} \Big[ \delta_{1a} \delta_{1b} \big( Y_d Y_u \big)_{IJ} + \delta_{2a} \delta_{2b} \big( Y_u Y_d \big)_{IJ} \Big] \bigg\}.
    \end{split}
\end{align}
The 4-dimensional non-invariant fermion--gauge interactions are governed by
\begin{equation}\label{Eq:SM-Fermion-Gauge-Interaction-4d-CTs}
    \begin{aligned}
        \delta\overline{\mathcal{X}}_{\overline{f}\vgb f}^{(1)} = 
        - \frac{4-\myxi}{4\epsilon} \mathcal{C}_2\big(G^{A'}\big).
    \end{aligned}
\end{equation}
The evanescent non-invariant quark--gauge interactions are
\begin{align}\label{Eq:SM-Evanescent-Quark-Gauge-Interaction-CTs}
        \delta&\widehat{\mathcal{X}}^{\overline{q}\vgb q,(1),A}_{ab,ij,IJ} = - k_{\mathrm{QCD}}\, \mathcal{C}_2\big(G^{A'}\big) \, \myGenQu^{A}_{R,ab,ij} \, \delta_{IJ}
        \nonumber\\
        &-\frac{1}{2} \myGenScl^{A}_{ce} \delta_{ij} \Big[ \varepsilon_{ea} \delta_{1c} \delta_{2b} \big(Y_uY_d\big)_{IJ} + \varepsilon_{2e} \delta_{ac} \delta_{1b} \big(Y_dY_u\big)_{IJ} \Big] 
        \times \begin{cases}
            (1-2 c_{\mathrm{QED}}) , \quad &A=1\\
            1, \quad &A\in\{2,3,4\}\\
            0, \quad &\mathrm{else}
        \end{cases}
        \nonumber\\
        &= -\frac{1}{\epsilon}
        \begin{cases}
            \frac{g_Y}{2} \mathcal{Y}_S (1-2 c_{\mathrm{QED}}) \delta_{ij} \Big[ \delta_{2a} \delta_{2b} \big(Y_uY_d\big)_{IJ} - \delta_{1a} \delta_{1b} \big(Y_dY_u\big)_{IJ} \Big], \quad &A=1\\
            \frac{g_W}{2} t^A_{L,ce} \delta_{ij} \Big[ \varepsilon_{ea} \delta_{1c} \delta_{2b} \big(Y_uY_d\big)_{IJ} + \varepsilon_{2e} \delta_{ac} \delta_{1b} \big(Y_dY_u\big)_{IJ} \Big], \quad &A\in\{2,3,4\}\\
            k_{\mathrm{QCD}} \, g_s^3 \, C_{2}(G_s) \, t_{s,ij}^A \, \delta_{ab} \, \delta_{IJ}, \quad &A\in\{5,\ldots,12\},
        \end{cases}
\end{align}
with $k_{\mathrm{QCD}}$ as defined in Eq.~\eqref{Eq:SM-k_QCD}.

\paragraph{Gauge Sector:}
The divergent gauge-sector counterterm action is presented in Eq.~\eqref{Eq:SM-S_sct_gauge}, with explicit coefficients given by
\begin{align}
    \begin{split}\label{Eq:SM-GaugeBoson-SE-D-dim-CTs}
        \delta \mathcal{Z}_{\vgb}^{(1)} &= \frac{1}{\epsilon} \bigg\{ 
        \frac{10+3\myxi}{6} \mathcal{C}_2\big(G^{A'}\big) - \frac{1}{3} \mathcal{S}_2\big(S^{A'}\big) 
        \bigg\},
    \end{split}\\[1.25ex]
    \begin{split}
        \delta\overline{\mathcal{Z}}_{\vgb}^{(1)} &= -\frac{2}{3 \epsilon}  N_{f} \sum_{f\in\{l,q\}} \Big[ \mathcal{S}_2\big(F^{f,A'}_R\big) + \mathcal{S}_2\big(F^{f,A'}_L\big) \Big],
    \end{split}\\[1.25ex]
    \begin{split}
        \delta\widehat{\mathcal{Z}}^{(1)}_{\vgb} &= -\frac{4}{3\epsilon} N_{f} \sum_{f\in\{l,q\}} \mathcal{S}_2\big(F^{f,A'}_{LR}\big),
    \end{split}\\[1.25ex]
    \begin{split}\label{Eq:SM-GaugeBoson-Evanescent-Breaking-CT-indep-of-evan-generators}
        \delta\widehat{\mathcal{X}}^{(1),AB}_{\vgb,12} &= - \frac{1}{3\epsilon} N_{f}
        \bigg\{
        \sum_{f\in\{l,q\}} \Big[ \mathcal{S}_2\big(F^{f,A'}_R\big) + \mathcal{S}_2\big(F^{f,A'}_L\big) \Big] \delta^{AB}\\
        &\qquad\qquad + \sum_{f\in\{l,q\}} \Big[ \mathcal{S}^{AB}_2\big(F^f_R,F^f_L\big) + \mathcal{S}^{AB}_2\big(F^f_L,F^f_R\big) \Big]
        \bigg\}\\
        &= \frac{1}{2} \, \delta\overline{\mathcal{Z}}_{\vgb}^{(1)} \delta^{AB} 
        - \frac{N_{f}}{3\epsilon} \sum_{f\in\{l,q\}} \Big[ \mathcal{S}^{AB}_2\big(F^f_R,F^f_L\big) + \mathcal{S}^{AB}_2\big(F^f_L,F^f_R\big) \Big]\\
        &= \delta\widehat{\mathcal{X}}^{(1),BA}_{\vgb,12} \, ,
    \end{split}\\[1.25ex]
    \begin{split}
        \delta\widehat{\mathcal{X}}^{(1),AB}_{\vgb,34} &= - \frac{2}{3\epsilon} N_{f} \sum_{f\in\{l,q\}} \Big[ \mathcal{S}_2\big(F^{f,A'}_R,F^{f,B'}_{LR}\big) \delta^{AB} + \mathcal{S}^{AB}_2\big(F^f_L,F^f_{LR}\big) \Big]
        = \delta\widehat{\mathcal{X}}^{(1),BA}_{\vgb,43} \, ,
    \end{split}\\[1.25ex]
    \begin{split}
        \delta\widehat{\mathcal{X}}^{(1),AB}_{\vgb,43} &= - \frac{2}{3\epsilon} N_{f} \sum_{f\in\{l,q\}} \Big[ \mathcal{S}_2\big(F^{f,A'}_{LR},F^{f,B'}_R\big) \delta^{AB} + \mathcal{S}^{AB}_2\big(F^{f}_{LR},F^f_L\big) \Big]
        = \delta\widehat{\mathcal{X}}^{(1),BA}_{\vgb,34} \, ,
    \end{split}
\end{align}
and triple gauge boson interaction counterterms
\begin{align}\label{Eq:SM-Triple-Gauge-Interaction-CTs}
    \begin{split}
        \delta \mathcal{Z}^{(1)}_{3\vgb} &=  \frac{1}{\epsilon} \bigg\{ 
        \frac{8+9\myxi}{12} \mathcal{C}_2\big(G^{A'}\big) - \frac{1}{3} \mathcal{S}_{2}\big(S^{A'}\big) \bigg\},
    \end{split}\\[1.25ex]
    \begin{split}
        \delta\widehat{\mathcal{X}}_{3\vgb}^{(1),ABC} &= - \frac{\mathscr{C}^{ABD}}{2}\big( \delta\widehat{\mathcal{X}}^{(1),DC}_{\vgb,34} + \delta\widehat{\mathcal{X}}^{(1),CD}_{\vgb,34}\big).
    \end{split}
\end{align}
The coefficients for quartic gauge boson self-interactions read
\begin{align}\label{Eq:SM-4-Gauge-Self-Interaction-CTs}
    \begin{split}
        \delta \mathcal{Z}^{(1)}_{4\vgb} 
        &= - \frac{1}{\epsilon} \bigg\{ \frac{1-3\myxi}{3} \, \mathcal{C}_{2}\big(G^{A'}\big) + \frac{1}{3} \, \mathcal{S}_{2}\big(S^{A'}\big) \bigg\},
    \end{split}\\[1.25ex]
    \begin{split}
        \delta \widehat{\mathcal{X}}^{(1),ABCD}_{4\vgb} &= - \frac{N_f}{3\epsilon} \sum_{f\in\{l,q\}} \mathcal{S}_{4}^{ABCD}\big(F^{f}_L+F_R^f,F^{f}_L+F_R^f,F_{LR}^f,F_{LR}^f\big),
    \end{split}
\end{align}
and, for the purely evanescent term in the quartic gauge boson self-interaction, we find
\begin{align}
        \delta \widehat{\mathcal{X}}^{(1),ABCD}_{4\vgb,\mathrm{ev}} &= - \frac{4 N_f}{3\epsilon} \sum_{f\in\{l,q\}} \mathcal{S}_{4}^{ABCD}\big(F_{LR}^f,F_{LR}^f,F_{LR}^f,F_{LR}^f\big)
        \nonumber\\
        &= -
        \begin{cases}
            \frac{4 N_f}{3\epsilon} \, c_{\mathrm{QCD}}^4 \, g_s^4 \, \big(f^{ACE}f^{BDE} + f^{ADE}f^{BCE}\big), \quad &A,B,C,D \in \{5,\ldots,12\}\\
            0, \quad &\mathrm{else}
        \end{cases}
        \nonumber\\
        &= c_{\mathrm{QCD}}^2 \, \delta\widehat{\mathcal{Z}}^{(1)}_{\vgb} \, \Big(\mathscr{C}^{ACE}\mathscr{C}^{BDE} + \mathscr{C}^{ADE}\mathscr{C}^{BCE}\Big)
\end{align}

\paragraph{Higgs Sector:}
The divergent 1-loop counterterm action of the Higgs sector is shown in Eq.~\eqref{Eq:SM-S_sct_Higgs}.
The coefficients of the scalar kinetic terms read
\begin{align}
    \begin{split}\label{Eq:SM-Scalar-SE-CTs-phidagger-phi-D-dim}
        \delta \mathcal{Z}^{(1)}_{\phi} &= \frac{1}{\epsilon} \Big[ (2+\myxi) \mathcal{C}_2\big(S\big) - \frac{2}{3} \, \mathcal{H}_2(Y) \Big],
    \end{split}\\[1.25ex]
    \begin{split}
        \delta\overline{\mathcal{Z}}^{(1)}_{\phi} &= - \frac{1}{3\epsilon} \, \mathcal{H}_2(Y),
    \end{split}\\[1.25ex]
    \begin{split}
        \delta\widehat{\mathcal{X}}^{(1)}_{\phi} \delta_{1a}\delta_{1b} &= - \frac{N_c}{3\epsilon} \Big[ \delta_{1b} \varepsilon_{2a} \trYY{u}{d} + \delta_{1a} \varepsilon_{2b} \trYbarYbar{u}{d} \Big] \\
        &= \frac{1}{3\epsilon} \mathcal{H}_2(D,U) \, \delta_{1a}\delta_{1b} \,,
    \end{split}\\[1.25ex]
    \begin{split}
        \delta \mathcal{Z}^{(1)}_{\mu} &= - \frac{1}{\epsilon} \Big[ (1-\myxi) \mathcal{C}_2\big(S\big) - 2 (\myNL+1) \lambda \Big],
    \end{split}\\[1.25ex]
    \begin{split}
        \delta\widehat{\mathcal{X}}^{(1)}_{\phi\phi} \delta_{2a}\delta_{2b} &= - \frac{1}{3\epsilon} \bigg\{ \Big[ \trYY{l}{l} + N_c \trYY{d}{d} \Big] \delta_{2a} \delta_{2b} + N_c \trYbarYbar{u}{u} \varepsilon_{1a} \varepsilon_{1b} \bigg\} \\
        &= - \frac{1}{3\epsilon} \mathcal{H}_2(L,D,\overline{U}) \, \delta_{2a}\delta_{2b} \, .
    \end{split}
\end{align}
For the scalar--gauge boson interactions, we find
\begin{align}\label{Eq:SM-Scalar-Gauge-Interaction-CTs}
    \begin{split}
        \delta \mathcal{Z}^{(1)}_{\phi^{\dagger}\phi\vgb} 
        &= \frac{1}{\epsilon} \bigg\{ (2+\myxi) \mathcal{C}_2(S) - \frac{4-\myxi}{4} \mathcal{C}_{2}\big(G^{A'}\big) \bigg\},
    \end{split}\\[1.25ex]
    \begin{split}
        \delta \overline{\mathcal{Z}}^{(1)}_{\phi^{\dagger}\phi\vgb} 
        &= - \frac{1}{\epsilon} \, \mathcal{H}_2(Y),
    \end{split}\\[1.25ex]
    \begin{split}
        \delta \widehat{\mathcal{X}}^{(1),A}_{\phi^{\dagger}\phi\vgb}
        &=
        \begin{cases}
            \begin{aligned}
        - \frac{4}{3\epsilon} c_{\mathrm{QED}} g_Y \mathcal{Y}_S \Big[ \mathcal{H}_2(Y) - \frac{1}{2} \mathcal{H}_2(D,U) \Big], 
        \end{aligned} \quad &A = 1\\
        0, \quad &\mathrm{else}.
        \end{cases}
    \end{split}
\end{align}
The coefficients of the double scalar--double gauge boson interactions are given by
\begin{align}\label{Eq:SM-2-Scalar-2-GaugeBoson-Interaction-CTs}
    \begin{split}
        \delta \mathcal{Z}^{(1)}_{\phi^{\dagger}\phi\vgb\vgb}
        &= - \frac{1}{\epsilon} 
        \bigg\{ 
        \frac{4-\myxi}{4} \Big[ \mathcal{C}_2\big(G^{A'}\big) + \mathcal{C}_2\big(G^{B'}\big) \Big] - (2+\myxi) \mathcal{C}_{2}\big(S\big)
        \bigg\},
    \end{split}\\[1.25ex]
    \begin{split}
        \delta \overline{\mathcal{Z}}^{(1)}_{\phi^{\dagger}\phi\vgb\vgb} &= - \frac{1}{\epsilon} \, \mathcal{H}_2(Y),
    \end{split}\\[1.25ex]
    \begin{split}
        \delta\widehat{\mathcal{X}}^{(1),AB}_{\phi^{\dagger}\phi\vgb\vgb} &= 
        \begin{cases}
            - \frac{16}{3\epsilon} \, c_{\mathrm{QED}}^2 \, g_Y^2 \, \mathcal{Y}_S^2 \Big[ \mathcal{H}_2(Y) - \frac{1}{2} \mathcal{H}_2(D,U) \Big], \quad &A,B=1\\
            0, \quad &\mathrm{else}.
        \end{cases}
    \end{split}
\end{align}
The divergent counterterm of the quartic scalar self-interaction is governed by
\begin{equation}
    \begin{aligned}
        \delta \mathcal{Z}^{(1)}_{\lambda} = \frac{1}{\epsilon} \Big[ 2 g_Y^4 \mathcal{Y}_S^4 + \big(\mathcal{C}_2(S)\big)^2 - 2(1-\myxi)\lambda\,\mathcal{C}_2(S) + 12 \lambda^2 - \mathcal{H}_4(Y) \Big].
    \end{aligned}
\end{equation}

\paragraph{Yukawa Sector:}
The divergent Yukawa counterterm action is given in Eq.~\eqref{Eq:SM-S_sct_Yukawa}.
For the lepton Yukawa vertex corrections, we obtain
\begin{equation}\label{Eq:SM-Lepton-Yukawa-Interaction-CTs}
    \begin{aligned}
        \delta \mathcal{Y}^{(1)}_{l,IJ} = - \frac{1}{\epsilon} \Big[ (1-\myxi) \mathcal{C}_2(S) + (4-\myxi) \mathcal{C}^{22}_2\big(F_R^l,F_L^l\big) \Big] Y_{l,IJ},
    \end{aligned}
\end{equation}
while the quark Yukawa corrections read
\begin{align}\label{Eq:SM-Quark-Yukawa-Interaction-CTs}
    \begin{split}
        \delta \mathcal{Y}^{(1)}_{d,IJ} &= - \frac{1}{\epsilon} \Big[ (1-\myxi) \mathcal{C}_2(S) \, \delta_{IK} + (4-\myxi) \mathcal{C}^{22}_2\big(F_R^q,F_L^q\big) \, \delta_{IK}
        + \big(\mathbf{Y}_2(\overline{U})\big)_{IK} \Big] Y_{d,KJ},
    \end{split}\\[1.25ex]
    \begin{split}
        \delta \mathcal{Y}^{(1)}_{u,IJ} &= - \frac{1}{\epsilon} \Big[ (1-\myxi) \mathcal{C}_2(S) \, \delta_{IK} + (4-\myxi) \mathcal{C}^{11}_2\big(F_R^q,F_L^q\big) \, \delta_{IK}
        + \big(\mathbf{Y}_2(\overline{D})\big)_{IK} \Big] Y_{u,KJ}.
    \end{split}
\end{align}

\paragraph{Ghosts and BRST Transformations:}
Finally, for the ghost sector, with counterterm action shown in Eq.~\eqref{Eq:SM-S_sct_ghost}, we find
\begin{align}
    \begin{split}\label{Eq:SM-Ghost-SE-CTs}
        \delta \mathcal{Z}^{(1)}_{c} &= \frac{2+\myxi}{4\epsilon} \mathcal{C}_2\big(G^{A'}\big),
    \end{split}\\[1.25ex]
    \begin{split}
        \delta \mathcal{Z}^{(1)}_{\overline{c}\vgb c} &= - \frac{1-\myxi}{2\epsilon} \mathcal{C}_2\big(G^{A'}\big).
    \end{split}
\end{align}

\subsection{Counterterm Coefficients of Finite 1-Loop Contributions}\label{App:SM-Finite1LoopCTCoeffs}

Here, we provide the coefficients of the finite symmetry-restoring 1-loop contributions.

\paragraph{Fermionic Sector:}
We begin with the fermionic sector.
The finite symmetry-restoring counterterm action is given in Eq.~\eqref{Eq:SM-S_fct_fermion}, and the corresponding lepton coefficients read
\begin{align}
    \begin{split}\label{Eq:SM-finite-CT-Leptons-KineticTerm}
        \delta \mathcal{F}^{(1)}_{l,ab} &= \frac{3-\xi_F}{3} \mathcal{C}_2^{ab}(F^l_{LR}) - \frac{\xi_F}{3}\mathcal{C}_2^{ab}(F_{L}^l, F^l_{LR}),
    \end{split}\\[1.25ex]
    \begin{split}\label{Eq:SM-finite-CT-Leptons-Right-handed-Interaction}
        \delta \mathcal{F}^{(1),A}_{\overline{l}\vgb l,R,ab} &=  
        - \frac{6-\xi_F}{6} \mathcal{C}_2^{ac}(F^l_R) \myGenLe^A_{R,cb} 
        - \frac{\xi_F}{3} \mathcal{C}_2^{ac}(F^l_R,F^l_{LR}) \myGenLe^A_{R,cb}
        - \frac{\xi_F}{6} \mathcal{C}_2^{ac}(F^l_{LR}) \myGenLe^A_{R,cb}\\ 
        &\quad\,\, 
        - \frac{\xi_F}{3} \mathcal{C}_2^{ac}(F^l_L,F^l_{LR}) \myGenLe^A_{R,cb}
        + \frac{6-\xi_F}{6}\big(\myGenLe^B_R \myGenLe^A_L \myGenLe^B_R\big)_{ab} 
        + \frac{6-\xi_F}{6}\big(\widehat{\myGenLe}^B \myGenLe^A_L \widehat{\myGenLe}^B\big)_{ab}\\
        &\quad\,\,  
        + \frac{\xi_F}{3}\big(\widehat{\myGenLe}^B \myGenLe^A_L \myGenLe^B_R\big)_{ab} ,
    \end{split}\\[1.25ex]
    \begin{split}\label{Eq:SM-finite-CT-Leptons-Left-handed-Interaction}
        \delta \mathcal{F}^{(1),A}_{\overline{l}\vgb l,L,ab,IJ} &= \bigg\{
        \frac{6-\xi_F}{6} \mathcal{C}_2^{ac}(F^l_{LR}) \myGenLe^A_{R,cb}
        + \frac{\xi_F}{3} \mathcal{C}_2^{ac}(F^l_L,F^l_{LR}) \myGenLe^A_{R,cb}
        - \frac{6-\xi_F}{6} \mathcal{C}_2(F^l_L) \myGenLe^A_{L,ab} \\ 
        &\quad\,\,
        - \frac{\xi_F}{3} \Big[ \mathcal{C}_2^{ac}(F^l_L,F^l_{LR}) \myGenLe^A_{L,cb} + \myGenLe^A_{L,ac} \mathcal{C}_2^{cb}(F^l_L,F^l_{LR}) \Big]\\
        &\quad\,\,
        + \frac{6-\xi_F}{6}\big(\myGenLe^B_L \myGenLe^A_R \myGenLe^B_L\big)_{ab}
        - \frac{\xi_F}{6}\big(\widehat{\myGenLe}^B \myGenLe^A_L \widehat{\myGenLe}^B\big)_{ab} \bigg\} \delta_{IJ}
        + \frac{1}{2} \myGenScl^A_{22} \delta_{ab} \big(\mathbf{Y}_2(\overline{L})\big)_{IJ},
    \end{split}
\end{align}
while the quark coefficients are
\begin{align}
    \begin{split}\label{Eq:SM-finite-CT-Quarks-KineticTerm}
        \delta \mathcal{F}^{(1)}_{q,ab} &= \frac{3-\xi_F}{3} \mathcal{C}_2^{ab}(F^q_{LR}) - \frac{\xi_F}{3}\mathcal{C}_2^{ab}(F_{L}^q, F^q_{LR}),
    \end{split}\\[1.25ex]
    \begin{split}\label{Eq:SM-finite-CT-Quarks-Right-handed-Interaction}
        \delta \mathcal{F}^{(1),A}_{\overline{q}\vgb q,R,ab,ij} &=  
        - \frac{6-\xi_F}{6} \mathcal{C}_2^{ac}(F^q_R) \myGenQu^A_{R,cb,ij} 
        - \frac{\xi_F}{3} \mathcal{C}_2^{ac}(F^q_R,F^q_{LR}) \myGenQu^A_{R,cb,ij}
        - \frac{\xi_F}{6} \mathcal{C}_2^{ac}(F^q_{LR}) \mathrlap{ \myGenQu^A_{R,cb,ij} }\\ 
        &\quad\,\, 
        - \frac{\xi_F}{3} \mathcal{C}_2^{ac}(F^q_L,F^q_{LR}) \myGenQu^A_{R,cb,ij}
        + \frac{6-\xi_F}{6}\big(\myGenQu^B_R \myGenQu^A_L \myGenQu^B_R\big)_{ab,ij} \\
        &\quad\,\,  
        + \frac{6-\xi_F}{6}\big(\widehat{\myGenQu}^B \myGenQu^A_L \widehat{\myGenQu}^B\big)_{ab,ij}
        + \frac{\xi_F}{3}\big(\widehat{\myGenQu}^B \myGenQu^A_L \myGenQu^B_R\big)_{ab,ij} \\
        &\quad\,\,  
        + \frac{1+c_\mathrm{QCD}}{2} \mathcal{C}_2( G^{A'} ) {\widehat{\myGenQu}}{}^A_{ab,ij},
    \end{split}\\[1.25ex]
    \begin{split}\label{Eq:SM-finite-CT-Quarks-Left-handed-Interaction}
        \delta \mathcal{F}^{(1),A}_{\overline{q}\vgb q,L,ab,ij,IJ} 
        &= \bigg\{
        \frac{6-\xi_F}{6} \mathcal{C}_2^{ac}(F^q_{LR}) \myGenQu^A_{R,cb,ij}
        + \frac{\xi_F}{3} \mathcal{C}_2^{ac}(F^q_L,F^q_{LR}) \myGenQu^A_{R,cb,ij} \\
        &\quad\,\,
        - \frac{6-\xi_F}{6} \mathcal{C}_2(F^q_L) \myGenQu^A_{L,ab,ij}
        + \frac{3-\xi_F}{6} \mathcal{C}_2( G^{A'} ) {\widehat{\myGenQu}}{}^A_{ab,ij} \\
        &\quad\,\,
        - \frac{\xi_F}{3} \Big[ \mathcal{C}_2^{ac}(F^q_L,F^q_{LR}) \myGenQu^A_{L,cb,ij} + \myGenQu^A_{L,ac,ij} \mathcal{C}_2^{cb}(F^q_L,F^q_{LR}) \Big]\\
        &\quad\,\,
        + \frac{6-\xi_F}{6}\big(\myGenQu^B_L \myGenQu^A_R \myGenQu^B_L\big)_{ab,ij}
        - \frac{\xi_F}{6}\big(\widehat{\myGenQu}^B \myGenQu^A_L \widehat{\myGenQu}^B\big)_{ab,ij} 
        \bigg\} \delta_{IJ} \\
        &\quad\,\,
        + \frac{1}{2} \myGenScl^A_{22} \delta_{ab} \delta_{ij} \Big[ 
        \big(\mathbf{Y}_2(\overline{D})\big)_{IJ}
        - \big(\mathbf{Y}_2(\overline{U})\big)_{IJ}
        \Big].
    \end{split}
\end{align}

\paragraph{Gauge Sector:}
Next, we turn to the gauge sector, with counterterm action given in Eq.~\eqref{Eq:SM-S_fct_gauge}.
The corresponding coefficients take the form
\begin{align}
    \begin{split}\label{Eq:SM-finite-CT-gauge-boson-SelfEnergy}
        \delta \mathcal{F}^{(1),AB}_{2\vgb} &= - \frac{N_f}{6} \!\! \sum_{f\in\{l,q\}} \!\!  \Big[ 
        \mathcal{S}_2\big(F_R^{f,A'}\big) \delta^{AB} + \mathcal{S}_2\big(F_L^{f,A'}\big) \delta^{AB}\\
        &\phantom{= \frac{N_f}{6} \!\! \sum_{f\in\{l,q\}} \!\!  \Big[} 
        - \mathcal{S}^{AB}_2\big(F^{f}_R,F^{f}_L\big) - \mathcal{S}^{BA}_2\big(F^{f}_R,F^{f}_L\big)
        \Big],
    \end{split}\\[1.25ex]
    \begin{split}\label{Eq:SM-finite-CT-gauge-boson-3Point-1}
        \delta \mathcal{F}^{(1),ABC}_{3\vgb\text{-1}} &= \frac{N_f}{3} \!\! \sum_{f\in\{l,q\}} \!\! \Big[ 
        \mathcal{S}^{CD}_2\big(F^{f}_R,F^{f}_L\big) \mathscr{C}^{DAB} 
        - \mathcal{S}^{BD}_2\big(F^{f}_R,F^{f}_L\big) \mathscr{C}^{DAC}\\
        &\hspace{1.92cm}
        - 2 \,\mathcal{S}_2\big(F_L^{f,C'}\big) \mathscr{C}^{ABC}
        \Big],
    \end{split}\\[1.25ex]
    \begin{split}\label{Eq:SM-finite-CT-gauge-boson-3Point-2}
        \delta \mathcal{F}^{(1),ABC}_{3\vgb\text{-2}} &= - \frac{N_f}{3} \!\! \sum_{f\in\{l,q\}} \!\!
        \mathcal{S}^{CD}_2\big(F^{f}_R,F^{f}_L\big) \mathscr{C}^{DAB},
    \end{split}\\[1.25ex]
    \begin{split}\label{Eq:SM-finite-CT-gauge-boson-4Point}
        \delta \mathcal{F}^{(1),ABCD}_{4\vgb} &= - \frac{N_f}{12} \!\! \sum_{f\in\{l,q\}} \!\! \Big[
        5 \myTrbig{F_L^{f,A}, F_L^{f,D}, F_L^{f,B}, F_L^{f,C} }
        - 4 \myTrbig{ F_R^{f,A}, F_L^{f,D}, F_L^{f,B}, F_L^{f,C} }\\
        &\hspace{1.75cm}
        - 2 \myTrbig{ F_R^{f,A}, F_L^{f,D}, F_R^{f,B}, F_L^{f,C} }
        + 2 \myTrbig{ F_R^{f,A}, F_R^{f,D}, F_R^{f,B}, F_L^{f,C} }\\
        &\hspace{1.75cm}
        - \myTrbig{ F_R^{f,A}, F_R^{f,D}, F_R^{f,B}, F_R^{f,C} }
        - 6 \myTrbig{ F_L^{f,A}, F_L^{f,D}, F_L^{f,C}, F_L^{f,B} }\\
        &\hspace{1.75cm}
        + 8 \myTrbig{ F_R^{f,A}, F_L^{f,D}, F_L^{f,C}, F_L^{f,B} }
        + 4 \myTrbig{ F_R^{f,A}, F_L^{f,D}, F_R^{f,C}, F_L^{f,B} }\\
        &\hspace{1.75cm}
        + 2 \myTrbig{ F_R^{f,A}, F_R^{f,D}, F_R^{f,C}, F_L^{f,B} }
        - 8 \myTrbig{ F_R^{f,A}, F_R^{f,D}, F_L^{f,C}, F_L^{f,B} }
        \Big],
    \end{split}
\end{align}
where in the last equation $F_Y^{f,A}$ denotes the fermion generator of chirality $Y\in\{R,L\}$ with adjoint colour index $A$, such that for instance
\begin{align}
    \myTrbig{ F_R^{q,A}, F_L^{q,D}, F_R^{q,B}, F_L^{q,C} } \equiv \myTrbig{ \myGenQu_R^A, \myGenQu_L^D, \myGenQu_R^B, \myGenQu_L^C }.
\end{align}

\paragraph{Higgs Sector:}
The finite symmetry-restoring restoring counterterm action of the Higgs sector is provided by Eq.~\eqref{Eq:SM-S_fct_Higgs}.
The global hypercharge conserving scalar and scalar--gauge boson counterterm coefficients read
\begin{align}
    \begin{split}\label{Eq:SM-finite-CT-Scalar-SelfEnergy}
        \delta \mathcal{F}^{(1)}_{\phi^\dagger\phi,ab} &= \frac{1}{6} \mathcal{H}_2(D,U) \sigma^3_{ab},
    \end{split}\\[1.25ex]
    \begin{split}
        \delta \mathcal{F}^{(1),A}_{-,ab} &= \frac{1}{6}\bigg[
        \mathcal{H}_2(Y) - \frac{1}{2}\mathcal{H}_2(D,U) 
        \bigg] \{\myGenScl^A, \sigma^3\}_{ab} 
        - \frac{1}{3} \mathcal{H}_2(Y) \myGenScl^A_{ab},
    \end{split}\\[1.25ex]
    \begin{split}
        \delta \mathcal{F}^{(1),A}_{+,ab} &= - \frac{1}{4} \mathcal{H}_2(D,U) [\myGenScl^A, \sigma^3]_{ab},
    \end{split}\\[1.25ex]
    \begin{split}\label{Eq:SM-finite-CT-Scalar-DoubleScalar-DoubleGaugeBoson}
        \delta \mathcal{F}^{(1),AB}_{\phi^\dagger\phi\vgb\vgb,ab} &= 
        h_1 \{\myGenScl^A, \myGenScl^B\}_{ab} \\
        &
        + h_2 \Big[ 
        (\myGenScl^A \sigma^3 \myGenScl^B)_{ab} 
        + (\myGenScl^B \sigma^3 \myGenScl^A)_{ab} 
        - (\sigma^3 \myGenScl^A \myGenScl^B \sigma^3)_{ab} 
        - (\sigma^3 \myGenScl^B \myGenScl^A \sigma^3)_{ab} \Big]\\
        &
        + h_3 \Big[ 
        (\sigma^3 \myGenScl^A \myGenScl^B)_{ab} 
        + (\sigma^3 \myGenScl^B \myGenScl^A)_{ab} 
        + (\myGenScl^A \myGenScl^B \sigma^3)_{ab} 
        + (\myGenScl^B \myGenScl^A \sigma^3)_{ab} \Big] \\
        &
        + h_4 \Big[ 
        (\sigma^3 \myGenScl^A \sigma^3 \myGenScl^B)_{ab} 
        + (\sigma^3 \myGenScl^B \sigma^3 \myGenScl^A)_{ab} 
        + (\myGenScl^A \sigma^3 \myGenScl^B \sigma^3)_{ab} 
        + (\myGenScl^B \sigma^3 \myGenScl^A \sigma^3)_{ab} \\
        &\qquad
        + (\sigma^3 \myGenScl^A \sigma^3 \myGenScl^B \sigma^3)_{ab} 
        + (\sigma^3 \myGenScl^B \sigma^3 \myGenScl^A \sigma^3)_{ab}\\
        &\qquad
        - 8 (\sigma_{12}^{+} \myGenScl^A \sigma_{03}^{-} \myGenScl^B \sigma_{12}^{-} )_{ab}
        - 8 (\sigma_{12}^{+} \myGenScl^B \sigma_{03}^{-} \myGenScl^A \sigma_{12}^{-} )_{ab} \Big],
    \end{split}
\end{align}
with $\sigma_{12}^\pm$ and $\sigma_{03}^\pm$ introduced in Eq.~\eqref{Eq:Pauli-Combi-Matrices}, and $h_i$ defined in Eq.~\eqref{Eq:Def-YukawaCoeffs-h_1--h_4}.

The global hypercharge violating contributions are governed by 
\begin{align}
    \begin{split}\label{Eq:SM-finite-CT-Scalar-phi-phi}
        \delta \mathcal{F}_{\phi \phi}^{(1)} &= - \frac{1}{6} \, \mathcal{H}_2(L,D,\overline{U}),
    \end{split}\\[1.25ex]
    \begin{split}
        \delta \mathcal{F}_{\phi \phi \vgb,ab}^{(1),A} &=
        -\frac{1}{2} \mathcal{H}_2(L,D,\overline{U}) \bigg[ 
        \myGenScl^A_{ab}
        - \frac{1}{2}\{\myGenScl^A, \sigma^3\}_{ab} 
        + \big(\sigma_{12}^{+} \myGenScl^A \sigma_{12}^{+}\big)_{ab}
        - \big(\sigma_{03}^{+} \myGenScl^A \sigma_{03}^{-}\big)_{ab}\bigg]\\
        &\quad
        + \varrho_0 \bigg[ 
        \frac{1}{2}[\myGenScl^A, \sigma^3]_{ab} 
        - \big(\sigma_{12}^{+} \myGenScl^A \sigma_{12}^{+}\big)_{ab}
        + \big(\sigma_{03}^{+} \myGenScl^A \sigma_{03}^{-}\big)_{ab} \bigg],
    \end{split}\\[1.25ex]
    \begin{split}\label{Eq:SM-finite-CT-Scalar-phi-phi-V-V}
        \delta \mathcal{F}_{\phi\phi \vgb \vgb,ab}^{(1),AB} &=
        \varrho_1 \Big[ 
        \big(\sigma_{12}^+ \myGenScl^A \myGenScl^B \sigma_{12}^+\big)_{ab} 
        + \big(\sigma_{12}^+ \myGenScl^B \myGenScl^A \sigma_{12}^+ \big)_{ab} 
        \Big] \\
        & 
        + \varrho_2 \Big[ 
        \big( \sigma_{03}^- \myGenScl^A \myGenScl^B \sigma_{03}^- \big)_{ab} 
        + \big( \sigma_{03}^- \myGenScl^B \myGenScl^A \sigma_{03}^- \big)_{ab}\\
        &\qquad
        + \big( \sigma_{03}^- \myGenScl^A \myGenScl^B \sigma_{03}^+ \big)_{ab} 
        + \big( \sigma_{03}^- \myGenScl^B \myGenScl^A \sigma_{03}^+ \big)_{ab} 
        \Big]\\
        & 
        + \varrho_3 \Big[ 
        \big( \sigma_{12}^+ \myGenScl^A \sigma_{12}^+ \myGenScl^B \big)_{ab} 
        + \big( \sigma_{12}^+ \myGenScl^B \sigma_{12}^+ \myGenScl^A \big)_{ab}\\
        &\qquad
        - \big( \sigma_{03}^- \myGenScl^A \sigma_{03}^+ \myGenScl^B \big)_{ab} 
        - \big( \sigma_{03}^- \myGenScl^B \sigma_{03}^+ \myGenScl^A \big)_{ab}
        \Big],
    \end{split}
\end{align}
with $\varrho_i$ as defined in Eq.~\eqref{Eq:Def-YukawaCoeffs-varrho_0--varrho_3}.

Finally, the coefficients of the finite quartic scalar self-interactions are given by 
\begin{align}
    \begin{split}\label{Eq:SM-finite-CT-Scalar-phidagger-phi-phidagger-phi}
        \delta \mathcal{F}^{(1)}_{\phi^\dagger\phi\phi^\dagger\phi,abcd} &=
        \frac{1}{6} \Big[ 
        \mathcal{H}_{4}^{(1)} (\sigma_{03}^{+})_{ab} (\sigma_{03}^{+})_{cd} 
        + \mathcal{H}_{4}^{(2)} (\sigma_{03}^{-})_{ab} (\sigma_{03}^{-})_{cd} \Big],
    \end{split}\\[1.25ex]
    \begin{split}
        \delta \mathcal{F}^{(1)}_{\phi^\dagger\phi\phi\phi,abcd} &= \frac{1}{3} \Big[ \mathcal{H}_4^{(3)} - \mathcal{H}_4^{(4)} \Big] \big( \sigma^+_{03} \big)_{ab} \big( \sigma^-_{03} \big)_{cd}
        + \frac{1}{3} \mathcal{H}_4^{(3)} \big( \sigma^-_{03} \big)_{ab} \big( \sigma^-_{03} \big)_{cd},
    \end{split}\\[1.25ex]
    \begin{split}\label{Eq:SM-finite-CT-Scalar-phi-phi-phi-phi}
        \delta \mathcal{F}^{(1)}_{\phi\phi\phi\phi} &=
        \frac{1}{12} \, \mathcal{H}_4(L,D,\overline{U})\,,
    \end{split}
\end{align}
with $\mathcal{H}_4^{(i)}$ defined in Eqs.~\eqref{Eq:Def-YukawaCoeffs-H4_1}--\eqref{Eq:Def-YukawaCoeffs-H4_4}.

\paragraph{Yukawa Sector:}
We continue with the Yukawa sector, with counterterm action given in Eq.~\eqref{Eq:SM-S_fct_Yukawa}.
The lepton contributions are given by
\begin{align}
    \begin{split}\label{Eq:SM-finite-CT-Yukawa-Lepton-F}
        \delta \mathcal{F}^{l,(1)}_{Y,abc} &= - \bigg[ 
        \frac{3+\xi_F}{12}\mathcal{C}_2^{ab}(F^l_{R}, F^l_{LR}) + \frac{\xi_F}{6} \mathcal{C}_2^{ab}(F^l_{LR})
        \bigg] \delta_{c2} \, ,
    \end{split}\\
    \begin{split}
        \delta \mathcal{G}^{l,(1)}_{Y,abc,IJ} &= Y_{l,JI}^* \bigg[
        \frac{6-\xi_F}{6} \mathcal{C}_2^{cb}(F^l_R)
        + \frac{\xi_F}{2}\mathcal{C}_2^{cb}(F^l_{R},F^l_{LR}) 
        + \frac{6-\xi_F}{3}\mathcal{C}_2^{cb}(F^l_{LR})
        \bigg] \delta_{a2} \, ,
    \end{split}
\end{align}
while the down-type quark contributions are governed by the coefficients
\begin{align}
    \begin{split}
        \delta \mathcal{F}^{d,(1)}_{Y,abc} &= \bigg[ 
        \frac{3+\xi_F}{12}\mathcal{C}_2^{ab}(F^q_{R}, F^q_{LR}) 
        + \frac{\xi_F}{6} \mathcal{C}_2^{ab}(F^q_{LR})
        + (1-\xi_F)\mathcal{C}_2^{ab}(F^q_{L}, F^q_{LR}) 
        \bigg] \delta_{c2} \, ,
    \end{split}\\
    \begin{split}
        \delta \mathcal{G}^{d,(1)}_{Y,abc,IJ} &=
        Y_{d,JI}^* \bigg[
        \frac{\xi_F}{3} \mathcal{C}_2^{cb}(F^q_{R},F^q_{LR})\delta_{a2} 
        - \frac{6-\xi_F}{3}\mathcal{C}_2^{cb}(F^q_{LR})\delta_{a2} 
        - \xi_F \mathcal{C}_2^{cb}(F^q_{L}, F^q_{LR})\delta_{a2}\\
        &\quad 
        + \frac{6-\xi_F}{3}\mathcal{C}_2^{cb}(F^q_{R}, F^q_L)\delta_{a2} 
        + \frac{4}{3}\xi_F\mathcal{C}_2^{2a}(F^q_L, F^q_{LR})\delta_{bc} 
        + 2 \, \mathcal{C}_2^{2a}(F^q_{R}, F^q_L)\delta_{bc}\\
        &\quad 
        + \frac{2}{3}(6-\xi_F)\mathcal{C}_2^{2b}(F^q_{LR})\delta_{ac} 
        - 2 \, \mathcal{C}_2^{2b}(F^q_{R}, F^q_L)\delta_{ac} \bigg]\\
        &\quad
        + \frac{1}{2} (\sigma_{12}^{-})_{bc} \delta_{a1} (Y_dY_uY_u)_{IJ}
        - \frac{1}{2} \Big[ (\sigma_{12}^{-})_{bc} \delta_{a1} - \delta_{bc} \delta_{a2} \Big] (Y_uY_uY_d)_{IJ} \, ,
    \end{split}
\end{align}
and the up-type quark coefficients take the form
\begin{align}
    \begin{split}
        \delta \mathcal{F}^{u,(1)}_{Y,abc} &= 
        \frac{\xi_F}{3} \mathcal{C}_2^{ea}(F^q_{LR}) \varepsilon_{be} \delta_{c1} 
        + \frac{9-\xi_F}{6}\mathcal{C}_2^{ea}(F^q_{L}, F^q_{LR}) \varepsilon_{be}\delta_{c1}
        + 2\,\mathcal{C}_2^{ea}(F^q_{R}, F^q_L)\varepsilon_{be} \delta_{c1}\\
        &\quad
        + \frac{3}{2}\mathcal{C}_2^{ce}(F^q_R)\varepsilon_{eb}\delta_{a1} 
        - \mathcal{C}_2^{ce}(F^q_{R}, F^q_L)\varepsilon_{eb}\delta_{a1}
        + \frac{1}{2}\mathcal{C}_2^{1a}(F^q_R)\varepsilon_{bc} \, ,
    \end{split}\\
    \begin{split}\label{Eq:SM-finite-CT-Yukawa-Up-Quark-G}
        \delta \mathcal{G}^{u,(1)}_{Y,abc,IJ} &=
        Y_{u,JI}^* \bigg[
        \frac{6-\xi_F}{3}\mathcal{C}_2^{ec}(F^q_R, F^q_L)\varepsilon_{eb}\delta_{a1}
        + \frac{6+5\xi_F}{6}\mathcal{C}_2^{ec}(F^q_R, F^q_{LR})\varepsilon_{eb}\delta_{a1} \\
        &\quad
        - \mathcal{C}_2^{ec}(F^q_{LR})\varepsilon_{eb}\delta_{a1} 
        - \frac{2+\xi_F}{2}\mathcal{C}_2^{a1}(F^q_R, F^q_{LR})\varepsilon_{cb} 
        - \frac{6-\xi_F}{3}\mathcal{C}_2^{a1}(F^q_{LR})\varepsilon_{cb}\\ 
        &\quad 
        + \frac{6+\xi_F}{3}\mathcal{C}_2^{a1}(F^q_L, F^q_{LR})\varepsilon_{cb}
        - \mathcal{C}_2^{ec}(F^q_{L}, F^q_{LR})\varepsilon_{eb}\delta_{a1} 
        + \frac{15-2\xi_F}{3}\mathcal{C}_2^{ac}(F^q_{LR})\varepsilon_{1b}\\ 
        &\quad 
        - \mathcal{C}_2^{ac}(F^q_{L}, F^q_{LR})\varepsilon_{1b} \bigg]
        - \frac{1}{2} (\sigma_{12}^{+})_{bc} \delta_{a1} (Y_uY_dY_d)_{IJ}
        + \frac{1}{2} (\sigma_{12}^{-})_{bc} \delta_{a1} (Y_dY_dY_u)_{IJ} \, .
    \end{split}
\end{align}

\paragraph{Ghosts and BRST Transformations:}
Finally, the finite symmetry-restoring counterterm contribution to the BRST transformations is given by Eq.~\eqref{Eq:SM-S_fct_ext}, with coefficient
\begin{align}\label{Eq:SM-finite-CT-external-Sources}
    \delta \mathcal{F}^{(1)}_{R_f} = - \frac{1-\xi_F}{4} \mathcal{C}_2(G^{A'}).
\end{align}

\section{Final Results for the Counterterm Coefficients} \label{App:CountertermCoeffs_Expanded}

Here, we list the explicit 1-loop results for the coefficients entering the sector-wise counterterm action of Sec.~\ref{Sec:SM_1-Loop_Renormalisation_Expanded}, expressed in terms of the individual gauge fields $B_\mu$, $W_\mu^A$ and $G_\mu^A$.
The coefficients of the divergent counterterms in Sec.~\ref{Sec:SM-S1_SingularCT} are given in App.~\ref{App:Divergent_CountertermCoeffs_Expanded}, and those of the finite symmetry-restoring counterterms in Sec.~\ref{Sec:SM-S1_FiniteCT} in App.~\ref{App:Finite_CountertermCoeffs_Expanded}.
The corresponding coefficients of the compact twelveplet notation are collected in App.~\ref{App:Results-TheStandardModel}.

All results are obtained as discussed in Sec.~\ref{Sec:SM_1-Loop_Renormalisation_Expanded}.
In principle, they follow from those of App.~\ref{App:Results-TheStandardModel} by decomposing the twelveplet quantities into the contributions of the individual gauge groups, up to rearrangements of the counterterms and the addition of finite symmetric counterterms (see the same section).
As before, they are expressed in terms of the modified gauge parameter $\xi_F=1-\xi$, and the group invariants and coupling structures defined in App.~\ref{App:SM-GroupInvariants-CouplingStructures}.

\subsection{Coefficients of Divergent 1-Loop Counterterms} \label{App:Divergent_CountertermCoeffs_Expanded}

We begin with the coefficients of the divergent 1-loop contributions.

\paragraph{Fermionic Sector:}
For the divergent fermion counterterms, we find for the leptons in Eq.~\eqref{Eq:SM-S1_sct_Lepton_expanded},
\begin{align}
    \begin{split}
        \delta \overline{Z}{}^{l,(1)}_{R,ab,IJ} &= - \frac{1}{\epsilon} \bigg\{ \big( 1 - \xi_F \big) g_Y^2 \big( \mathcal{Y}^l_R \mathcal{Y}^l_R \big)_{ab} \delta_{IJ} + \frac{N_L}{2} \delta_{2a} \delta_{2b} \big( Y_l^\dagger Y_l \big)_{IJ} \bigg\},
    \end{split}\\[1.25ex]
    \begin{split}
        \delta \overline{Z}{}^{l,(1)}_{L,IJ} &= - \frac{1}{\epsilon} \bigg\{ \big( 1 - \xi_F \big) \Big[ g_Y^2 \mathcal{Y}_l^2 + g_W^2 C_2(F_L) \Big] \delta_{IJ} + \frac{1}{2} \big( Y_l Y_l^\dagger \big)_{IJ} \bigg\},
    \end{split}\\[1.25ex]
    \begin{split}
        \delta {\overline{X}}{}^{(1)}_{\overline{l}Wl} &= - \frac{4 - \xi_F}{4\epsilon} \, g_W^2 C_2(G_L),
    \end{split}\\[1.25ex]
    \begin{split}
        \delta {\widehat{X}}{}^{l,(1)}_{ab} &= - \frac{g_Y^2}{\epsilon} \bigg[ \frac{6 - (2 + c_{\mathrm{QED}}) \xi_F}{3} \big( \mathcal{Y}^l_L \mathcal{Y}^l_R \big)_{ab} - c_{\mathrm{QED}} \Big( \frac{\xi_F}{3} + c_{\mathrm{QED}} \frac{3 - \xi_F}{3} \Big) \big( \mathcal{Y}^l_R \mathcal{Y}^l_R \big)_{ab} \bigg],
    \end{split}
\end{align}
and for the quarks in Eq.~\eqref{Eq:SM-S1_sct_Quark_expanded},
\begin{align}
    \begin{split}
        \delta \overline{Z}{}^{q,(1)}_{R,ab,IJ} 
        &= - \frac{1}{\epsilon} \bigg\{ \big( 1 - \xi_F \big) \Big[ g_Y^2 \big( \mathcal{Y}^q_R \mathcal{Y}^q_R \big)_{ab} + g_s^2 C_2(F_s) \delta_{ab} \Big] \delta_{IJ}\\
        &\qquad\quad + \frac{N_L}{2} \delta_{2a} \delta_{2b} \big( Y_d^\dagger Y_d \big)_{IJ} + \delta_{1a} \delta_{1b} \big( Y_u^\dagger Y_u \big)_{IJ} \bigg\},
    \end{split}\\[1.25ex]
    \begin{split}
        \delta \overline{Z}{}^{q,(1)}_{L,IJ} &= - \frac{1}{\epsilon} \bigg\{ \big( 1 - \xi_F \big) \Big[ g_Y^2 \mathcal{Y}_q^2 + g_W^2 C_2(F_L) + g_s^2 C_2(F_s) \Big] \delta_{IJ} \\
        &\qquad\quad + \frac{1}{2} \Big[ \big( Y_d Y_d^\dagger \big)_{IJ} + \big( Y_u Y_u^\dagger \big)_{IJ} \Big] \bigg\},
    \end{split}\\[1.25ex]
    \begin{split}
        \delta {\overline{X}}{}^{(1)}_{\overline{q}Wq} &= - \frac{4 - \xi_F}{4\epsilon} \, g_W^2 C_2(G_L), 
    \end{split}\\[1.25ex]
    \begin{split}
        \delta {\overline{X}}{}^{(1)}_{\overline{q}Gq} &= - \frac{4 - \xi_F}{4\epsilon} \, g_s^2 C_2(G_s),
    \end{split}\\[1.25ex]
    \begin{split}
        \delta {\widehat{X}}{}^{q,(1)}_{ab} &= - \frac{1}{\epsilon} \bigg\{ g_Y^2 \bigg[ \frac{6 - (2 + c_{\mathrm{QED}}) \xi_F}{3} \big( \mathcal{Y}^q_L \mathcal{Y}^q_R \big)_{ab} - c_{\mathrm{QED}} \bigg( \frac{\xi_F}{3} + c_{\mathrm{QED}} \frac{3 - \xi_F}{3} \bigg) \big( \mathcal{Y}^q_R \mathcal{Y}^q_R \big)_{ab} \bigg]\\
        &\qquad\quad + g_s^2 \bigg[ \frac{6 - 2\xi_F}{3} - c_{\mathrm{QCD}} \bigg( \frac{2\xi_F}{3} + c_{\mathrm{QCD}} \frac{3 - \xi_F}{3} \bigg) \bigg] C_2(F_s) \delta_{ab} \bigg\},
    \end{split}
\end{align}
\begin{align}
    \begin{split}
        \delta {\widehat{X}}{}^{q,(1)}_{Y,ab,IJ} &= \frac{1}{2\epsilon} \Big[ \delta_{1a} \delta_{1b} \big( Y_d Y_u \big)_{IJ} + \delta_{2a} \delta_{2b} \big( Y_u Y_d \big)_{IJ} \Big],
    \end{split}\\[1.25ex]
    \begin{split}
        \delta {\widehat{X}}{}^{(1)}_{\overline{q}Bq,ab,IJ} &= \frac{1 - 2 c_{\mathrm{QED}}}{2\epsilon} \mathcal{Y}_{S} \Big[ \delta_{1a} \delta_{1b} \big( Y_d Y_u \big)_{IJ} - \delta_{2a} \delta_{2b} \big( Y_u Y_d \big)_{IJ} \Big],
    \end{split}\\[1.25ex]
    \begin{split}
        \delta {\widehat{X}}{}^{(1),A}_{\overline{q}Wq,ab,IJ} &= \frac{1}{2\epsilon} \Big[ \big( t^A_{L,11} \delta_{1a} \delta_{1b} + t^A_{L,21} \delta_{2a} \delta_{1b} \big) \big( Y_d Y_u \big)_{IJ} + \big( t^A_{L,22} \delta_{2a} \delta_{2b} + t^A_{L,12} \delta_{1a} \delta_{2b} \big) \big( Y_u Y_d \big)_{IJ} \Big],
    \end{split}\\[1.25ex]
    \begin{split}
        \delta {\widehat{X}}{}^{(1)}_{\overline{q}Gq} &= - \frac{k_{\mathrm{QCD}}}{\epsilon} \, g_s^2 C_2(G_s),
    \end{split}
\end{align}
with $k_{\mathrm{QCD}}$ given in Eq.~\eqref{Eq:SM-k_QCD}.

\paragraph{Gauge Sector:}
The electroweak part of the divergent gauge-sector counterterms are given in Eq.~\eqref{Eq:SM-S1_sct_EW-gauge_expanded}, which coefficients
\begin{align}
    \begin{split}
        \delta Z^{(1)}_{B} &= - \frac{N_L}{3 \epsilon} \, g_Y^2 \mathcal{Y}_S^2,
    \end{split}\\[1.25ex]
    \begin{split}
        \delta \overline{Z}{}^{(1)}_{B} &= - \frac{2}{3\epsilon} N_f \, g_Y^2 \Big\{ \mathrm{Tr} \Big[ \big( \mathcal{Y}^l_R \big)^2 + \big( \mathcal{Y}^l_L \big)^2 \Big] + N_c \, \mathrm{Tr} \Big[ \big( \mathcal{Y}^q_R \big)^2 + \big( \mathcal{Y}^q_L \big)^2 \Big] \Big\},
    \end{split}\\[1.25ex]
    \begin{split}
        \delta \widehat{Z}{}^{(1)}_{B} &= - \frac{4}{3\epsilon} N_f \, g_Y^2 \Big\{ \mathrm{Tr} \Big[ \big( \mathcal{Y}^l_R \big)^2 \Big] + N_c \, \mathrm{Tr} \Big[ \big( \mathcal{Y}^q_R \big)^2 \Big] \Big\},
    \end{split}\\[1.25ex]
    \begin{split}
        \delta \widehat{X}{}^{(1)}_{B,12} &= - \frac{1}{3\epsilon} N_f \, g_Y^2 \Big\{ \mathrm{Tr} \Big[ \big( \mathcal{Y}^l_R + \mathcal{Y}^l_L \big)^2 \Big] + N_c \, \mathrm{Tr} \Big[ \big( \mathcal{Y}^q_R + \mathcal{Y}^q_L \big)^2 \Big] \Big\},
    \end{split}\\[1.25ex]
    \begin{split}
        \delta \widehat{X}{}^{(1)}_{B,34} &= - \frac{2}{3\epsilon} N_f \, g_Y^2 \Big\{ \mathrm{Tr} \big[ \mathcal{Y}^l_R \big( \mathcal{Y}^l_R + \mathcal{Y}^l_L \big) \big] + N_c \, \mathrm{Tr} \big[ \mathcal{Y}^q_R \big( \mathcal{Y}^q_R + \mathcal{Y}^q_L \big) \big] \Big\},
    \end{split}\\[1.25ex]
    \begin{split}
        \delta \widehat{X}{}^{(1),A}_{BW,12} &= - \frac{2}{3\epsilon} N_f \, g_Y g_W \Big\{ \mathrm{Tr} \big( \mathcal{Y}^l_R t^A_L \big) + N_c \, \mathrm{Tr} \big( \mathcal{Y}^q_R t^A_L \big) \Big\},
    \end{split}\\[1.25ex]
    \begin{split}
        \delta \widehat{X}{}^{(1),AB}_{BBWW} &= - \frac{2}{3\epsilon} N_f \, g_Y g_W \Big\{ \mathrm{Tr} \big( \mathcal{Y}^l_R \mathcal{Y}^l_R t^A_L t^B_L \big) + \mathrm{Tr} \big( \mathcal{Y}^l_R \mathcal{Y}^l_R t^B_L t^A_L \big) - 2 \, \mathrm{Tr} \big( \mathcal{Y}^l_R t^A_L \mathcal{Y}^l_R t^B_L \big)\\
        &\qquad\qquad\quad\! \ + N_c \Big[ \mathrm{Tr} \big( \mathcal{Y}^q_R \mathcal{Y}^q_R t^A_L t^B_L \big) + \mathrm{Tr} \big( \mathcal{Y}^q_R \mathcal{Y}^q_R t^B_L t^A_L \big) - 2 \, \mathrm{Tr} \big( \mathcal{Y}^q_R t^A_L \mathcal{Y}^q_R t^B_L \big) \Big] \Big\},
    \end{split}\\[1.25ex]
    \begin{split}
        \delta Z^{(1)}_{W} &= \frac{g_W^2}{\epsilon} \bigg[ \frac{10 + 3\xi_F}{6} C_2(G_L) - \frac{1}{3} S_2(F_L) \bigg],
    \end{split}\\[1.25ex]
    \begin{split}
        \delta Z^{(1)}_{3W} &= \frac{g_W^2}{\epsilon} \bigg[ \frac{8 + 9\xi_F}{12} C_2(G_L) - \frac{1}{3} S_2(F_L) \bigg],
    \end{split}\\[1.25ex]
    \begin{split}
        \delta Z^{(1)}_{4W} &= - \frac{g_W^2}{\epsilon} \bigg[ \frac{1 - 3\xi_F}{3} C_2(G_L) + \frac{1}{3} S_2(F_L) \bigg],
    \end{split}\\[1.25ex]
    \begin{split}
        \delta \overline{Z}{}^{(1)}_{W} &= - \frac{2}{3\epsilon} N_f ( 1 + N_c ) g_W^2 S_2(F_L),
    \end{split}
\end{align}
and the strong part is provided in Eq.~\eqref{Eq:SM-S1_sct_QCD-gauge_expanded}, with coefficients
\begin{align}
    \begin{split}
        \delta Z^{(1)}_{G} &= \frac{g_s^2}{\epsilon} \, \frac{10 + 3\xi_F}{6} C_2(G_s),
    \end{split}\\[1.25ex]
    \begin{split}
        \delta Z^{(1)}_{3G} &= \frac{g_s^2}{\epsilon} \, \frac{8 + 9\xi_F}{12} C_2(G_s),
    \end{split}\\[1.25ex]
    \begin{split}
        \delta Z^{(1)}_{4G} &= - \frac{g_s^2}{\epsilon} \, \frac{1 - 3\xi_F}{3} C_2(G_s),
    \end{split}\\[1.25ex]
    \begin{split}\label{Eq:SCT-dZGbar1}
        \delta \overline{Z}{}^{(1)}_{G} &= - \frac{4}{3\epsilon} N_f N_L \, g_s^2 S_2(F_s).
    \end{split}
\end{align}

\paragraph{Higgs Sector:}
The divergent 1-loop Higgs-sector counterterms are presented in Eq.~\eqref{Eq:SM-S1_sct_Higgs_expanded} and their coefficients read
\begin{align}
    \begin{split}
        \delta Z^{(1)}_{\phi} &= \frac{1}{\epsilon} ( 2 + \xi_F ) \Big[ g_Y^2 \mathcal{Y}_S^2 + g_W^2 C_2(F_L) \Big],
    \end{split}\\[1.25ex]
    \begin{split}
        \delta Z^{(1)}_{\phi^\dagger\phi W} &= \frac{1}{\epsilon} \bigg\{ ( 2 + \xi_F ) \Big[ g_Y^2 \mathcal{Y}_S^2 + g_W^2 C_2(F_L) \Big] - \frac{4 - \xi_F}{4} g_W^2 C_2(G_L) \bigg\},
    \end{split}\\[1.25ex]
    \begin{split}
        \delta Z^{(1)}_{\phi^\dagger\phi WW} &= \frac{1}{\epsilon} \bigg\{ ( 2 + \xi_F ) \Big[ g_Y^2 \mathcal{Y}_S^2 + g_W^2 C_2(F_L) \Big] - \frac{4 - \xi_F}{2} g_W^2 C_2(G_L) \bigg\},
    \end{split}\\[1.25ex]
    \begin{split}
        \delta \overline{Z}{}^{(1)}_{\phi} &= - \frac{1}{\epsilon} \mathcal{H}_2(Y),
    \end{split}\\[1.25ex]
    \begin{split}
        \delta \widehat{X}{}^{(1)}_{\phi,ab} &= - \frac{2}{3\epsilon} \Big[ \mathcal{H}_2(Y) \delta_{ab} - \frac{1}{2} \mathcal{H}_2(D,U) \delta_{1a} \delta_{1b} \Big],
    \end{split}\\[1.25ex]
    \begin{split}
        \delta \widehat{X}{}^{(1)}_{\phi^\dagger\phi B} &= - \frac{4}{3\epsilon} \Big[ \mathcal{H}_2(Y) - \frac{1}{2} \mathcal{H}_2(D,U) \Big],
    \end{split}\\[1.25ex]
    \begin{split}
        \delta \widehat{X}{}^{(1)}_{\phi\phi} &= - \frac{1}{3\epsilon} \mathcal{H}_2(L,D,\overline{U}),
    \end{split}\\[1.25ex]
    \begin{split}
        \delta Z^{(1)}_{\mu} &= - \frac{1}{\epsilon} \bigg\{ ( 1 - \xi_F ) \Big[ g_Y^2 \mathcal{Y}_S^2 + g_W^2 C_2(F_L) \Big] - 2 \big( 1 + N_L \big) \lambda \bigg\},
    \end{split}\\[1.25ex]
    \begin{split}
        \delta Z^{(1)}_{\lambda} &= \frac{1}{\epsilon} \bigg\{ 12 \lambda^2 - 2 ( 1 - \xi_F ) \Big[ g_Y^2 \mathcal{Y}_S^2 + g_W^2 C_2(F_L) \Big] \lambda\\
        &\qquad\quad + 3 g_Y^4 \mathcal{Y}_S^4 + 2 g_Y^2 g_W^2 \mathcal{Y}_S^2 C_2(F_L) + g_W^4 \big(C_2(F_L)\big)^2 - \mathcal{H}_4(Y) \bigg\}.
    \end{split}
\end{align}

\paragraph{Yukawa Sector:}
The divergent Yukawa 1-loop counterterm action is given in Eq.~\eqref{Eq:SM-S1_sct_Yukawa_expanded}. For its coefficients, we obtain
\begin{align}
    \begin{split}
        \delta Y^{(1)}_{l,IJ} &= - \frac{1}{\epsilon} \bigg\{ ( 1 - \xi_F ) \Big[ g_Y^2 \mathcal{Y}_S^2 + g_W^2 C_2(F_L) \Big] + ( 4 - \xi_F ) g_Y^2 \big( \mathcal{Y}^l_L \mathcal{Y}^l_R \big)_{22} \bigg\} Y_{l,IJ},
    \end{split}\\[1.25ex]
    \begin{split}
        \delta Y^{(1)}_{d,IJ} &= - \frac{1}{\epsilon} \bigg\{ 
        ( 1 - \xi_F ) \Big[ g_Y^2 \mathcal{Y}_S^2 + g_W^2 C_2(F_L) \Big] \delta_{IK}\\ 
        &\qquad\quad + ( 4 - \xi_F ) \Big[ g_Y^2 \big( \mathcal{Y}^q_L \mathcal{Y}^q_R \big)_{22} + g_s^2 C_2(F_s) \Big] \delta_{IK}
        + \big( Y_u Y_u^\dagger \big)_{IK} \bigg\} Y_{d,KJ} ,
    \end{split}\\[1.25ex]
    \begin{split}
        \delta Y^{(1)}_{u,IJ} &= - \frac{1}{\epsilon} \bigg\{ ( 1 - \xi_F ) \Big[ g_Y^2 \mathcal{Y}_S^2 + g_W^2 C_2(F_L) \Big] \delta_{IK}\\
        &\qquad\quad 
        + ( 4 - \xi_F ) \Big[ g_Y^2 \big( \mathcal{Y}^q_L \mathcal{Y}^q_R \big)_{11} + g_s^2 C_2(F_s) \Big] \delta_{IK}
        + \big( Y_d Y_d^\dagger \big)_{IK} \bigg\} Y_{u,KJ}.
    \end{split}
\end{align}

\paragraph{Ghosts and BRST Transformations:}
The singular 1-loop counterterms of the ghost sector are given in Eq.~\eqref{Eq:SM-S1_sct_ghost_expanded}, and those of the BRST transformations in Eq.~\eqref{Eq:SM-S1_sct_ext_expanded}.
These counterterms are governed by the coefficients,
\begin{align}
    \begin{split}
        \delta Z^{(1)}_{c_W} &= \frac{2 + \xi_F}{4\epsilon} \, g_W^2 C_2(G_L),
    \end{split}\\[1.25ex]
    \begin{split}
        \delta Z^{(1)}_{c_G} &= \frac{2 + \xi_F}{4\epsilon} \, g_s^2 C_2(G_s),
    \end{split}\\[1.25ex]
    \begin{split}
        \delta Z^{(1)}_{\overline{c}Wc} &= - \frac{1 - \xi_F}{2\epsilon} \, g_W^2 C_2(G_L),
    \end{split}\\[1.25ex]
    \begin{split}
        \delta Z^{(1)}_{\overline{c}Gc} &= - \frac{1 - \xi_F}{2\epsilon} \, g_s^2 C_2(G_s).
    \end{split}
\end{align}

\subsection{Coefficients of Finite Symmetry-Restoring 1-Loop Counterterms} \label{App:Finite_CountertermCoeffs_Expanded}

Here, we provide the coefficients of the finite symmetry-restoring 1-loop contributions.

\paragraph{Fermionic Sector:}
The finite symmetry-restoring 1-loop lepton counterterms of Eq.~\eqref{Eq:SM-S1_fct_Lepton_expanded} are governed by 
\begin{align}
    \begin{split}
        \delta F^{(1)}_{l,ab} &= - c_{\mathrm{QED}} \, g_Y^2 \bigg[ \frac{\xi_F}{3} \big( \mathcal{Y}^l_L \mathcal{Y}^l_R \big)_{ab} - c_{\mathrm{QED}} \frac{3 - \xi_F}{3} \big( \mathcal{Y}^l_R \mathcal{Y}^l_R \big)_{ab} \bigg],
    \end{split}\\[1.25ex]
    \begin{split}
        \delta F^{(1)}_{\overline{l}Bl,R,ab} &= 
        g_Y^2 \bigg[ ( 1 + c_{\mathrm{QED}}^2 ) \frac{6 - \xi_F}{6} \big( \mathcal{Y}^l_L \mathcal{Y}^l_R \big)_{ab} 
        - b_\mathrm{QED} \big( \mathcal{Y}^l_R \mathcal{Y}^l_R \big)_{ab} \bigg],
    \end{split}\\[1.25ex]
    \begin{split}
        \delta F^{(1)}_{\overline{l}Wl,R} &= g_Y^2 \bigg[ \frac{6 - \xi_F}{6} + c_{\mathrm{QED}} \bigg( \frac{\xi_F}{3} + c_{\mathrm{QED}} \frac{6 - \xi_F}{6} \bigg) \bigg],
    \end{split}\\[1.25ex]
    \begin{split}
        \delta F^{(1)}_{\overline{l}Bl,L,ab,IJ} &= 
        - \bigg\{ 
        \frac{6 - \xi_F}{6} 
        \Big[ g_Y^2 \mathcal{Y}_l^2 + g_W^2 C_2(F_L) \Big] \mathcal{Y}^l_{L,ab}
        - a_\mathrm{QED} \, g_Y^2 \big( \mathcal{Y}^l_L \mathcal{Y}^l_L \mathcal{Y}^l_R \big)_{ab}\\
        &\qquad
        - c_{\mathrm{QED}} \frac{2 - c_{\mathrm{QED}}}{6} \xi_F \, g_Y^2 \big( \mathcal{Y}^l_L \mathcal{Y}^l_R \mathcal{Y}^l_R \big)_{ab}
        - c_{\mathrm{QED}}^2 \frac{6 - \xi_F}{6} g_Y^2 \big( (\mathcal{Y}^l_R)^3 \big)_{ab}\\
        &\qquad
        - \frac{6 - \xi_F}{6} g_W^2 \big( t^B_L \mathcal{Y}^l_R t^B_L \big)_{ab} \bigg\} \delta_{IJ}
        + \frac{1}{2} \mathcal{Y}_S \, \delta_{ab} \big( Y_l Y_l^\dagger \big)_{IJ},
    \end{split}\\[1.25ex]
    \begin{split}
        \delta F^{(1),A}_{\overline{l}Wl,L,ab,IJ} &= - \bigg\{ \frac{6 - \xi_F}{6} \Big[ g_Y^2 \mathcal{Y}_l^2 + g_W^2 C_2(F_L) \Big] t^A_{L,ab}\\
        &\qquad 
        + c_{\mathrm{QED}} \frac{\xi_F}{3} g_Y^2 \Big[ \big( \mathcal{Y}^l_L \mathcal{Y}^l_R t^A_L \big)_{ab} + \big( \mathcal{Y}^l_L t^A_L \mathcal{Y}^l_R \big)_{ab} + \frac{c_{\mathrm{QED}}}{2} \big( \mathcal{Y}^l_R t^A_L \mathcal{Y}^l_R \big)_{ab} \Big] \bigg\} \delta_{IJ}\\
        &\quad\, + \frac{1}{2} t^A_{L,22} \, \delta_{ab} \big( Y_l Y_l^\dagger \big)_{IJ},
    \end{split}
\end{align}
and for the quark 1-loop counterterms of Eq.~\eqref{Eq:SM-S1_fct_Quark_expanded}, the coefficients read
\begin{align}
    \begin{split}
        \delta F^{(1)}_{q,ab} &= - c_{\mathrm{QED}} \, g_Y^2 \bigg[ \frac{\xi_F}{3} \big( \mathcal{Y}^q_L \mathcal{Y}^q_R \big)_{ab} - c_{\mathrm{QED}} \frac{3 - \xi_F}{3} \big( \mathcal{Y}^q_R \mathcal{Y}^q_R \big)_{ab} \bigg],
    \end{split}\\[1.25ex]
    \begin{split}
        \delta F^{(1)}_{\overline{q}Bq,R,ab} &= 
        g_Y^2 \bigg[ ( 1 + c_{\mathrm{QED}}^2 ) \frac{6 - \xi_F}{6} \big( \mathcal{Y}^q_L \mathcal{Y}^q_R \mathcal{Y}^q_R \big)_{ab}
        - b_\mathrm{QED} \big( ( \mathcal{Y}^q_R )^3 \big)_{ab} \bigg]\\
        &\quad + u_{\mathrm{QCD}} \, g_s^2 C_2(F_s) \Big[ \mathcal{Y}^q_{L,ab} - \mathcal{Y}^q_{R,ab} \Big],
    \end{split}\\[1.25ex]
    \begin{split}
        \delta F^{(1),A}_{\overline{q}Wq,R,ab} &= u_{\mathrm{QED}} \, g_Y^2 \big( \mathcal{Y}^q_R t^A_L \mathcal{Y}^q_R \big)_{ab} + u_{\mathrm{QCD}} \, g_s^2 C_2(F_s) \, t^A_{L,ab},
    \end{split}\\[1.25ex]
    \begin{split}
        \delta F^{(1)}_{\overline{q}Bq,L,ab,IJ} &= 
        - \bigg\{
        \frac{6 - \xi_F}{6} \Big[ g_Y^2 \mathcal{Y}_q^2 + g_W^2 C_2(F_L) \Big] \mathcal{Y}^q_{L,ab}
        - a_\mathrm{QED} \, g_Y^2 \big( \mathcal{Y}^q_L \mathcal{Y}^q_L \mathcal{Y}^q_R \big)_{ab}\\
        &\qquad
        - c_{\mathrm{QED}}  \frac{2 - c_{\mathrm{QED}}}{6} \xi_F \, g_Y^2 \big( \mathcal{Y}^q_L \mathcal{Y}^q_R \mathcal{Y}^q_R \big)_{ab}
        - c_{\mathrm{QED}}^2 \frac{6 - \xi_F}{6} \, g_Y^2 \big( (\mathcal{Y}^q_R)^3 \big)_{ab}\\
        &\qquad
        - \frac{6 - \xi_F}{6} g_W^2 \big( t^B_L \mathcal{Y}^q_R t^B_L \big)_{ab}
        + u_{\mathrm{QCD}} \, g_s^2 C_2(F_s) \Big[ \mathcal{Y}^q_{L,ab} - \mathcal{Y}^q_{R,ab} \Big] 
        \bigg\} \delta_{IJ}\\
        &\quad\, + \frac{1}{2} \mathcal{Y}_S \, \delta_{ab} \Big[ \big( Y_d Y_d^\dagger \big)_{IJ} - \big( Y_u Y_u^\dagger \big)_{IJ} \Big],
    \end{split}\\[1.25ex]
    \begin{split}
        \delta F^{(1),A}_{\overline{q}Wq,L,ab,IJ} &= 
        - \bigg\{ 
        \frac{6 - \xi_F}{6} \Big[ g_Y^2 \mathcal{Y}_q^2 + g_W^2 C_2(F_L) \Big] t^A_{L,ab}
        + u_{\mathrm{QCD}} \, g_s^2 C_2(F_s) \, t^A_{L,ab}\\
        &\qquad + c_{\mathrm{QED}} \, \frac{\xi_F}{3} g_Y^2 \Big[ \big( \mathcal{Y}^q_L \mathcal{Y}^q_R t^A_L \big)_{ab} + \big( \mathcal{Y}^q_L t^A_L \mathcal{Y}^q_R \big)_{ab} + \frac{c_{\mathrm{QED}}}{2} \big( \mathcal{Y}^q_R t^A_L \mathcal{Y}^q_R \big)_{ab} \Big] \bigg\} \delta_{IJ}\\
        &\quad\, + \frac{1}{2} t^A_{L,22} \, \delta_{ab} \Big[ \big( Y_d Y_d^\dagger \big)_{IJ} - \big( Y_u Y_u^\dagger \big)_{IJ} \Big],
    \end{split}\\[1.25ex]
    \begin{split}
        \delta F^{(1)}_{\overline{q}Gq,ab} &= - c_{\mathrm{QED}} \, g_Y^2 \bigg[ \frac{\xi_F}{3} \big( \mathcal{Y}^q_L \mathcal{Y}^q_R \big)_{ab} - c_{\mathrm{QED}} \frac{3 - \xi_F}{3} \big( \mathcal{Y}^q_R \mathcal{Y}^q_R \big)_{ab} \bigg]\\
        &\quad - w_{\mathrm{QCD}} \, g_s^2 C_2(G_s) \delta_{ab}.
    \end{split}
\end{align}
Here, we used $a_\mathrm{QED}$, $b_\mathrm{QED}$, $u_\mathrm{QED/QCD}$, $w_\mathrm{QCD}$ and $k_\mathrm{QCD}$ defined in Eqs.~\eqref{Eq:SM-a_QED-b_QED}--\eqref{Eq:SM-k_QCD} to simplify the resulting expressions of the fermionic counterterm coefficients.

\paragraph{Gauge Sector:}
For the gauge sector, the coefficients of the finite symmetry-restoring 1-loop counterterms presented in Eq.~\eqref{Eq:SM-S1_fct_gauge_expanded} read
\begin{align}
    \begin{split}
        \delta F^{(1)}_{B} &= - \frac{N_f}{3} g_Y^2 \Big\{ \mathrm{Tr} \Big[ \big( \mathcal{Y}^l_L - \mathcal{Y}^l_R \big)^2 \Big] + N_c \, \mathrm{Tr} \Big[ \big( \mathcal{Y}^q_L - \mathcal{Y}^q_R \big)^2 \Big] \Big\},
    \end{split}\\[1.25ex]
    \begin{split}
        \delta F^{(1),A}_{BW} &= \frac{N_f}{3} g_Y g_W \Big\{ \mathrm{Tr} \big( \mathcal{Y}^l_R t^A_L \big) + N_c \, \mathrm{Tr} \big( \mathcal{Y}^q_R t^A_L \big) \Big\},
    \end{split}\\[1.25ex]
    \begin{split}
        \delta F^{(1)}_{W} &= - \frac{N_f}{3} ( 1 + N_c ) g_W^2 S_2(F_L),
    \end{split}\\[1.25ex]
    \begin{split}
        \delta F^{(1)}_{4B} &= \frac{2 N_f}{3} g_Y^2 \Big\{ \mathrm{Tr} \Big[ \big( \mathcal{Y}^l_L - \mathcal{Y}^l_R \big)^4 \Big] + N_c \, \mathrm{Tr} \Big[ \big( \mathcal{Y}^q_L - \mathcal{Y}^q_R \big)^4 \Big] \Big\},
    \end{split}\\[1.25ex]
    \begin{split}
        \delta F^{(1),A}_{BBBW} &= \frac{N_f}{3} g_Y^3 g_W \Big\{ \mathrm{Tr} \Big[ \big( \mathcal{Y}^l_L - \mathcal{Y}^l_R \big)^3 t^A_L \Big] + N_c \, \mathrm{Tr} \Big[ \big( \mathcal{Y}^q_L - \mathcal{Y}^q_R \big)^3 t^A_L \Big] \Big\},
    \end{split}\\[1.25ex]
    \begin{split}
        \delta F^{(1),AB}_{BBWW\text{-1}} &= \frac{N_f}{6} g_Y^2 g_W^2 \Big\{ 
        \mathrm{Tr} \big[ \big( \mathcal{Y}^l_L - \mathcal{Y}^l_R \big) t^A_L \big( \mathcal{Y}^l_L - \mathcal{Y}^l_R \big) t^B_L \big]\\
        &\qquad\qquad\quad + N_c \, \mathrm{Tr} \big[ \big( \mathcal{Y}^q_L - \mathcal{Y}^q_R \big) t^A_L \big( \mathcal{Y}^q_L - \mathcal{Y}^q_R \big) t^B_L \big] \Big\},
    \end{split}\\[1.25ex]
    \begin{split}
        \delta F^{(1),AB}_{BBWW\text{-2}} &= \frac{N_f}{3} g_Y^2 g_W^2 \Big\{ 2 \, \mathrm{Tr} \Big[ \big( \mathcal{Y}^l_L - \mathcal{Y}^l_R \big)^2 t^A_L t^B_L \Big] - \mathrm{Tr} \Big[ \big( \mathcal{Y}^l_L - \mathcal{Y}^l_R \big) t^A_L \big( \mathcal{Y}^l_L - \mathcal{Y}^l_R \big) t^B_L \Big]\\
        &\qquad\quad + N_c \Big( 2 \, \mathrm{Tr} \Big[ \big( \mathcal{Y}^q_L - \mathcal{Y}^q_R \big)^2 t^A_L t^B_L \Big] - \mathrm{Tr} \Big[ \big( \mathcal{Y}^q_L - \mathcal{Y}^q_R \big) t^A_L \big( \mathcal{Y}^q_L - \mathcal{Y}^q_R \big) t^B_L \Big] \Big) \Big\},
    \end{split}\\[1.25ex]
    \begin{split}
        \delta F^{(1),ABC}_{BWWW} &= \frac{N_f}{3} g_Y g_W^3 \Big\{ 2 \, \mathrm{Tr} \Big[ \big( \mathcal{Y}^l_L - \mathcal{Y}^l_R \big) t^A_L t^C_L t^B_L \Big] - \mathrm{Tr} \Big[ \big( \mathcal{Y}^l_L - \mathcal{Y}^l_R \big) t^A_L t^B_L t^C_L \Big]\\
        &\qquad\quad + N_c \Big( 2 \, \mathrm{Tr} \Big[ \big( \mathcal{Y}^q_L - \mathcal{Y}^q_R \big) t^A_L t^C_L t^B_L \Big] - \mathrm{Tr} \Big[ \big( \mathcal{Y}^q_L - \mathcal{Y}^q_R \big) t^A_L t^B_L t^C_L \Big] \Big) \Big\},
    \end{split}\\[1.25ex]
    \begin{split}
        \delta F^{(1),ABCD}_{4W} &= - \frac{N_f}{6} ( 1 + N_c ) g_W^2 S_2(F_L) \Big[ 5 \, \mathrm{Tr} \big( t^A_L t^B_L t^C_L t^D_L \big) - 6 \, \mathrm{Tr} \big( t^A_L t^B_L t^D_L t^C_L \big) \Big].
    \end{split}
\end{align}

\paragraph{Higgs Sector:}
The finite symmetry-restoring counterterm action of the Higgs sector is given in Eq.~\eqref{Eq:SM-S1_fct_Higgs_expanded}.
The coefficients of the counterterms that preserve global hypercharge conservation read
\begin{align}
    \begin{split}
        \delta F^{(1)}_{\phi,ab} &= \frac{1}{6} \, \mathcal{H}_2(D,U) \, \sigma^3_{ab},
    \end{split}\\[1.25ex]
    \begin{split}
        \delta F^{(1)}_{B,-,ab} &= - \frac{2}{3} g_Y \mathcal{Y}_S \Big[ \mathcal{H}_2(Y) \big(\sigma^-_{03}\big)_{ab} + \frac{1}{4} \mathcal{H}_2(D,U) \, \sigma^3_{ab} \Big],
    \end{split}\\[1.25ex]
    \begin{split}
        \delta F^{(1),A}_{W,-,ab} &= \frac{g_W}{6} \bigg\{ 
        \Big[ \mathcal{H}_2(Y) - \frac{1}{2} \mathcal{H}_2(D,U) \Big] \big\{ t^A_L, \sigma^3 \big\}_{ab} 
        - 2 \, \mathcal{H}_2(Y) \, t^A_{L,ab}
        \bigg\},
    \end{split}\\[1.25ex]
    \begin{split}
        \delta F^{(1),A}_{W,+,ab} &= - \frac{g_W}{4} \mathcal{H}_2(D,U) \big[ t^A_L, \sigma^3 \big]_{ab},
    \end{split}\\[1.25ex]
    \begin{split}
        \delta F^{(1)}_{\phi^\dagger\phi BB,ab} &= 2 g_Y^2 \mathcal{Y}_S^2 \Big[ h_1 \delta_{ab} - 2 h_2 \big( \sigma^-_{03} \big)_{ab} + 2 h_3 \sigma^3_{ab} + h_4 \Big( \delta_{ab} - 6 \big( \sigma^+_{03} \big)_{ab} \Big) \Big],
    \end{split}\\[1.25ex]
    \begin{split}
        \delta F^{(1),A}_{\phi^\dagger\phi BW,ab} &= 2 g_Y g_W \mathcal{Y}_S 
        \Big[ 
        2 h_1 t^A_{L,ab} + h_2 \Big( \big\{ t^A_L, \sigma^3 \big\} - 2 \sigma^3 t^A_L \sigma^3 \Big)_{ab} + 2 h_3 \big\{ t^A_L, \sigma^3 \big\}_{ab}\\
        &\qquad\qquad\quad + h_4 \Big( 2 t^A_L + \big\{ t^A_L, \sigma^3 \big\} + 2 \sigma^3 t^A_L \sigma^3 - 16 \sigma^+_{12} t^A_L \sigma^-_{12} \Big)_{ab} \Big],
    \end{split}\\[1.25ex]
    \begin{split}
        \delta F^{(1),AB}_{\phi^\dagger\phi WW,ab} &= 
        g_W^2 \Big[ h_1 \big\{ t^A_L, t^B_L \big\}_{ab}
        + h_2 \Big( t^A_L \sigma^3 t^B_L + t^B_L \sigma^3 t^A_L - \sigma^3 \big\{ t^A_L, t^B_L \big\} \sigma^3 \Big)_{ab}
        \\
        &\qquad
        + h_3 \Big( \sigma^3 \big\{ t^A_L, t^B_L \big\} + \big\{ t^A_L, t^B_L \big\} \sigma^3 \Big)_{ab}
        + h_4 \Big( \sigma^3 t^A_L \sigma^3 t^B_L + \sigma^3 t^B_L \sigma^3 t^A_L
        \\
        &\qquad\qquad
        + t^A_L \sigma^3 t^B_L \sigma^3 + t^B_L \sigma^3 t^A_L \sigma^3
        + \sigma^3 t^A_L \sigma^3 t^B_L \sigma^3 + \sigma^3 t^B_L \sigma^3 t^A_L \sigma^3
        \\
        &\qquad\qquad
        - 8 \, \sigma^+_{12} t^A_L \sigma^-_{03} t^B_L \sigma^-_{12} - 8 \, \sigma^+_{12} t^B_L \sigma^-_{03} t^A_L \sigma^-_{12} \Big)_{ab} \Big],
    \end{split}
\end{align}
with $\sigma_{12}^\pm$ and $\sigma_{03}^\pm$ given in Eq.~\eqref{Eq:Pauli-Combi-Matrices}, and $h_i$ introduced in Eq.~\eqref{Eq:Def-YukawaCoeffs-h_1--h_4}.
For the hypercharge-violating counterterms, we find
\begin{align}
    \begin{split}
        \delta F^{(1)}_{\phi\phi} &= - \frac{1}{6} \, \mathcal{H}_2(L,D,\overline{U}),
    \end{split}\\[1.25ex]
    \begin{split}
        \delta F^{(1)}_{\phi\phi B} &= - g_Y \mathcal{Y}_S \, \mathcal{H}_2(L,D,\overline{U}),
    \end{split}\\[1.25ex]
    \begin{split}
        \delta F^{(1),A}_{\phi\phi W,ab} &= - \frac{g_W}{2} \mathcal{H}_2(L,D,\overline{U}) \Big[ t^A_L - \frac{1}{2} \big\{ t^A_L, \sigma^3 \big\} + \sigma^+_{12} t^A_L \sigma^+_{12} - \sigma^+_{03} t^A_L \sigma^-_{03} \Big]_{ab}\\
        &\quad\, + g_W \varrho_0 \Big( \frac{1}{2} \big[ t^A_L, \sigma^3 \big] - \sigma^+_{12} t^A_L \sigma^+_{12} + \sigma^+_{03} t^A_L \sigma^-_{03} \Big)_{ab},
    \end{split}\\[1.25ex]
    \begin{split}
        \delta F^{(1)}_{\phi\phi BB} &= 2 g_Y^2 \mathcal{Y}_S^2 \, \varrho_2,
    \end{split}\\[1.25ex]
    \begin{split}
        \delta F^{(1),A}_{\phi\phi BW,ab} &= 2 g_Y g_W \mathcal{Y}_S \Big[ 
        2 \varrho_1 \big( \sigma^+_{12} t^A_L \sigma^+_{12} \big)_{ab}
        + 2 \varrho_2 \Big( \sigma^-_{03} t^A_L \sigma^-_{03} + \sigma^-_{03} t^A_L \sigma^+_{03} \Big)_{ab}
        \\
        &\qquad\qquad\quad\,\,
        + \varrho_3 \Big( \sigma^+_{12} t^A_L \sigma^+_{12} - \sigma^-_{03} t^A_L \sigma^+_{03} \Big)_{ab} \Big],
    \end{split}\\[1.25ex]
    \begin{split}
        \delta F^{(1),AB}_{\phi\phi WW,ab} &= g_W^2 \Big[ 
        \varrho_1 \Big( \sigma^+_{12} \big\{ t^A_L, t^B_L \big\} \sigma^+_{12} \Big)_{ab}
        + \varrho_2 \Big( \sigma^-_{03} \big\{ t^A_L, t^B_L \big\} \sigma^-_{03} + \sigma^-_{03} \big\{ t^A_L, t^B_L \big\} \sigma^+_{03} \Big)_{ab}\\
        &\qquad\,\, + \varrho_3 \Big( \sigma^+_{12} t^A_L \sigma^+_{12} t^B_L + \sigma^+_{12} t^B_L \sigma^+_{12} t^A_L - \sigma^-_{03} t^A_L \sigma^+_{03} t^B_L - \sigma^-_{03} t^B_L \sigma^+_{03} t^A_L \Big)_{ab} \Big],
    \end{split}
\end{align}
where $\varrho_i$ are defined in Eq.~\eqref{Eq:Def-YukawaCoeffs-varrho_0--varrho_3}.
The finite quartic breakings of the scalar sector are governed by
\begin{align}
    \begin{split}
        \delta F^{(1)}_{\phi^\dagger\phi\phi^\dagger\phi,abcd} &= \frac{1}{6} \Big[ \mathcal{H}_4^{(1)} \big( \sigma^+_{03} \big)_{ab} \big( \sigma^+_{03} \big)_{cd} + \mathcal{H}_4^{(2)} \big( \sigma^-_{03} \big)_{ab} \big( \sigma^-_{03} \big)_{cd} \Big],
    \end{split}\\[1.25ex]
    \begin{split}
        \delta F^{(1)}_{\phi^\dagger\phi\phi\phi,abcd} &= \frac{1}{3} \Big[ \mathcal{H}_4^{(3)} - \mathcal{H}_4^{(4)} \Big] \big( \sigma^+_{03} \big)_{ab} \big( \sigma^-_{03} \big)_{cd}
        + \frac{1}{3} \mathcal{H}_4^{(3)} \big( \sigma^-_{03} \big)_{ab} \big( \sigma^-_{03} \big)_{cd},
    \end{split}\\[1.25ex]
    \begin{split}
        \delta F^{(1)}_{\phi\phi\phi\phi} &= \frac{1}{12} \, \mathcal{H}_4(L,D,\overline{U}),
    \end{split}
\end{align}
with $\mathcal{H}_4^{(i)}$ as provided in Eqs.~\eqref{Eq:Def-YukawaCoeffs-H4_1}--\eqref{Eq:Def-YukawaCoeffs-H4_4}.

\paragraph{Yukawa Sector:}
The finite symmetry-restoring 1-loop Yukawa counterterms are provided in Eq.~\eqref{Eq:SM-S1_fct_Yukawa_expanded}.
For the global hypercharge-conserving counterterms, the $\delta F_Y$-type counterterms, we obtain
\begin{align}
    \begin{split}
        \delta F^{l,(1)}_{Y,ab} &= - g_Y^2 \bigg( \frac{3 + \xi_F}{12} + c_{\mathrm{QED}} \frac{\xi_F}{6} \bigg) \big( \mathcal{Y}^l_R \mathcal{Y}^l_R \big)_{ab} ,
    \end{split}\\[1.25ex]
    \begin{split}
        \delta F^{d,(1)}_{Y,ab} &= g_Y^2 \bigg[ \bigg( \frac{3 + \xi_F}{12} + c_{\mathrm{QED}} \frac{\xi_F}{6} \bigg) \big( \mathcal{Y}^q_R \mathcal{Y}^q_R \big)_{ab} + ( 1 - \xi_F ) \big( \mathcal{Y}^q_L \mathcal{Y}^q_R \big)_{ab} \bigg] ,
    \end{split}\\[1.25ex]
    \begin{split}
        \delta F^{u,(1)}_{Y,ab} &= - g_Y^2 \bigg[ \frac{9 - \xi_F}{6} \big( \mathcal{Y}^q_L \mathcal{Y}^q_R \big)_{ac} + c_{\mathrm{QED}} \, \frac{\xi_F}{3} \big( \mathcal{Y}^q_R \mathcal{Y}^q_R \big)_{ac} \bigg] \varepsilon_{cb} ,
    \end{split}
\end{align}
while the hypercharge-violating $\delta G_Y$-type counterterms are governed by
\begin{align}
    \begin{split}
        \delta G^{l,(1)}_{Y,abc,IJ} &= g_Y^2 \bigg[ \frac{6 - \xi_F}{6} + c_{\mathrm{QED}} \, \frac{\xi_F}{2} + c_{\mathrm{QED}}^2 \frac{6 - \xi_F}{3} \bigg] \big( \mathcal{Y}^l_R \mathcal{Y}^l_R \big)_{cb} \delta_{2a} \big( Y_l^\dagger \big)_{IJ},
    \end{split}\\[1.25ex]
    \begin{split}
        \delta G^{d,(1)}_{Y,abc,IJ} &= \bigg\{ g_Y^2 \bigg[ \bigg( \frac{6 - \xi_F}{3} - c_{\mathrm{QED}} \, \xi_F \bigg) \big( \mathcal{Y}^q_L \mathcal{Y}^q_R \big)_{cb} \delta_{2a}
        - \frac{1}{9} \bigg( 1 + \frac{2}{3} c_{\mathrm{QED}} \xi_F \bigg) \delta_{cb} \delta_{2a}\\
        &\qquad\quad + c_{\mathrm{QED}} \bigg( \frac{\xi_F}{3} - c_{\mathrm{QED}} \frac{6 - \xi_F}{3} \bigg) \big( \mathcal{Y}^q_R \mathcal{Y}^q_R \big)_{cb} \delta_{2a}
        + \frac{1}{9} \bigg( 1 + 2 c_{\mathrm{QED}}^2 \frac{6 - \xi_F}{3} \bigg) \delta_{ac} \delta_{2b} \bigg] \\
        &\qquad + g_s^2 C_2(F_s) \bigg[ \bigg( \frac{12 - \xi_F}{3} + \frac{2}{3} c_{\mathrm{QCD}} \xi_F - c_{\mathrm{QCD}}^2 \frac{6 - \xi_F}{3} \bigg) \delta_{2a} \delta_{bc}\\
        &\qquad\qquad\qquad\quad - 2 \bigg( 1 - c_{\mathrm{QCD}}^2 \frac{6 - \xi_F}{3} \bigg) \delta_{ac} \delta_{2b} \bigg] \bigg\} \big( Y_d^\dagger \big)_{IJ}\\
        &\quad + \frac{1}{2} \delta_{1a} \delta_{2b} \delta_{1c} \big( Y_d Y_u Y_u \big)_{IJ}
        - \frac{1}{2} \Big( \delta_{1a} \delta_{2b} \delta_{1c} - \delta_{2a} \delta_{bc} \Big) \big( Y_u Y_u Y_d \big)_{IJ},
    \end{split}\\[1.25ex]
    \begin{split}
        \delta G^{u,(1)}_{Y,abc,IJ} &= \bigg\{ g_Y^2 \bigg[ \bigg( \frac{6 - \xi_F}{3} + c_{\mathrm{QED}} \frac{3 + 5\xi_F}{3} + 4 \, c_{\mathrm{QED}}^2 \frac{6 - \xi_F}{3} \bigg) \big( \mathcal{Y}^q_L \mathcal{Y}^q_R \big)_{ce} \varepsilon_{eb} \delta_{1a}
        \\
        &\qquad\quad
        - c_{\mathrm{QED}}^2 \frac{15 - 2\xi_F}{3} \big( \mathcal{Y}^q_R \mathcal{Y}^q_R \big)_{ce} \varepsilon_{eb} \delta_{1a}
        \\
        &\qquad\quad
        - c_{\mathrm{QED}} \big( \mathcal{Y}^q_L \mathcal{Y}^q_R \big)_{ac} \delta_{2b}
        + c_{\mathrm{QED}}^2 \frac{15 - 2\xi_F}{3} \big( \mathcal{Y}^q_R \mathcal{Y}^q_R \big)_{ac} \delta_{2b} \bigg]\\
        &\qquad + g_s^2 C_2(F_s) \bigg[ \bigg( \frac{6 - \xi_F}{3} + c_{\mathrm{QCD}} \frac{3 + 2\xi_F}{3} - c_{\mathrm{QCD}}^2 \frac{9 - \xi_F}{3} \bigg) \varepsilon_{cb} \delta_{1a}
        \\
        &\qquad\qquad\qquad\quad
        - c_{\mathrm{QCD}} \bigg( 1 - c_{\mathrm{QCD}} \frac{15 - 2\xi_F}{3} \bigg) \delta_{ac} \delta_{2b} \bigg] \bigg\} \big( Y_u^\dagger \big)_{IJ}\\
        &\quad - \frac{1}{2} \delta_{1a} \delta_{1b} \delta_{2c} \big( Y_u Y_d Y_d \big)_{IJ}
        + \frac{1}{2} \delta_{1a} \delta_{2b} \delta_{1c} \big( Y_d Y_d Y_u \big)_{IJ} \ .
    \end{split}
\end{align}

\paragraph{Ghosts and BRST Transformations:}
The finite symmetry-restoring 1-loop counterterms of the BRST transformations in Eq.~\eqref{Eq:SM-S1_fct_ext_expanded} are governed by the coefficients
\begin{align}
    \begin{split}
        \delta F^{(1)}_{R,W} &= \frac{1 - \xi_F}{4} \, g_W^2 C_2(G_L),
    \end{split}\\[1.25ex]
    \begin{split}\label{Eq:FCT-dFRG1-ExternalFermionSource}
        \delta F^{(1)}_{R,G} &= \frac{1 - \xi_F}{4} \, g_s^2 C_2(G_s).
    \end{split}
\end{align}

\end{appendices}

\printbibliography

\end{document}